\documentclass[JMPS,onecolumn,noshowpacs,superscriptaddress,groupedaddress]{elsarticle}  
\usepackage[paperwidth=210mm,paperheight=297mm,centering,hmargin=2.5cm,vmargin=2.5cm]{geometry}

\usepackage{graphicx}  
\usepackage{dcolumn}   
\usepackage{bm}        
\usepackage{amssymb}   
\usepackage{amsmath}
\usepackage{float}
\usepackage{xcolor}
\usepackage{mathtools}    
\usepackage{mathrsfs}  
\usepackage{fourier}

\newlength{\bibitemsep}
\newlength{\bibparskip}
\let\oldthebibliography\thebibliography
\renewcommand\thebibliography[1]{%
  \oldthebibliography{#1}%
  \setlength{\parskip}{\bibitemsep}%
  \setlength{\itemsep}{\bibparskip}%
}
\renewcommand{\thefigure}{\arabic{figure}}

  \newcommand{\be}{\begin{equation}}
\newcommand{\ee}{\end{equation}}
\newcommand{\ba}{\begin{array}}
\newcommand{\ea}{\end{array}}

\usepackage{float}
\usepackage{subfloat}

\begin{document}




\title{
Quantized plastic deformation
}

\bigskip 

\author{ N. Perchikov and  L. Truskinovsky}
\address{\small 
Physique et M\'ecanique des Milieux H\'et\'erog\`enes, 
 ESPCI
 , 
CNRS ,\\  Quai Saint-Bernard 9, 75005  
 Paris, 
 France} 

%

\begin{abstract}
In engineering crystal plasticity inelastic mechanisms correspond to tensorial zero-energy valleys in the space of macroscopic strains. The flat nature of such valleys  is in contradiction with the fact that plastic slips, mimicking lattice-invariant shears, are inherently discrete. A reconciliation has recently been achieved in the mesoscopic tensorial model (MTM) of crystal plasticity, which introduces periodically modulated energy valleys while also capturing in a geometrically exact way the crystallographically-specific aspects of plastic slips. In this paper, we extend the MTM framework, which in its original form  had  the  appearance of a discretized nonlinear elasticity theory, by  explicitly introducing  the concept of plastic deformation. The ensuing  model contains a novel matrix-valued  spin variable, representing the quantized  plastic distortion, whose rate-independent evolution can be described by a discrete (quasi-)automaton. The proposed  reformulation of the MTM leads to a considerable computational speedup  associated with the use of a robust and efficient hybrid Gauss-Newton--Cauchy energy minimization algorithm. To illustrate the effectiveness of the new approach, we present  a detailed case-study focusing on the aspects of crystal plasticity that are beyond reach for the classical continuum theory. Thus, we provide compelling evidence that the re-formulated MTM  is fully adequate to deal with the intermittency of plastic response under quasi-static loading. In particular,  our numerical experiments  show that the statistics of dislocational avalanches, associated with plastic yield in 2D square crystals, exhibits a power-law tail with a critical exponent matching the value predicted by general theoretical considerations and also independently observed in discrete-dislocation-dynamics (DDD) simulations.

 \smallskip 
\smallskip 
\smallskip 
\smallskip 
\smallskip 
\smallskip
\end{abstract}


\maketitle

\section{Introduction}
\label{sec1}

Recent experiments  provided  evidence  that, at least in some types of crystals, quasi-static plastic flows involve intermittent, power-law distributed dislocation avalanches which generate   scale-free dislocational patterns \cite{Weiss2021}.  The associated plastic deformation   cannot be simply homogenized, as it is not a sequence of uncorrelated events of similar size  and is instead a highly   correlated process spanning a broad range  of spatial and temporal scales  \cite{Weiss2015}. 

To rationalize this complex behavior, one has to go beyond the conventional continuum models   and account adequately for the inherently  discrete  nature of plastic flows. The latter  is  neglected in the macroscopic engineering crystal plasticity (CP), where inelastic deformations are modeled  by smooth  elastically neutral  mechanisms, operating in the space of macroscopic strains.

An alternative, strongly discrete perspective on crystal plasticity emerges  from the microscopic perspective  where  plastic deformation is  viewed as a collection of interdependent lattice-invariant shears. Such shears  represent  
cooperative displacements of  atoms on crystallographically specific lattice planes and   have a distinctly \emph{quantized} nature. 

Several modeling tools have been developed  to provide a lattice-scale description of crystal plasticity, ranging from microscopic molecular dynamics \cite{NatureMD} and crystal phase-field approaches \cite{Elder2004} to mesoscopic phase-field dislocation dynamics \cite{Koslowski2002}, discrete dislocation dynamics \cite{Devincre2001}, and continuum dislocation dynamics \cite{Zaiser2014}.  
An attempt to bridge the entire range of scales from microscopic to macroscopic, while fully  accounting for each of them, was made within the quasi-continuum framework \cite{Tadmor}. However, despite many successes in resolving specific sub-continuum features of crystal plasticity, see, for instance, \cite{Krebs}, the ability of the existing approaches to capture the correct statistics of plastic fluctuations, while also accounting for large lattice rotations and adequately representing the crystallographic nature of lattice-invariant shears, remains rather  limited.  

An efficient conceptual interpolation between the microscopic, dislocation-based, and the  macroscopic, continuum mechanics-based,  models, has been recently proposed in the form of the  mesoscopic tensorial model (MTM) of crystal plasticity \cite{Umut2011,Umut2012,Zhangetal,Roberta, BaggioST2023a,SalmanBBZGT21,BaggioST2023b}. It   operates with the macroscopic notions of stress and strain while capturing, in a geometrically exact way, the slip-generating mappings of crystal  lattices  onto themselves. 

Interestingly, while the microscopic models are based on   a system of ODEs and while the macroscopic models are formulated in terms of a system of PDEs,  {the} approach of MTM is  intermediate  as it operates with   mesoscopic discrete elements (DEs){.} The latter deform according to elastic constitutive {relations} while interacting through conventional mechanical forces as in models of granular and particulate materials {\cite{Cundall1979,Wang2022, Rorato2021,Guo2015}}. The idea is that the  deformation below a certain mesoscopic   length scale  can be  coarse{-}grained,  while still capturing correctly the   intermittency and  fractality  above the cut-off.

The MTM approach  can be  viewed as a far reaching generalization of the  `purely elastic' toy models of crystal plasticity \cite{Peierls1940size,Nabarro1947dislocations,Frenkel1939theory,Prandtl}. Using the  depinning-type friction  as the main prototype \cite{Popov2012}, such models  typically postulate  that the elastic energy landscape is wiggly and associate  dissipation with   snap-through instabilities, which typically  accompany  {quasi-static} driving in the  systems of this type \cite{Puglisi}. 

The development of the tensorial version of such `purely elastic' models was inspired by the pioneering insights of J. Ericksen {\cite{Ericksen1970,Ericksen1977}}, which were almost immediately endorsed by R. Hill {\cite{Hill1979}}.  The  MTM  represents  an important   step in the development of this class of models, offering their geometrically exact, finite-strain  generalization.  Its main achievement is in  turning prototypical schemes into fully comprehensive and technologically relevant  engineering tools.  In the spirit of  continuum  crystal plasticity, the MTM   links plastic deformation  with crystallographically-specific  valleys in the elastic-energy landscape, however, it replaces the `flat' valleys of the CP theory {by} `periodically modulated' ones.  Instead of being postulated, these valleys emerge in the MTM  as natural elements of the globally-periodic configurational energy landscape constructed to respect the tensorial symmetries of lattice-invariant shears {\cite{Ericksen,Boyer,Folkins,Wang,Waal,Parry1976,Parry1977,Parry, Pitteri1984, CZ,Pitteri2002}}. The downscaling from macro to meso scale  is   accompanied in the MTM by two types of discretization.

The first one concerns the \emph{configurational space} of strain tensors. It  involves the quantization of plastic deformation{,} which is natural in view of the implied periodicity of the associated  Landau-type elastic energy density characterized by {an} infinite number of equivalent energy wells. The second one concerns  the \emph{physical space} and amounts to the representation  of an inhomogeneously deformed macroscopic body  as a collection of homogeneously deformed mesoscopic  elements. This step is formally necessary as a regularization {since}   the configurational nonconvexity of the  energy density creates {an} unphysical degeneracy \cite{Fonseca}. 
  
Note that none of the emerging  discretization scales  is a part of classical continuum CP, which  can then be viewed as a macroscopic analogue of the MTM.    The implicit underlying  coarse-graining  comes  with the loss  of  information, in particular,  the short  range interactions  involved in topological transitions   have to be brought into CP through {\emph{ad hoc}} phenomenological assumptions  prescribing, for instance, the rate of dislocation nucleation and other dislocation reactions. Still, despite the underlined profound differences between {CP} and the MTM, we attempt in this paper to build a new conceptual bridge between the two theories.

%

 Our main idea is  the  re-introduction into the MTM of the crucial concept of plastic deformation. As we have already mentioned, in its original formulation, the MTM had the guise of a  `purely elastic' theory \cite{BaggioST2023a}. While this  may  create  an impression that the notion of plastic strain is completely {foreign} to the MTM, the concept of plastic strain has already appeared explicitly in the scalar version of the MTM in the form of an integer-valued  (quantized) order parameter \cite{Umut2011,Umut2012,Zhangetal}.  Interestingly, the  reason for its introduction  in such a restricted setting  was not conceptual but technical,  as it {enabled the} explicit reduction of the description of plastic flow to a simple integer-valued automaton. 
   
Here we generalize the idea of a quantized plastic strain, implicit in \cite{Umut2011,Umut2012,Zhangetal}, by   placing it in the fully tensorial MTM framework. We show that in the  finite strain   setting, plastic strain can be  naturally associated with a locally-defined integer-valued matrix, whose role is to select one among the infinitely many equivalent  wells of the globally periodic constitutive energy landscape. In this perspective, crystal plasticity emerges as a physical theory where the order parameter  is   a matrix-valued spin variable. The proposed {re-formulation} of the MTM  can be then viewed as a transition from the soft-spin, `purely elastic' description,  to the hard-spin,  elasto-plastic description.  

By enabling  explicit  access to plastic strain in a DE-based model, the proposed re-formulation of the MTM opens a possibility  to follow the   history of the quantized  lattice-invariant  deformations, which remains hidden  behind the  smoothness of macroscopic plastic flows.  The reference to the discreteness of slip  also creates a new link between the   molecular representation of lattice-invariant deformations and the engineering, continuum level, description of plastic fluctuations in terms of macroscopically observable variables. 

The most important technical advantage of  introducing the explicit distinction between elastic and plastic contributions to deformation is the possibility to effectively separate (in the case of quasi-static driving)  the \emph{continuous} elasticity problem from the \emph{discrete} plasticity problem.  We show that when  such  a  separation is complete, the   well-posed elastic problem can be solved `on the fly', allowing one to relegate the evolution of the plastic strain to a tensorial integer-valued  automaton. In other words, in the resulting `condensed' description {\cite{Junker2013,Carstensen2008,Jezdan2023}} the dissipation  is effectively removed from the continuous problem, which reduces to energy minimization, while  dissipation remains   associated only with a discrete  sequence of discontinuous elastic instabilities{.} Each of those is accompanied by a finite energy drop,    representing quantized advance  of plastic strain. The  associated updates of the automaton mimic discontinuous elastic-branch-switching events    associated, for instance, with  collective  bond breaking and attendant bond reforming.


It is important to stress that, outside the simplest scalar setting, the numerical algorithm solving the `condensed' MTM problem can be interpreted only as a  \emph{quasi-automaton}. While, indeed, it describes a sequence of load-driven updates of quantized plastic variables, each of those updates necessarily contains an embedded elastic energy-minimization step. Behind the incomplete separability of the elastic and plastic problems  lies the geometric nonlinearity of the elastic problem, which cannot be neglected due to the ubiquitous presence in elastoplastic flows of large rotations  {\cite{ BaggioST2023a, BaggioST2023b,Dafalias1998,Arminjon1993,Boyce1989,Levitas1998, Fathallah2019}}. 

Still, we show that  in the new  version of the MTM, the elastic energy density can  be chosen in such a way that the solution of the elastic problem is practically straightforward even if it cannot be expressed in terms of an explicit `elastic propagator' as in the   scalar  problem \cite{Umut2011,Umut2012,Zhangetal}. Therefore, even in the absence of a complete separation of elastic and plastic problems, a detailed estimate of the algorithmic complexity of the new approach suggests up to two orders of  magnitude reduction of the computational time   vis-a-vis   the original `purely elastic' approach.  As we show, {significant} acceleration potential of the proposed re-formulation of the MTM {lies} in the possibility to perform only local elastic {adjustments} in response to discrete plastic updates.  

To illustrate the effectiveness of the elasto-plastic version of the  MTM, we present in this paper a detailed case study of  the emergence of plastic yield in (almost) pristine  2D square  crystals subjected to homogeneous loading in a hard device. We present a series of numerical experiments  designed to reveal  the  aspects of crystal plasticity that are beyond reach for the classical CP approach and  our main interest  is in the critical nature of the emerging   plastic flow. The latter     is revealed through a peculiar  statistical signature of plastic fluctuations representing   dislocation avalanches. Specifically, our study  provides a definitive  evidence for the non-Gaussian character of plastic flows with the emergence of spatial and temporal scaling.

 In particular, we show that the statistics of the intermittent energy-dissipating dislocational avalanches accompanying plastic yield exhibits a power-law tail with a critical exponent matching the value predicted by some theoretical considerations and independently observed in discrete-dislocation-dynamics simulations. We also show that the spatial-distribution of slip zones is characterized by a non-integer fractal dimension, which, again, closely matches the value  anticipated by the general theory. The computed relation between the macroscopic stress and macroscopic {plastic} strain is found to be  well  approximated by the  empirical Johnson-Cook  strain-hardening scaling law  with a hardening exponent remarkably close to the one found experimentally for a material with similar parameters. 
 
The paper is organized as follows. In Section 2, we motivate our approach by considering an oversimplified zero-dimensional model. The detailed fully analytical study of this model  makes the main ideas of the  subsequent development explicit and transparent. The 2D version of the MTM, containing  continuum elastic  and  discrete plastic strain measures, is formulated in Sections 3, where we  also  address various aspects of the numerical implementation of the model. 
In Section 4, we present a case study of a fluctuating plastic flow in a   model crystal and rationalize the emerging scale-free statistics. Our results are summarized in Section 5.  Several appendixes contain discussions of a more technical nature.

\section{Preliminaries}
\label{sec2}

\paragraph{Continuum plasticity} 
The standard assumption in finite-strain plasticity is that the compatible deformation gradient $\textbf{F}=\nabla\textbf{y}^{\top}$ corresponding to the total deformation $\textbf{y}(\textbf{x})$ can be multiplicatively decomposed into a product of  (potentially incompatible)   elastic, $\textbf{F}_e$, and plastic, $\textbf{F}_p$ contributions, such that  {\cite{Clayton2010, Reina1,Roters2010}} $$\textbf{F}=\textbf{F}_e\textbf{F}_p .$$  The  objective elastic energy density per reference volume  
 can be then written in the form $\psi=\psi(\textbf{C}_e)$, where $\textbf{C}_e=\textbf{F}_e^{\top}\textbf{F}_e$ is the metric tensor.  If the loading is fixed and 
 the plastic distortion $\textbf{F}_p$ is known, the deformation $\textbf{y}(\textbf{x})$ can    be found by solving the equilibrium equations 
$$\nabla\cdot\textbf{P}^{\top}=0,$$ 
where
$
\textbf{P}=2\textbf{F}_e\frac{\partial\psi}{\partial\textbf{C}_e} \textbf{F}_p^{-\top}
$
{is} the (first) Piola-Kirchhoff stress. From the  equilibrium equations complemented by the appropriate boundary conditions one obtains the solution of the `elastic' part of the problem \cite{Simo2006,Lubliner2008,Han2012,Schroder2013}. 

The `plastic'   problem, which reduces to  finding   $\textbf{F}_p$, is solved concurrently with the `elastic' problem.  The corresponding equations are usually formulated in incremental form and involve   constitutive assumptions, which identify plastic `mechanisms'  while also specifying the corresponding yield conditions, flow rules and hardening  laws. The only fundamental constraint imposed on such  phenomenological assumptions is the incremental condition of the overall dissipativity of plastic deformation,
$\textbf{tr}(\textbf{P}^{\top}  \textbf{F}_e\dot{\textbf{F}}_p)\ge0$  {\cite{Roters2010,McDowell2018}}.

To {better} understand the implied separation of elastic and plastic problems, we observe that in such a continuum theory the elastic energy density can be   written in the  form $$\psi=\psi(\textbf{F}\textbf{F}_p^{-1}).$$ This representation suggests  that if  $\textbf{F}_p$ is compatible, the  distortions with $\textbf{F}=\textbf{F}_p$ appear as elastically `soft' modes.  In other words, the  elastic energy density of continuum CP would be degenerate along such tensorial directions.  The  associated   `plastic mechanisms' , which can be  prescribed by the ansatzes imposed on $\textbf{F}_p$,  are usually indeed chosen to be compatible, crystallographically-specific simple shears (rank-one directions in the space of deformation gradients) \cite{Asaro1983,Gurtin2000,McHugh2004}.

To ensure that  the plastic  flow along the implied  flat energy valleys, cutting through the configurational elastic energy landscape,  is  dissipative,  continuum theories of {the} CP type introduce an effective (dry) friction operating along the valleys  and  encapsulated in the corresponding yield thresholds and flow rules. Those are chosen to be necessarily compatible with the dissipativity inequality {\cite{Gurtin2000,Rice1971,Mielke2003, Petryk2005, Svendsen2010}}.

\paragraph{Ideas behind the MTM} To {avoid} redundant CP phenomenology without giving up the macroscopic concepts of stress and strain, the MTM approach transforms the rigid constitutive assumptions of CP regarding the `plastic mechanisms' {into} constraints of {a} `soft' nature emerging naturally from the geometrically necessary structural {properties} of the configurational energy landscape $\psi(\textbf{C})$. 

Moreover, in its `purely elastic' form the MTM anticipates that the  analogues of the conventional concepts of yield surface and flow rule are  generated automatically after the  configurational  landscape is prescribed. In particular,  it interprets the yield surface as an extended  corridor  in the configurational space where elastic instabilities in individual elastic elements are activated  as the system is driven away from an energy well. Similarly, it views the flow rule as a post-instability response aimed at the stabilization of the individual destabilized  elements in  new  energy wells.  In the MTM, it  is also connoted that such a stabilization  is guided by an overdamped dynamics of {the} viscous type.  In the case of quasi-static driving, the latter takes the form of continuous  local energy minimization with occasional discontinuous transitions between the neighboring energy wells. It is also anticipated that  the geometrically necessary  structure of the configurational energy landscape ensures that  all such  discontinuous  transitions are accompanied by energy loss. This means, in particular,  that  the dissipativity condition is always satisfied, despite the fact that in a quasi-static setting  the effective (normalized) viscosity is effectively  equal to zero. 
 
As far as the  inner structure of the energy valleys is concerned, the MTM approach anticipates that  the `frictional'  dissipation mechanisms, operating at  a  macro-scale,  can be modeled  by  the  periodic  modulation of the  elastic energy landscape inside the  valleys. Such a modulation is then viewed  as a signature of the meso-scale  description{,} which disappears in the process of macro-scopic  coarse graining \cite{Mielke2012}. 

The first step in the implementation {of} all these ideas  is to specify the configurational energy landscape $\psi(\textbf{C})${, representing,} in a geometrically precise way{,} the {relevant} lattice-invariant shears. The corresponding quantized strains    would then be represented by the    equivalent  energy minima  in such a  tensorial landscape. If the latter is constructed to respect the crystallographic symmetry of the crystal lattice, the modulated energy valleys, imitating `plastic mechanisms'{,} would  appear automatically{,} and plastic dissipation {would} emerge as a sheer consequence of the energy minimization \cite{Roberta}.

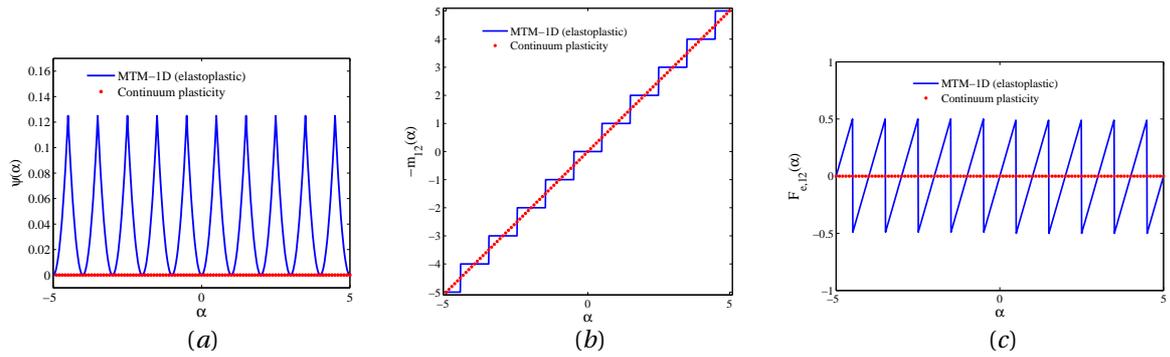
\begin{figure}[h!]
\centering
{
\includegraphics[scale=0.34,trim={0cm -0.5cm 0.5cm 0.0cm},clip]{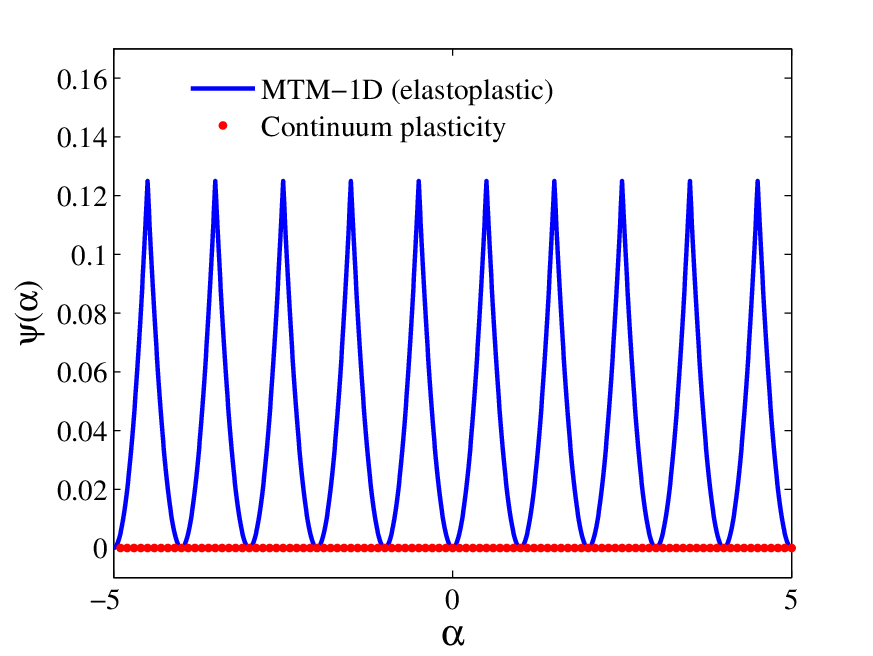} 
\includegraphics[scale=0.31,trim={2.5cm 0.0cm 2.5cm 0.5cm},clip]{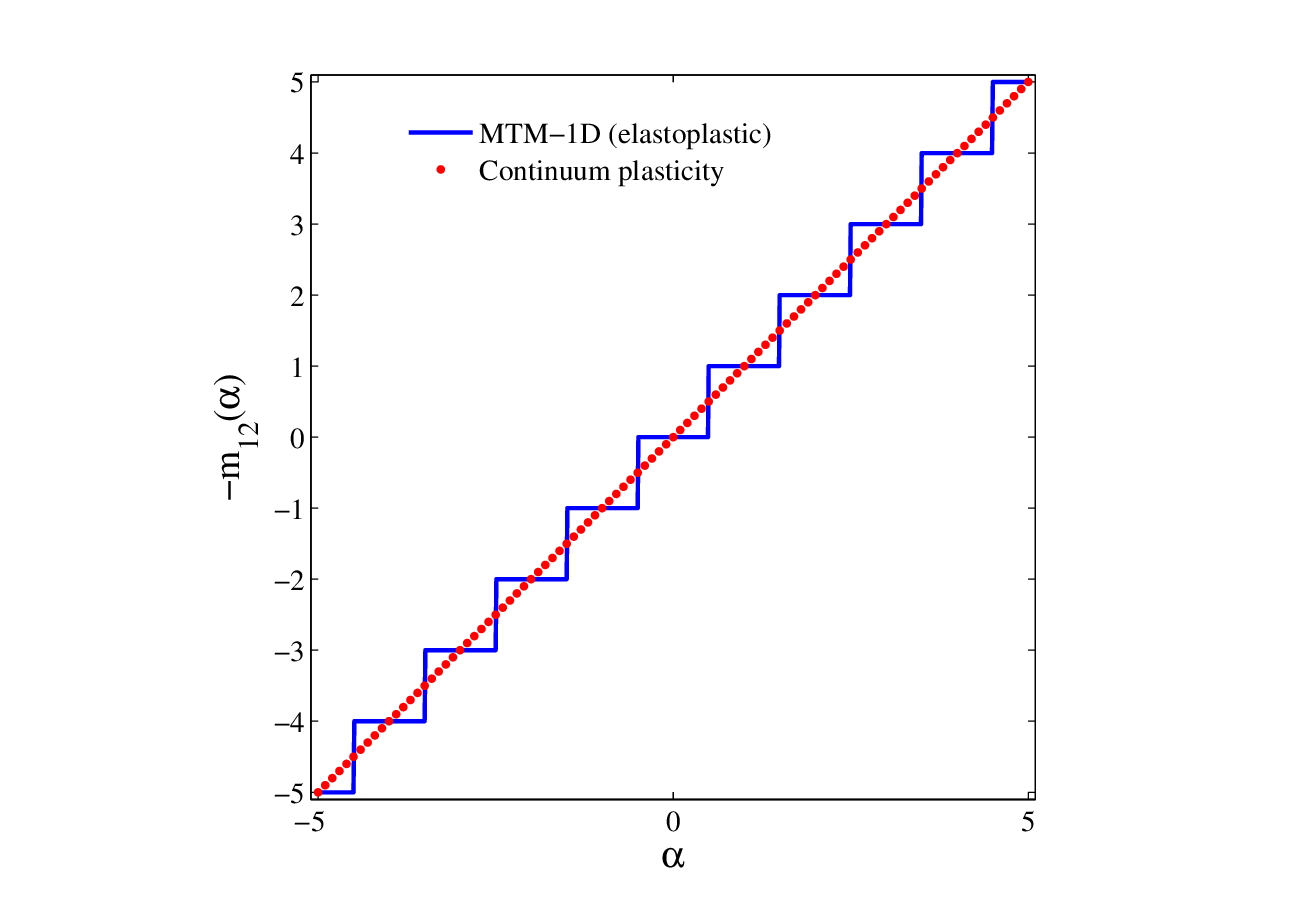}  
\includegraphics[scale=0.31 ,trim={2.0cm 1.0cm 2.5cm 0.5cm},clip]{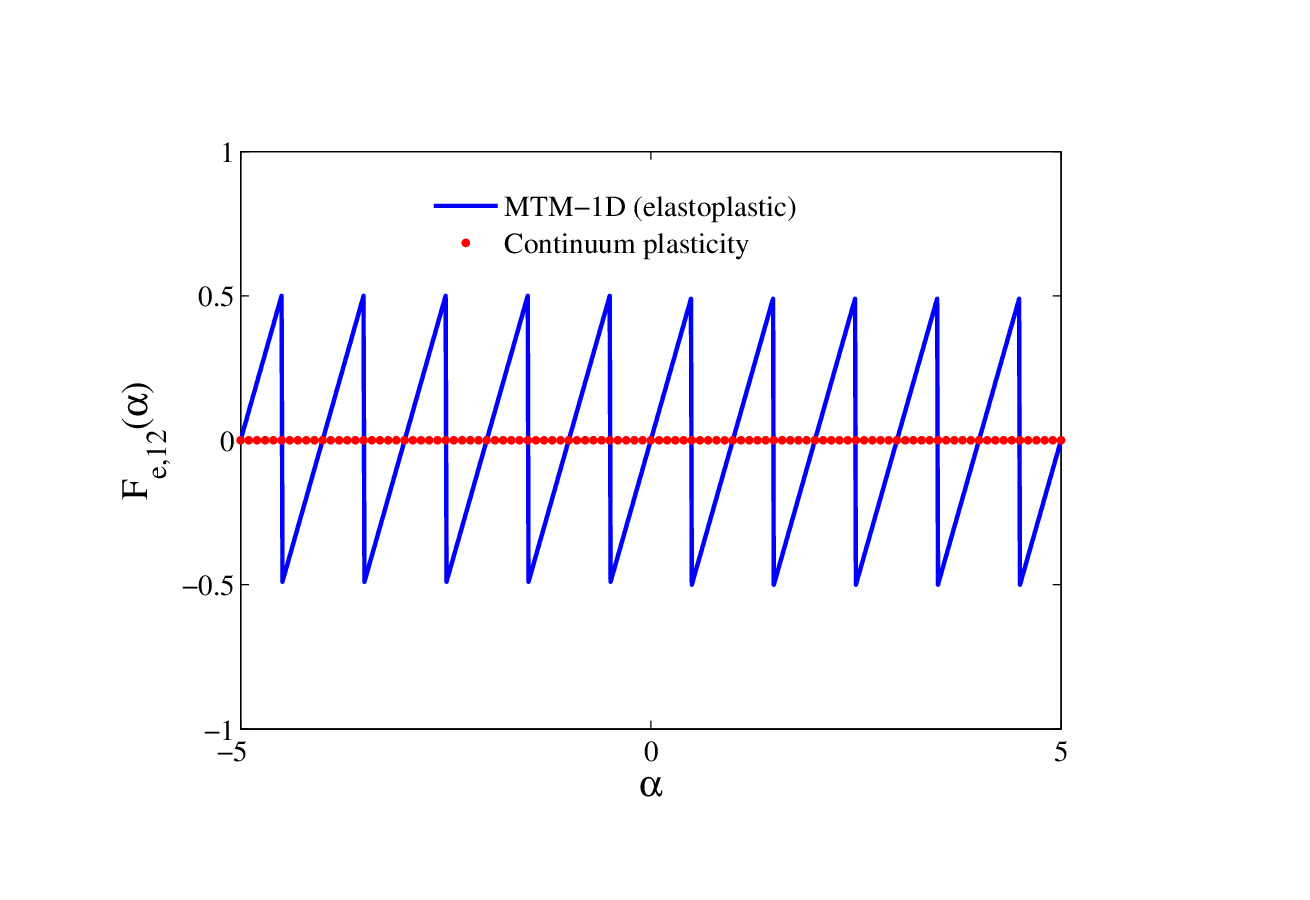}
}
  \begin{picture}(0,0)
  \put(-385,-5){$(a)$ }
    \put(-240,-5){$(b)$ }
\put(-85,-5){$(c)$ }
  \end{picture}
 \caption{Schematic comparison of the MTM (blue) and the CP ( red) based response under one-parametric  simple shear loading in a hard device: (a) energy density, (b) plastic  distortion  $F_{p,12}=-m_{12}$, (c) elastic distortion. } 
\label{Fig1Dpsi}
\end{figure}
 
\paragraph{Quantized plastic strain} To illustrate in {the} simplest form the idea that the introduction of a periodically modulated valley inside the configurational energy landscape can lead to the description of lattice-invariant shears, we now consider {the} elementary shear deformation path
\begin{equation}
\begin{split}
\textbf{F}=\begin{bmatrix}
1 & \alpha\\
0 & 1 
\end{bmatrix},    
\end{split}
\label{eq21fh}
\end{equation}
where $\alpha$ will be  interpreted as a loading parameter (in a hard loading device). Suppose that the path \eqref{eq21fh} extends along the floor of a valley inside the energy landscape $\psi(\textbf{F})${,} which would correspond in the CP framework to a `plastic mechanism'. While in CP the elastic energy along such a path would be identically equal to zero, see the red line $\psi(\textbf{F}( \alpha) )   \equiv 0$  in Fig. \ref{Fig1Dpsi}(a), in the MTM we postulate that the energy $\psi(\textbf{F}( \alpha) )$  is periodic. For simplicity we can further assume that the corresponding one{-}dimensional energy landscape is described by {the} piecewise-quadratic function
\begin{equation} 
 \psi(\textbf{F}( \alpha) )  =\frac{\mu}{2}\left[\alpha-\left\lfloor|\alpha|+\frac{1}{2}\right\rfloor\text{sgn}(\alpha)\right]^2,
\label{eq1DEn}
\end{equation}
where $\lfloor \cdot \rfloor$  is the `floor' function (the greatest integer less than or equal to  the argument) and $\mu$ is the effective shear modulus. Note that while in the CP framework we could assume  that all the distortions $\psi(\textbf{F}( \alpha) )$ are purely plastic, in the MTM framework the distinction between the elastic and the plastic distortion becomes more subtle.
 
In view of the inherent discreteness of plastic deformation, the natural choice of  the meso-scale   plastic distortion 
 would be the  integer-valued matrix 
\begin{equation}
\begin{split}
  \textbf{m}( \alpha)  =\begin{bmatrix}1 & -\left\lfloor|\alpha|+\frac{1}{2}\right\rfloor\text{sgn}(\alpha)\\
0 & 1 
\end{bmatrix}. 
\end{split}
\label{eq21fh2}
\end{equation}
Here we implicitly assume that  ${\textbf{F}}_p^{-1}( \alpha)  =\textbf{m} ( \alpha)  $. In other words, the meso-scale plastic distortion would correspond to the  blue line in Fig. \ref{Fig1Dpsi}(b){,} which is different from the red line in Fig. \ref{Fig1Dpsi}(b){,} representing {the} conventional macro-scale assumption of the CP approach.
One can see that the meso-scale   plastic distortion exhibits a  ``staircase'' structure, where vertical segments, indicating discrete  integer-valued slips, correspond  to quantized advances of plastic  deformation, while the horizontal segments   describe  the complementary, purely elastic, deformation. Furthermore,  while in the CP appproach we have $  \textbf{F}_e( \alpha) \equiv 0$ (red horizontal line in Fig. \ref{Fig1Dpsi}(c)), our quantized assumption  \eqref{eq21fh2} combined with  the multiplicative decomposition $  \textbf{F}_e( \alpha) ={\textbf{F}}( \alpha) \textbf{m}( \alpha) $,  gives a different  expression for the elastic  distortion (blue zigzag line in Fig. \ref{Fig1Dpsi}(c)):
\begin{equation}
\begin{split}
  \textbf{F}_e( \alpha) =\begin{bmatrix}
1 & \alpha-\left\lfloor|\alpha|+\frac{1}{2}\right\rfloor\text{sgn}(\alpha)\\
0 & 1 
\end{bmatrix}
\end{split}
\label{eq21fh3}
\end{equation}
As we see,  an   elastic distortion, associated with the activation of  a meso-scopic  `plastic mechanism' , is  described by a piece-wise continuous function oscillating around  zero. 
 
This  simple  example  suggests that the quantized increments of plastic distortion occur when the driven system reaches the {boundaries} of one of the equivalent elastic domains, each containing a  replica of one elastic energy well.  Inside each of such domains the mechanical response is purely elastic and  
{the} transitions between neighboring domains can be viewed as `elastic instabilities'. The corresponding strain threshold can be interpreted as  a simplistic `yield criterion' and since  the system can advance only between the neighboring   domains,  the  associated (quantized) advances  of  plastic strain are then governed by   an equally  oversimplified  `flow rule'.
 
 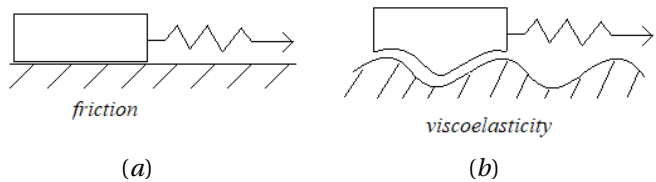
\begin{figure}[h!]
\centering
{\includegraphics[scale=0.78,trim={0cm 0cm 0cm 0.5cm},clip]{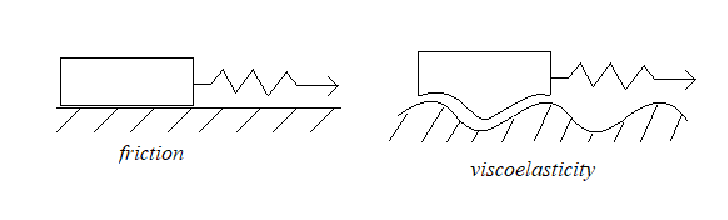} 
 }
 \begin{picture}(0,0)
  \put(-215,0){$(a)$ }
    \put(-85,0){$(b)$ }
  \end{picture}
\caption{Schematic representation of a zero-dimensional  model as it  operates at the macro-scale (a)  together with its analogue operating at the meso-scale (b) }
\label{ToyScheme}
\end{figure}

\paragraph{Zero-dimensional model} The above discussion of a simple example can serve as an illustration of the idea of periodically modulated valleys in the MTM. However, such a discussion remains largely kinematic as no equilibrium equations are  involved, and since no dissipation is associated  with `elastic instabilities'. To address these and other related issues, we need to augment  our model mechanically. At the same time it should be simplified  geometrically if the goal is  to maintain the same level of analytical transparency. With this idea in view we now consider another simple example explaining how in the MTM the discrete (quantized) evolution of plastic strain can be driven externally. We use the same example to also explain why such an evolution is caused by elastic instabilities, which are necessarily dissipative.

The zero-dimensional toy model which serves our purposes was first proposed by L. Prandtl \cite{Prandtl,Popov2012,Abeyaratne1996,Choski1999,Heslot1994}, see Fig. \ref{ToyScheme}.  It  provides an elementary representation of a \emph{depinning} phenomenon dealing with an elastic object (manifold) being dragged along  a modulated background;  the latter   allows for  a large number of equilibrium configurations where the elastic manifold can be pinned \cite{fisher1998collective,kardar1998nonequilibrium}. Prandtl was the first to realize  that depinning  could be  the right qualitative (but not necessarily quantitative) metaphor for crystal plasticity, where the manifold is a network of  elastically linked defects,   the role  of   modulations  is  played by the system of lattice-invariant shears and  dragging ensured by the applied loads.

Below we use this prototypical model   to illustrate the fact that   the  dry-friction-type dissipation on a flat surface   can emerge as a homogenized description of  an externally driven   viscous-type  dynamics   on a  periodically modulated surface, which is exemplified by the depinning phenomenon. More broadly, our goal is to show   that  the  classical ideally plastic dissipative response, described  in CP by a homogeneous dissipative function of degree one, can be viewed as an averaged description of an MTM-type response with viscoelastic (quadratic) dissipation. Our analysis shows that   behind the nontrivial nature of such coarse graining  lies  a cascade of sub-critical elastic instabilities leading to fast and abrupt embedded  rearrangements, which remain dissipative even in the case of quasi-static driving \cite{Puglisi, Mielke2012}.


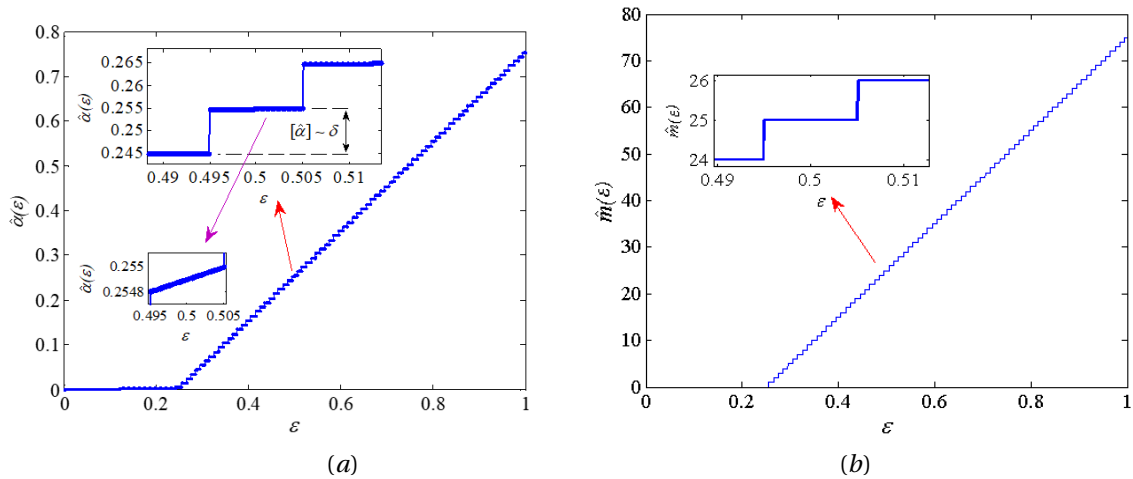
\begin{figure}[h!]
\centering
\centerline{
\includegraphics[scale=0.55]{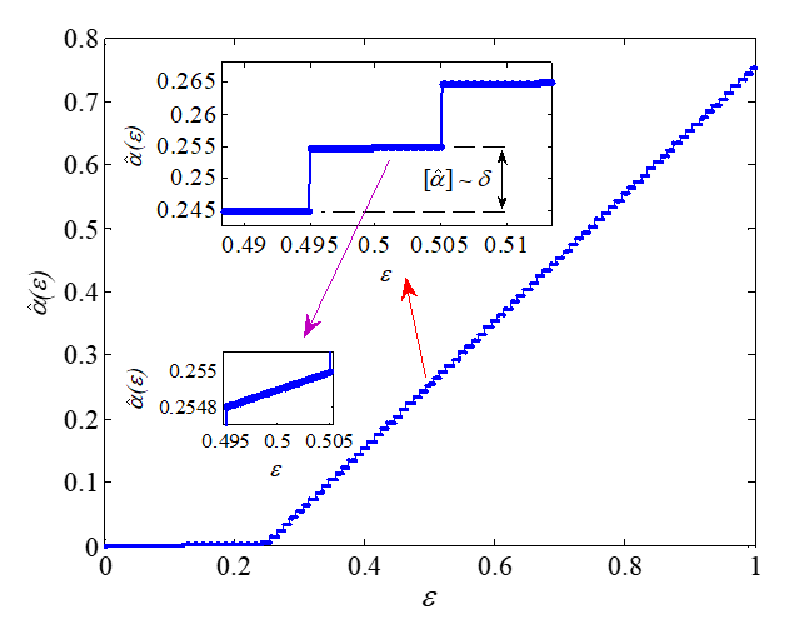}
\includegraphics[scale=0.55]{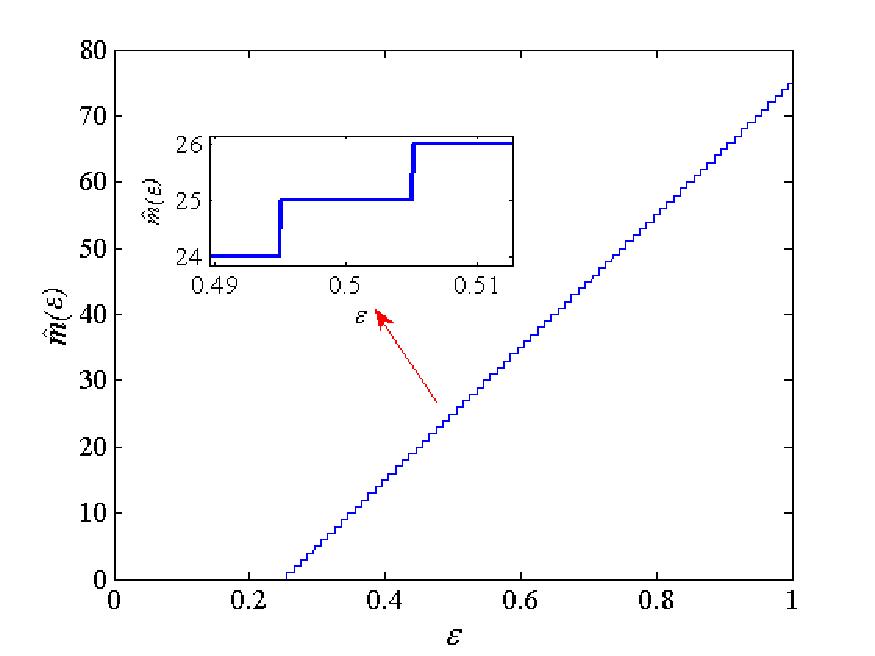}}
 \begin{picture}(0,0)
  \put(-97,0){$(a)$ }
    \put(95,0){$(b)$ }
  \end{picture}
 \caption{`Equilibrium' responses of (a) total strain, $\hat \alpha (\varepsilon)$, and (b) plastic strain, $\hat {m} (\varepsilon)$, produced by the zero-dimensional model for $k/E=1/2, \ \delta=0.01, \nu \to 0 .$}
\label{FiglandCor}
\end{figure}

The model shown in Fig. \ref{ToyScheme} (b) is characterized by the  energy
\begin{equation} \label{toy}
 f(\varepsilon,\alpha)=-k \delta \psi\left(\frac{\alpha}{\delta}\right)+\frac{E}{2} (\alpha-\varepsilon)^2.
 \end{equation} 
Here the first term on the right  represents the  MTM-type energy landscape   presumed to be operative  inside the (periodically modulated) plastic energy valley.  The second term  describes  the elasticity {of the  environment,} operative
outside such a valley{. O}ne can view this term  as a description of the elastic coupling  between  the `plastic mechanism' and the loading device. 

In view of this interpretation, the variable $\alpha$  emerges as  the representation of plastic  distortion,  while the role of the loading parameter is played  by $\varepsilon$. For simplicity, we assume that the periodic potential $\psi$ is  piecewise quadratic and write it first in the form 
$\psi  = -\frac{1}{2}\left (z-m \right )^2$, where 
 $m$ is the integer-valued  parameter.
   To ensure the periodicity of $\psi$ and  remain {in} the `purely elastic' setting, we complement the definition by the assumption that $  m=\left\lfloor{|z|}+\frac{1}{2}\right\rfloor\text{sgn}(z) $. 

The resulting model is then characterized by the energy 
\begin{equation} 
\label{toyPQ61}
\psi(z)=1-\frac{1}{2}\left[ z-\left\lfloor {|z|}+\frac{1}{2}\right\rfloor\text{sgn}(z)\right]^2.
\end{equation}
 We observe  that the quantized variable  $m$ selects at a given $z$ a particular (energy minimizing)  parabola, however, being enslaved to  $z$, it does not play an independent role in the response of the system.  The implied adiabatic elimination of the $m$-type {variable} represents a defining feature of the current version of the MTM{,} where plastic strain is not {present} explicitly.
 
To complete the introduction of the Prandtl model, we {should} mention that the small parameter $\delta$ is {a} measure {of} discreteness{,} which {points to} the meso-{scopic} nature of this model. We show below that the  analogue of the macro-scale CP model emerges from \eqref{toy}  in the limit $\delta \to 0$; the remaining parameters $k$ and $E$  characterize in the ensuing  coarse-grained  continuum  model the yield stress and the  dissipative potential.

Suppose next  that the dynamics in  the system with the energy (\ref{toy}, \ref{toyPQ61}) is overdamped-viscous and is described by   the gradient-flow type kinetic equation $$\dot{\alpha} =-\gamma
   \partial f(\varepsilon,\alpha)/\partial \alpha,$$ where $\gamma$ is  a measure of the relaxation time. We also assume  that the system is driven in a hard device with a rate of $$\dot{\varepsilon}=v>0.$$  In the quasi-static limit, $$\nu = v/\gamma \rightarrow 0,$$ the energy  in such a system  is locally minimized almost always and we can use the equilibrium condition  $\partial f(\varepsilon,\alpha)/\partial \alpha  =0$ to  define the locus of the equilibrium response  $\hat \alpha (\varepsilon)$.
    Assuming that $\delta \ll k/E$ (strong pinning condition)  we obtain an explicit formula:
  \begin{equation} 
\label{toyPQ6}
\begin{split}
\hat \alpha (\varepsilon)=\frac{ k\delta}{k+E\delta}\text{max}\left[0,\text{min}\left(\left\lfloor\frac{\varepsilon}{\delta}\right\rfloor,\left\lceil\frac{\varepsilon}{\delta}-\frac{1}{2}-\frac{1}{2}\frac{k}{E\delta}\right\rceil\right)\right]+\frac{E\delta\varepsilon}{k+E\delta}.
\end{split}
\end{equation}
The computed elastic response, shown in  Fig. \ref{FiglandCor} (a),  is characterized by a prolonged initial elastic range where the  system remains in the original energy-well and  where the elastic strain $\hat \alpha$ changes with $\varepsilon$ linearly. The subsequent evolution is a repetition of {the} same pattern: a succession of linear segments, signaling that the system remains  inside  the same  energy well,    interrupted by  the     transitions   to  the neighboring wells {occurring} during the short {increments} of the loading `time'  $\varepsilon${,} which scale with the small parameter $\delta$.   

We can now argue that the loading level  when the system reaches the boundary of the original  energy-well indicates  the beginning of plastic yield  and that the {corresponding (identical)} jump events can be associated with plastic slips. If we now reconstruct the evolution of the integer-valued variable  $m$, we obtain:
\begin{equation} 
\label{toyPQ5}
\begin{split}
\hat m (\varepsilon)=\text{max}\left[0,\text{min}\left(\left\lfloor\frac{\varepsilon}{\delta}\right\rfloor,\left\lceil\frac{\varepsilon}{\delta}-\frac{1}{2}-\frac{1}{2}\frac{k}{E\delta}\right\rceil\right)\right].
\end{split}
\end{equation}
The emerging response $\hat m (\varepsilon)$, illustrated  in Fig. \ref{FiglandCor} (b), suggests that the variable $m$ can be interpreted as the integer-valued measure of plastic strain.  Indeed, the  prolonged pre-yield elastic  range is characterized by the condition $\hat m =0$, while after the yield we observe the repeating `staircase'-type   pattern  of elastic horizontal steps and (almost) discontinuous transitions of the type $\hat m  \to \hat m +1$. The conceptual problem with this interpretation is that the variable $m$ remains  implicit while  both {the} elastic and {the} plastic phases of the response are still fully  characterized by a single strain variable $\alpha$. Before we turn the variable $m$ into an independent `player' in the model, it is instructive to  rationalize the obtained  elasto-plastic response  in energetic terms.

%
%


  \begin{figure}[h!]
\centering
\centerline{
\includegraphics[scale=0.58]{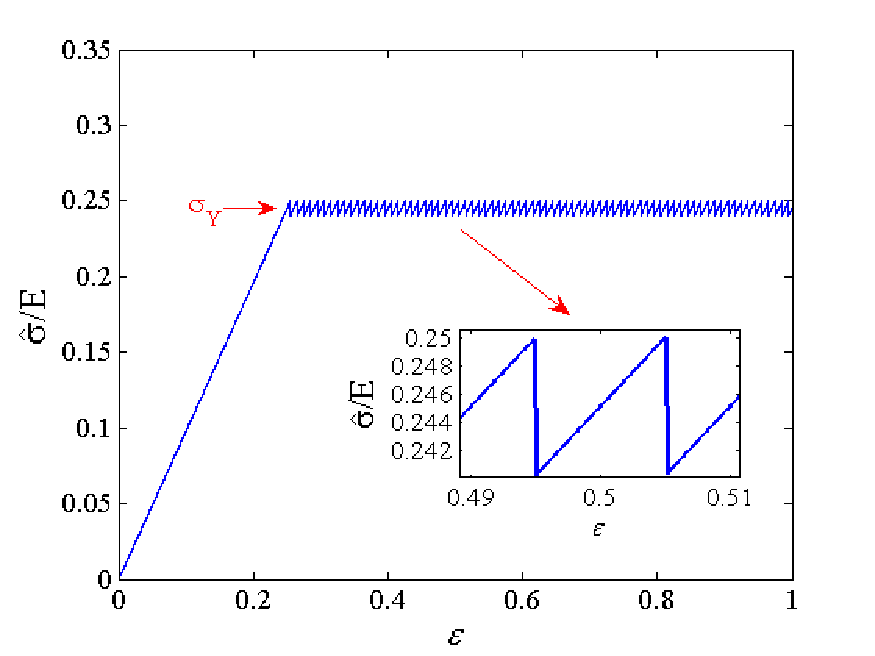}}
 \caption{Macroscopic   stress-strain relation   $\hat \sigma(\varepsilon)$   in  the zero-dimensional model  with   $k/E=1/2, \delta=0.01, \nu \to 0.$
} 
\label{FiglandCor10}
\end{figure}
  
\begin{figure}[h!]
\centering
\centerline{
\includegraphics[scale=0.6]{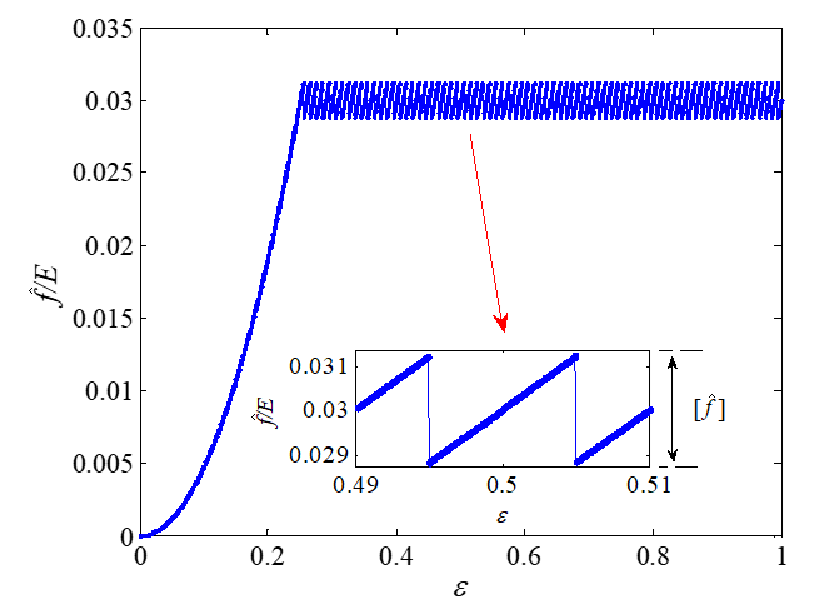}}
 \caption{Macroscopic energy-strain relation $\hat{f} (\varepsilon)$  in  the zero-dimensional model  with   $k/E=1/2, \delta=0.01, \nu \to 0.$
} 
\label{FiglandCor1}
\end{figure}

We first recapitulate  that the   quasi-static evolution of our `purely elastic'   system takes place at the slow  time scale of the loading, $\tau=v t$. This slow evolution is periodically interrupted by the  fast events taking place at the  relaxational time scale $t$. During the slow (elastic, equilibrium) stages of the evolution, the elastic strain $\hat \alpha $ varies continuously on the  scale $\tau$, while  the system stays on  the same  equilibrium branch of the elastic response. {On the other hand}, during the fast (plastic, nonequilibrium) stages of the evolution, {the} elastic strain $\hat \alpha(\varepsilon)$ changes on the scale $t$, and the system {transitions between} neighboring  branches of elastic equilibria.   In the quasi-static limit $\nu \to 0 $, such transitions can be viewed as occurring instantaneously and  we can associate them with  (over-simplistic) plastic avalanches.  In this  limit,  plastic avalanches are accompanied by the discontinuities of  both the strain and the energy, taking place  at a fixed value  of the loading parameter.  The  periodically-spaced   stress drops  can be seen on  the macroscopic stress-strain curve  $\hat \sigma(\varepsilon)=   d  \hat f/d\varepsilon$, shown {in} Fig. \ref{FiglandCor10}.   The energy jumps are also clearly visible on the  macroscopic energy-strain curve     $\hat
f(\varepsilon)=f(\varepsilon, \hat \alpha(\varepsilon))$, shown in   
 Fig. \ref{FiglandCor1}.  We note {that} the formally computed area under the graph $\hat \sigma(\varepsilon)$  has  nothing to do with the elastic energy stored in the system, as the  measure-valued fluctuations of stress associated with abrupt drops of energy $\hat
f(\varepsilon)$ would  not be properly accounted for in such a {na\"ive} computation, see \cite{Puglisi} for more details.

It is clear{,} though{,} that during the elastic stages of the deformation all the work of the loading {is} indeed stored in the system, as the dissipation associated with the phenomena {occurring} in slow time {is} absent in the limit $\nu \to 0 $.  Instead, the plastic-correction events, occurring in fast time, remain dissipative, notwithstanding the quasi-static nature of the loading,  as {can be seen clearly} from the plot {of} the energy $\hat f(\varepsilon)$ {in} Fig. \ref{FiglandCor1}. {One therefore finds} that after the yield stress is reached {the} energy-strain relation is punctuated by abrupt energy drops{, attesting} to the fact  that the associated branch-switching transitions are dissipative.  While in our overdamped model the released energy simply disappears, in {more realistic} systems it is either converted to lattice{-}scale vibrations {\cite{Efendiev2010}}, {or} emitted in the form of acoustic waves {\cite{Weiss2021}}.  

The amount of energy dissipated  in  each of the  elastic branch-switching events is
\begin{equation}
\label{dis}
 -\int (\partial f/\partial \alpha) \dot \alpha dt=
 -\llbracket \hat
f\rrbracket \geq 0,
\end{equation}
 where  we introduced  the notation  $\llbracket  
 A\rrbracket = A_+-A_-$  for  the difference  between the values after and before the jump (fast event). We emphasize that the integration in  \eqref{dis} is over the duration of the transition in fast time and the integral is different from zero in the limit $\nu \to 0 $ because  as duration tends to zero, the rate of deformation tends to infinity. 

As we see, the system experiences a cascade {of} dissipative events, {separated} by regularly spaced dissipation-free elastic  stages.  The total dissipation  is then $-\sum \llbracket \hat f\rrbracket$ and it is clear that it depends not on the rate of loading but only on the number of occurred jumps.   This suggests that  in  the continuum limit $$\delta\rightarrow0$$  the resultant continuous dissipation will be  rate-independent.  Indeed, in this limit, which {symbolizes the} transition from meso to macro  scale,  the magnitudes of the jumps tend to zero while, due to a particular  choice of the scaling in our zero-dimensional model,   their number  increases indefinitely so that  the total dissipation remains finite. 
 
 More specifically, one can show  \cite{Puglisi} that the limiting continu{ous} dissipative potential  is described by a homogeneous function of degree one:   $$\mathcal{D}= \sigma_Y |\dot{\hat \alpha}|,$$   where $\sigma_Y$  is
 the effective `yield stress'. To compute the value of $\sigma_Y$ we observe that  under the assumption that  $\delta \ll k/E$, each plastic avalanche produces {a} strain jump {with the} $\delta \to 0 $ asymptotics {of}
$
\llbracket \hat\alpha\rrbracket=\hat \alpha^{(m+1)}-\hat\alpha^{(m)}
\sim  \delta
$. 
To evaluate the magnitude of the corresponding  energy drops  we   write   $\llbracket f(\varepsilon,\hat \alpha )\rrbracket =-k\delta{\llbracket\psi(\hat \alpha /\delta)\rrbracket} + \frac{E}{2}\llbracket(\hat \alpha -\varepsilon)^2\rrbracket$ where   $ \frac{E}{2}\llbracket(\hat \alpha -\varepsilon)^2\rrbracket \sim -{\frac{k}{2} \delta} $ and 
$-k\delta{\llbracket\psi(\hat \alpha /\delta)\rrbracket} \sim -\frac{1}{2}E\delta^2$. Therefore 
$ \llbracket f(\varepsilon,\alpha_0)\rrbracket \sim -\frac{1}{2}{k\delta}$ and   we finally obtain   an explicit relation for the  effective yield stress:  
 \begin{equation} 
\label{toyPQ3lim7}
\sigma_Y=-\underset{\delta\to0}{\lim}\frac{\llbracket f(\varepsilon,\hat \alpha)\rrbracket}{\llbracket \hat\alpha\rrbracket} = \frac{1}{2}k.
\end{equation}

%

To summarize, the  zero-dimensional model of Prandtl conveys the main ideas behind  the  `purely elastic' version of the MTM and {adequately reproduces} the main targeted empirics. Thus, according to this toy model, an extended range of initial purely elastic response ends with the emergence of  plastic yield. The latter reveals itself through the {periodic fluctuation} of the stress around the yield value {$\sigma_Y$}. {The} fluctuations take the form of a repeating pattern of  elastic `prediction' segments, representing continuous deformation inside one energy well {(with} $\varepsilon$ changing at fixed $m$){, interrupted} by the abrupt drops, representing plastic `correction' events  (singular quantized advances of $\hat m$ at fixed $\varepsilon$). 
 
While the obtained fluctuation structure is still hardly realistic (trivial statistics of plastic avalanches, no intermittency), it is clear that the model captures the fundamentals of rate-independent plasticity (elastic range, plastic yield, presence of fast motions embedded in the otherwise quasi-static evolution, etc.). Even the so-called nucleation peak, a quasi-brittle event (avalanche) associated with the initial massive slip nucleation in a pristine crystal, can be easily recovered already in a one-dimensional generalization of this  zero-dimensional model \cite{TruskinovskyVainchtein2004}.  This does not mean, of course, that {such} schematic {a} representation of crystal plasticity is adequate. The task of implementing the same basic ideas  but in the full tensorial setting and  in arbitrary dimension  with the goal of comprehesive description of plastic fluctuations has been accomplished through the development of the `purely elastic' version of the  MTM \cite{Roberta, BaggioST2023a,SalmanBBZGT21,BaggioST2023b}.
 
Here we attempt to go a bit further{. T}he main ideas of the proposed re-formulation of the MTM can be {explained already} at the level of the zero-dimensional model. As we have {previously} mentioned, the elastic and plastic problems in the original Prandtl model are not  separated and although in the above discussion we  referred to the variable $m$ as describing plastic slip, in such  `purely elastic' {a} model the variable $m$ did not play {an} essential role. That is why we consider below another version of the Prandtl model, where the elastic and the plastic problems are fully  separated, and where the variable $m$ acquires {independent} significance. The main advantage of such a reformulation is that the elastic field can be minimized out while  the `condensed' description of plastic response in terms of the field $m$ {alone} can be reduced to an integer-valued discrete automaton.


\paragraph{Discrete  automaton representation} In the zero-dimensional toy model described above the elastic and the plastic problems could not be addressed independently. Although we formally introduced in this model the quantized plastic strain, the variable $m$ potentially representing ${\textbf{F}}_p$,  could be effectively enslaved to the variable $\alpha$, representing the elasto-plastic  strain, which we still denote by ${\textbf{F}_e}$, given that in this  zero-dimensional setting the role of the  deformation gradient ${\textbf{F}}$ was played by the loading parameter $\varepsilon$. Therefore, in such a  representation we are essentially dealing with  a quasistatically driven nonlinear-elastic system exhibiting a regular succession of internal snap-through instabilities. 



\begin{figure}[htpb!]
\centering
{
\includegraphics[scale=0.4]{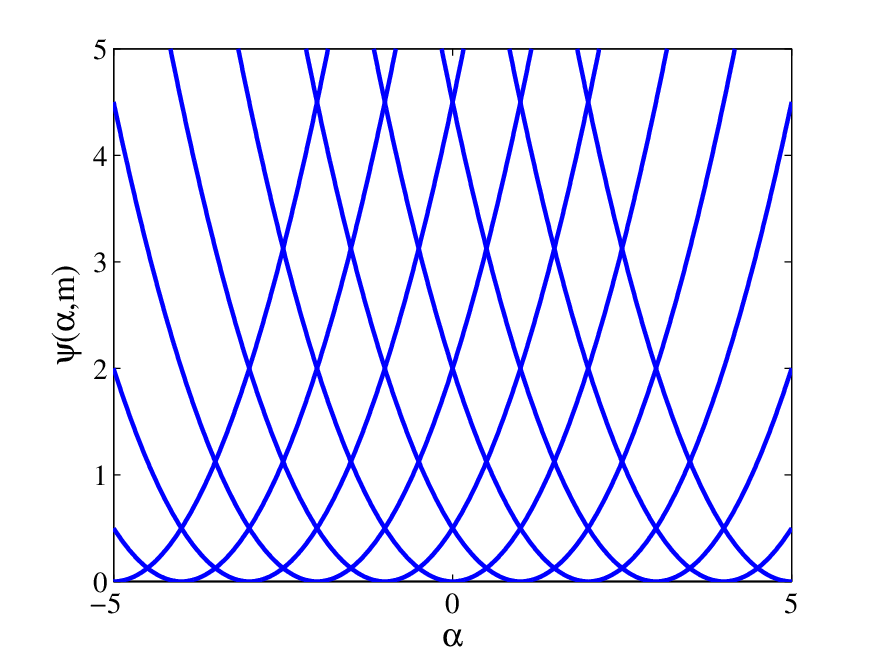} 
}
 \caption{Schematic representation of the   multivalued function $\psi(\alpha)$.} 
\label{Forest}
\end{figure}

The goal of the proposed reformulation of the Prandtl model  is to give the plastic distortion $m$  full  autonomy. To circumvent the `purely elastic' structure of the original  Prandtl model, we now abandon the idea of performing (adiabatic) elimination of the plastic strain $m$. Instead,  we follow the general approach  of classical continuum CP and interpret the integer-valued parameter $m$ as an independent variable. Staying within the same  piecewise-quadratic paradigm as before in describing elasticity, we can introduce the two-parametric energy density 
\begin{equation}
\label{lin}
\psi(\alpha,m)=-\frac{1}{2} \left(\frac{\alpha - m}{\delta}\right)^2 ,
\end{equation}
 where now the continuous variable  `$\alpha(m,\varepsilon)-m$' represents  the elastic strain ${\textbf{F}}_e$, and the integer-valued variable  $m$  represents  the plastic strain ${\textbf{F}}_p$. Since the two variables,  $\alpha $ and $m$, are now considered independent, the elastic energy density dependence on  $\alpha$ becomes  formally multivalued, as it is now represented by an infinite number of discrete branches parameterized by the value of the variable $m$, see Fig. \ref{Forest}. However, the nonlinearity of the problem and the single-valuedness of the energy can be   recovered if we place an additional constraint, namely, that the elastic strain is confined between the limits
\begin{equation}
\label{limits}
m-1/2 \leq \alpha-m \leq m+1/2. 
\end{equation}
These inequalities can be interpreted as defining the `elastic domain' corresponding to a given value of $m$. The whole range of possible values of $\alpha$ can be then viewed as being tessellated by the infinite number of equivalent `elastic domains' of this type{, representing} the periodicity domains of the elastic energy.
In other words, with the inequalities  \eqref{limits} imposed,  the model becomes again `purely elastic' as the plastic strain variable $m$ is  then effectively minimized out. As we show in what follows, the re-formulation of the MTM proposed {here} anticipates that the analog{ue}s of the  inequalities \eqref{limits} {can} be omitted at the stage of elastic `prediction' and {it would be sufficient to} implement them only at the stage of `elasto-plastic' correction. This  basic idea is illustrated below in the framework of our zero-dimensional model.

 
Observe that with inequalities \eqref{limits} suppressed, the equilibrium elastic response ${\alpha}(m,\varepsilon)$ at a given value of $m$ {can} be found (in the model with the energy  (\ref{toy}, \ref{lin})) explicitly{, as}: 
\begin{equation}
\label{adiab}
\alpha (m, \varepsilon)=\frac{k m+E\delta\varepsilon}{k+E\delta}.
\end{equation}
In other words, the assumption of linear elasticity at a fixed $m$ allows one to minimize out the variable $\alpha$   and opens the  possibility of obtaining a `condensed' description in terms of the measure of plastic strain $m$ only. However, we reiterate that since  the resulting elastic strain, $\alpha(m,\varepsilon)-m$,  may   not be located  inside the  corresponding  `elastic  domain' $(m-1/2,m+1/2)$,  the expression  \eqref{adiab} does not  represent the  solution of the  elastic problem and, given the value   of the  loading parameter $\varepsilon$,  can be at most interpreted as an elastic prediction.

Saying it differently, if the value of $m$ at the  given   $\varepsilon$ agrees with the `equilibrium' expression $\hat m(\varepsilon)$ given in \eqref{toyPQ5}, then the condition \eqref{limits}  is automatically satisfied and no elasto-plastic correction is necessary. If, however,  the predicted elastic strain at the given $m$ and $\varepsilon$ is located outside of the corresponding  `elastic domain', the  configuration \eqref{adiab}  should be rendered unstable, which means that an elasto-plastic correction should be performed with the idea that as a result of such correction the corresponding value $\hat m(\varepsilon)$ is attained. More specifically, since in this case the effective `yield condition' \eqref{limits} is violated, the value of the plastic strain $m$ must undergo a quantum update, $$m \to m \pm1,$$  one or several times. An  elementary update  of this kind (elementary plastic correction) can be then  interpreted as the action of a  primitive `flow rule'.  After an elementary update of this type is completed,  the equilibrium elastic problem for $\alpha$ can be  solved again producing an elastic correction.  The implied  iterative process can continue until the plastic strain  $m$ is  finally compatible with the `equilibrium' response $\hat m(\varepsilon)$.


The main advantage of our simplified description of elastic response inside each `elasticity domain' is that the numerical  algorithm alluded above can be  formulated without any reference to the variable $\alpha$, which, given that  the  equilibrium equation $\partial   f/\partial \alpha =0$ is  solvable explicitly,   could be thus considered as effectively eliminated.   The equations governing the evolution of the plastic strain $m$ in the resulting  `condensed'  problem reduce to a discrete dynamical system {that} can be formulatedas an integer-valued automaton:
\begin{equation}
\label{AutomatonEq10}
\varepsilon\to \varepsilon+\Delta\varepsilon \ \ \Rightarrow m\to m+\Delta m, \ 
\end{equation}
where 
\begin{equation}
\label{AutomatonEq1}
\begin{split}
\Delta m= \begin{cases} \ \ 0, & \frac{\varepsilon}{\delta}-\frac{1}{2}-\frac{1}{2}\frac{k}{E\delta}+1 <1 \\ -1, & 1\le \frac{\varepsilon}{\delta}-\frac{1}{2}-\frac{1}{2}\frac{k}{E\delta}+1 \le m \\ \ \ 1, & \frac{\varepsilon}{\delta}-\frac{1}{2}-\frac{1}{2}\frac{k}{E\delta} >m \ \ \end{cases}.
\end{split}
\end{equation}

Generically, the discrete mapping (\ref{AutomatonEq10},  \ref{AutomatonEq1}) performs the update from $m$ to $m+\Delta m$ in response to the load increment $\Delta \varepsilon$, however,  at a given value of $  \varepsilon$, depending on the value of  $ \Delta \varepsilon${,} such an update of  $m$  may or may not be warranted. Indeed,  the quantized variable $m$  is  updated by an elementary correction step $ \pm 1$ only if one of {the} thresholds  in \eqref{limits}  is violated.  In the `condensed' formulation \eqref{AutomatonEq1} the solution of the elastic problem  is, of course,  hidden behind the expressions for  the threshold. 
Note that in the limit $\Delta\varepsilon\to 0$ {(and} in our numerical experiments{,} with the automaton {in} (\ref{AutomatonEq10},  \ref{AutomatonEq1}){, and} the {parameter values} {given} in Fig. \ref{FiglandCor},
already {for} $\Delta\varepsilon<10^{-4})$, no elastic correction is ever needed and the automaton {in} (\ref{AutomatonEq10},  \ref{AutomatonEq1}) reproduces exactly the equilibrium response {given in} \eqref{toyPQ5}.

{Comparison} of {the} two prototypical zero-dimensional models {discussed above} suggests that, despite the differences {in} appearance{,} the discrete dynamics of the plastic strain $m$ under sufficiently small increments of the loading parameter  is  basically the same  in the two models. The main  advantage of  the  introduction of the simple  multi-valued elastic  energy is that the    elastic problem becomes fully separated and {could be} solved implicitly `on the fly'. Then, under the assumption that the load{ing} $\varepsilon$ {changs} quasistatically, the evolution of plastic strain in the `condensed'  problem reduces to an integer-valued automaton whose structure was not apparent in the `purely elastic' setting of Prandtl.  The  complexity of the nonlinear evolution problem in the Prandtl model is then relegated to  a simple  integer-valued discrete dynamical system governing the dynamics of {a} quantized plastic variable. The ensuing automaton description is fully adequate and  in the limit of slow driving it perfectly recovers  the staircase-like piece-wise smooth equilibrium response  $\hat m(\varepsilon)$. 
  
We reiterate that using a similar  re-formulation of  the existing version of the MTM approach, one can  expect to be able to overcome the computationally challenging nonlinearity of a `purely elastic'   problem,  whose numerical solution  presents  major computational difficulties in the fully tensorial multidimensional case. 
Note that in such a  reformulation one can also expect to {encounter} the analogues of the familiar notions of a yield surface, represented by the ridges of the equilibrium-energy landscape, and of a flow rule, represented by a  post-bifurcational response bringing the elastically destabilized system into the neighboring energy well. However, if in our zero-dimensional toy model those notions could be introduced by explicit conditions, similar notions  in the multidimensional tensorial case can be expected to be much more  obscure. Some glimpses of the emerging complexity can be seen already in a 2D scalar version of the MTM,  where even if the elimination of the elastic fields and the explicit reduction to a discrete automaton prove possible, the `diffuse{d}' analogues of the yield surface and the flow rule remain largely implicit \cite{Umut2011,Umut2012}.

 \section{Mesoscopic tensorial model (MTM)}
 \label{sec3}
 
In this  section we finally outline the proposed  CP-type version of  the geometrically nonlinear tensorial MTM, which can serve as a  generalization of  the   `condensed' model of toy crystal plasticity presented in Sec. \ref{sec2}. 
 
We follow the  classical continuum CP, and replace the additive decomposition of the scalar strain into elastic and plastic  contributions, adopted in   Sec. \ref{sec2},   by the multiplicative decomposition ${\textbf{F}} =  {\textbf{F}}_e{\textbf{F}}_p$. However, instead of the  continuous plastic distortion  ${\textbf{F}}_p$ of  CP, we use  its  discrete (quantized) analog{ue}. The function $\psi$, which depended in the toy model on a scalar variable, will be replaced by a rank-one convex function of the tensorial variable ${\textbf{F}}_e$ \cite{Mielke2002}. In view of the mesoscopic nature of the MTM, the parameter analogous to the meso-scopic length scale $\delta$ in  the toy model will be kept small but finite. To make the presentation of the ensuing  computationally   efficient  version of the MTM  more transparent, we limit our attention  to the simplest 2D Bravais crystals{, those} with square symmetry.

\paragraph{Matrix-valued spin variable} The central new  step  in the reformulation of the MTM is the introduction of the quantized plastic strain $ {\textbf{F}}_p$. From the discussion of the zero-dimensional model in Sec. \ref{sec2} it is clear  that the tensorial analogue of the scalar plastic strain $m$ should  be discrete,  but now the implied quantization should reflect  the structure of the lattice-invariant  transformations of  2D Bravais lattices onto themselves \cite{Wang,Waal,Kaxiras1994}.

 As we have already mentioned, the global symmetry of such lattices is described by the infinite discrete group $\text{GL}(2,\mathbb{Z})$ \cite{Ericksen,Folkins,Parry,Engel,Michel}. In fact, for the purpose of identifying plastic distortions, we can   neglect reflections and limit our attention to the infinite discrete group $\text{SL}(2,\mathbb{Z})$.  The implied global symmetry entails a tensorial periodicity of the energy density and  points to  the existence of an infinite number of equivalent energy minima describing  lattice-invariant shears \cite{ Roberta,CZ,Pitteri2002}. It is also  known that different domains of periodicity and the associated minima of the energy can  be parametrized by {a matrix} $\textbf{m}$ having integer-valued entries $ m_{ij} \ \in \mathbb{Z}$ and  $\text{det}({\textbf{m}})= 1$.   

Given such a characterization of the {matrix} $\textbf{m}$, we can now state the main assumption of the re-formulated MTM approach:
\begin{equation}
\label{slip}
 {\textbf{m}}= {\textbf{F}}_p^{-1}.
\end{equation}
The idea of using ${\textbf{F}}_p^{-1}$ as a measure of plastic distortion instead of the more conventional ${\textbf{F}}_p$ (see, for instance, {\cite{Clayton2010}}) goes back to Noll {\cite{Noll1967}}. The corresponding conceptual framework  was further developed  in {\cite{Epstein1996, Davini2001,Gupta2007}}; for recent reviews  see {\cite{Epstein2007,Steigmann2023}}. What makes  our mesoscopic approach different from all this work, aimed at the macroscopic description of material inhomogeneity,  is that we abandon the idea that  ${\textbf{F}}_p^{-1}$ is a continuous tensorial variable and restrict it only to discretized (quantized) values. The main role of  the resulting  `spin'-type plastic distortion  ${\textbf{F}}_p$ is  to select one  `energy well' among the infinitely many equivalent replicas inside the globally periodic energy landscape.

In view of {the} frame-indifference requirement, the globally periodic energy density $\psi( \textbf{C})$  is  defined  in the configurational space  of  objective metric tensors {$\textbf{C}=\textbf{F}^{\top}\textbf{F}$}. 

Due to the implied symmetry imposed by the action of the infinite discrete group, we must require that $$\psi( \textbf{C})=
\psi( \textbf{m}^{\top}\textbf{C}\textbf{m})$$ for all $m \in \text{SL}(2,\mathbb{Z})$.  Such symmetry  tessellates  the configuration space   into equivalent periodicity domains inside which the  response can be considered  as purely elastic. The discreteness of plastic strain is then simply a reflection of the discreteness  of the tessellation of the configurational space.



We conclude that  under the assumption  \eqref{slip}, a relevant measure of plastic strain emerges as a matrix-valued  spin  variable. The necessity of using a discrete measure of plastic slip is a manifestation of the sub-continuum, meso-scopic  nature of the MTM. Note that the coarse-graining of plastic strain, implied  in the CP approach, invalidates this discreteness and therefore abandons the idea of resolving the associated meso-scopic phenomena.

In contrast,  the main goal of the MTM  is to capture at least some of  the effects that become invisible  in the continuum limit.  Therefore it places considerable emphasis on the fact that  plastic flow  advances via a succession of `strain quanta'.  This idea is exemplified in the present re-formulation of the MTM{,} which introduces such configurational discreteness explicitly.  It should  be noted, however,   that the implied quantization of the configurational space is  scale-free, as it reflects only the periodicity of the atomic lattice, and not the existence  of any super-atomic internal length scale.  

Yet,  the  observed structure of plastic fluctuations apparently points to the existence in crystal plasticity of an  internal meso-scopic length scale. The introduction of an internal length scale  also plays an important  role of regularizing the  otherwise ill-posed mathematical problem. Thus, it has been long realized {\cite{Ericksen1973}} that in view of {the} severe nonconvexity of the globally periodic configurational landscape energy, the  energy  minimization   implicated  in the case of  quasi-static driving   would   necessarily  lead in the continuum (scale-free) setting  to a degenerate (fluid-like) mechanical response. 

In view of all this and despite its continuum appearance, the MTM approach is designed to carry  a finite  internal  scale.  As we have already mentioned, this is done  by complementing the configurational discreteness  with  spatial discreteness. The latter  is brought in through the  meso-scopic elements inside which the deformation is considered to be  affine.  The refusal of the MTM to resolve inhomogeneities below  a given cut-off length scale leads to the necessary loss of some microscopic information,    conceivably without compromising the main meso-scopic phenomena of interest. 

Finally we mention that,  characteristically,  neither configurational nor spatial   discreteness is a part of the conventional  macroscopic   CP approach to crystal plasticity.

%
%
%
%

 
%


\paragraph{Fundamental elastic domain (FED)} The first step in the proposed re-formulation of the MTM is to define the effective `yield criterion'.  In our setting this can be done by associating the   discrete upgrades of the quantized plastic strain ${\textbf{m}}$ with the crossing of the boundaries between the periodicity  domains in the configurational space, each containing a single replica of the same elastic  `energy well' \cite{Roberta}. 

The geometrically minimal  domain of periodicity in the above sense  is known as the `fundamental domain' (FD) \cite{Folkins,Parry,Engel,Michel}. For our purposes it is more convenient to deal with the FD extended by including its independent  images  under the action of the finite  point group  of the lattice \cite{Pitteri2002}. Such an extended fundamental domain is known as the Ericksen-Pitteri neighborhood \cite{CZ,Ericksen89,Duistermaat,Pitteri84}; in what follows  we refer to it  simply as the `fundamental elastic domain' (FED). 

If configurational displacements are confined  strictly {to} the interior of the FED (or {to} one of its equivalent images under the global symmetry transformation), {then} the deformation can be considered {purely} elastic. Instead, crossing the boundary of the FED implies the activation of plastic slip. 

\begin{figure}[h!]
\centering
\includegraphics[scale=0.61 ,trim={0cm 0cm 0cm 0cm},clip]{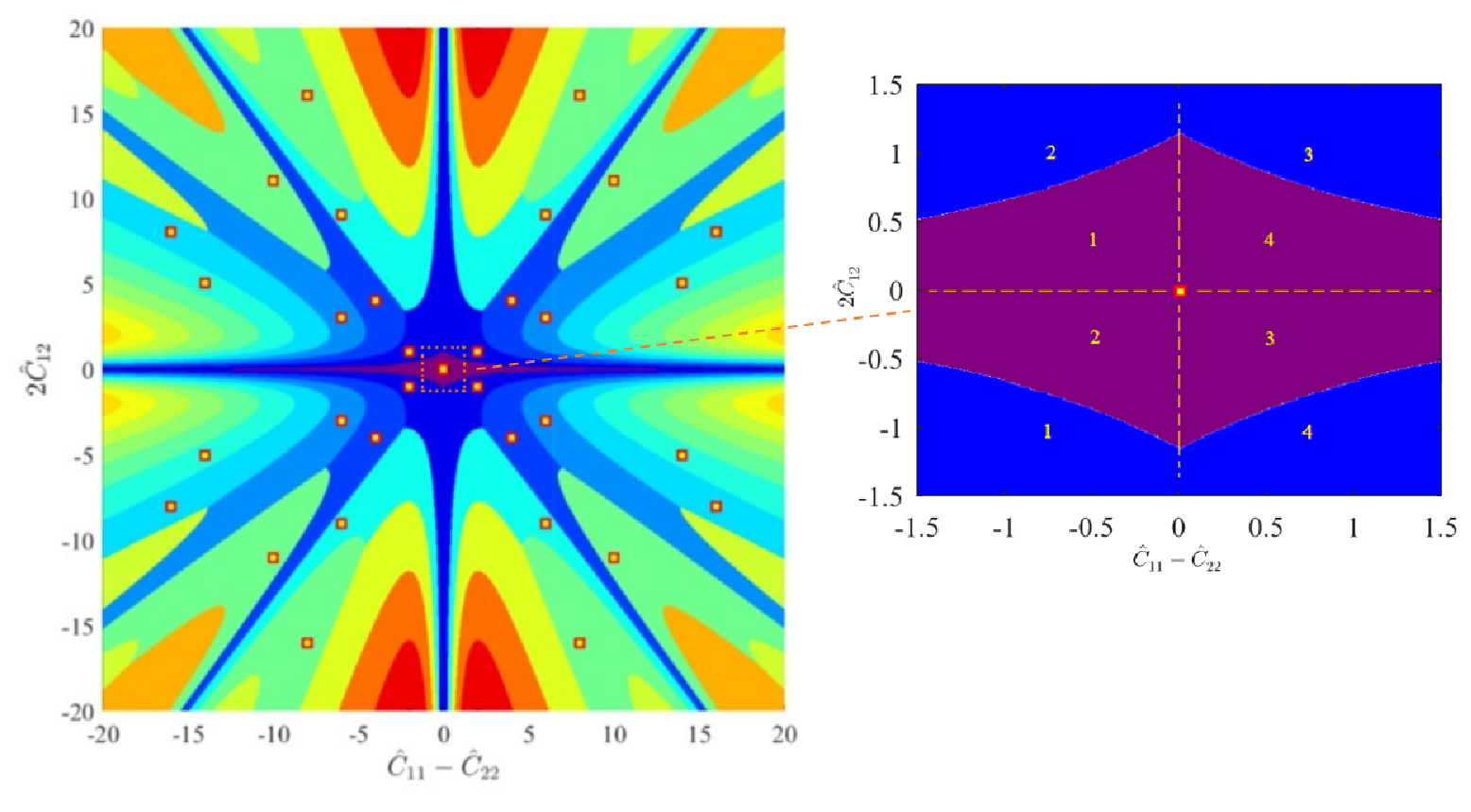} 
 \caption{The  tessellation of the space of metric tensors $\hat{C}$ into equivalent elastic domains, each containing a single energy minimum. Thirty three minima of this type   are identified by square markers. The inset on the right shows the  FED (in magenta);  for $\hat{C}_{12}\to0$  the boundaries of the FED meet at $\hat{C}_{11}-\hat{C}_{22} = \pm\infty$. The numbers 1--4 correspond to four possible choices of  FDs inside each choice of  the FED.} 
\label{FigEnergyMap3}
\end{figure}

To locate this boundary we now  turn to our special case of a simple square lattice defined by the basis vectors $\hat{\textbf{e}}_{1 }=(1,0)^{\top}$ and $\hat{\textbf{e}}_{ 2} =(0,1)^{\top}$. Then {the} FED is characterized by the condition that  in simple shear loading   its  boundary is reached when  an elementary-lattice cell reaches the kinematic mid-way towards the neighboring cell \cite{ CZ}. Therefore, outside  such a  domain the following inequalities are violated:  $
\hat{\textbf{e}}_1^{\top}\textbf{C}\hat{\textbf{e}}_2\le   \frac{1}{2}\hat{\textbf{e}}_1^{\top}\textbf{C}\hat{\textbf{e}}_1, $ and $ \hat{\textbf{e}}_1^{\top}\textbf{C}\hat{\textbf{e}}_2\le  \frac{1}{2}\hat{\textbf{e}}_2^{\top}\textbf{C}\hat{\textbf{e}}_2
$
or, equivalently,  $ 
 {2|\hat{C}_{12}| \leq \text{min}(\hat{C}_{11}, \hat{C}_{22})},
$  
where $
\hat{\textbf{C}} = \textbf{C} /\text{det}({\textbf{C} })$.  Therefore the  interior  of the FED in this case  
is  defined by the conditions:
\begin{equation}
\begin{split}
\left|\hat{C}_{11}-\hat{C}_{22}\right|\le\frac{1}{2}\left( \frac{1}{\left|\hat{C}_{12}\right|}-3\left|\hat{C}_{12}\right| \right).
\end{split}
\label{eq37c3}
\end{equation}
 The ensuing global tessellation of the configurational space of metric tensors (symmetric 2D matrices with determinant equal to one) into equivalent elastic domains is illustrated in  Fig. \ref{FigEnergyMap3}.  

Here different colors correspond to different but equivalent, symmetry-related, elastic domains, each containing a single energy-minimum.   
To the right of it we  show {a} zoom on the FED{.} 
Note that it is divided in turn into four equivalent subdomains, each of them representing a  replica of the geometrically minimal FD. 
Note also that in Fig. \ref{FigEnergyMap3} we used a particular parametrization of the configurational space  using as coordinates  two independent components of the metric tensor, {namely,} $\hat{C}_{12}$ and $\hat{C}_{11}-\hat{C}_{22}$. {Various} other possible parameterizations of the same space are discussed in \cite{Roberta, SalmanBBZGT21,  BaggioST2023a}.

\paragraph{The update rules} We have seen that  under the condition of  quasi-static driving the  evolution of our quantized plasticity measure is  expected to be  a combination of  two types of phenomena. First, there should be `slow-time' stages, when plastic strain remains unchanged while elastic strain increases (quiescent periods). Second{,} there should b{e }`fast-time' stages, representing {the} transitions between different  elastic{-}equilibrium branches (plastic events). 

\begin{figure}[h!]
\centering
\includegraphics[scale=0.63 ,trim={0cm 0cm 0cm -1cm},clip]{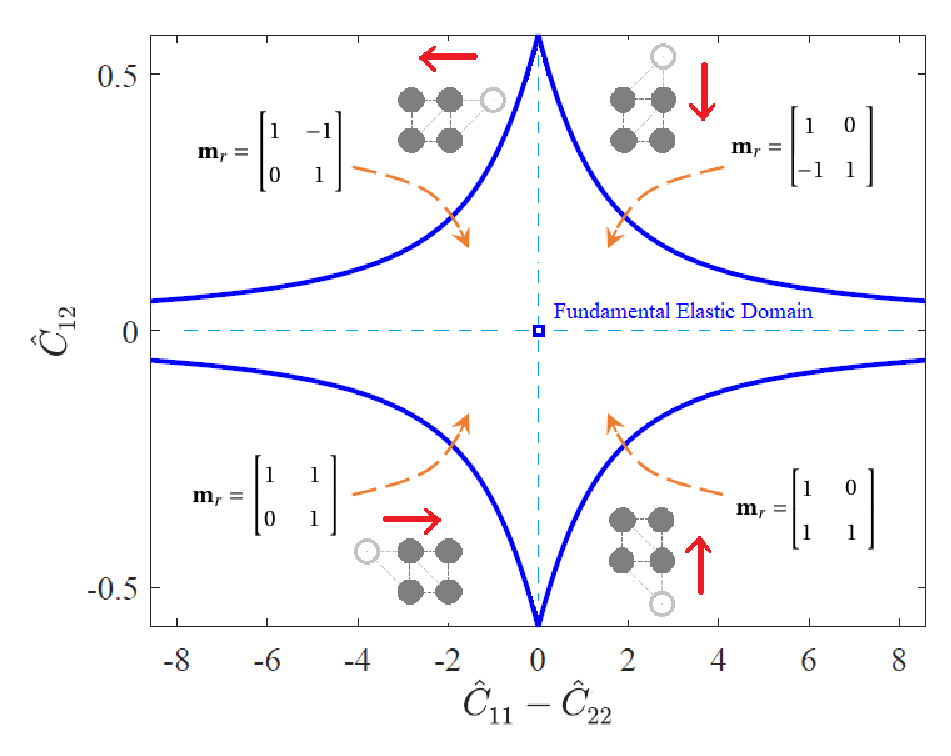}
 \caption{A schematic figure linking different matrices ${\textbf{m}}_r$ with different  boundaries of the FED. The crystallographic structures illustrate the corresponding lattice-invariant shears.}
\label{FigSchematicBound_mr} 
\end{figure}

Observe that in the tensorial setting, plastic events  necessarily imply  the crossing of a boundary of  one of the equivalent elastic domains.  In the case of  a 2D square lattice, such a boundary includes  four sub-boundaries  which correspond to the two slips (forward and backward) along  the two main slip directions (vertical and horizontal). In Fig. \ref{FigSchematicBound_mr}  we  present a schematic picture where we associate    these four elementary   transformations  with the crossing of the corresponding sub-boundaries of {the} FED.  

Suppose{,} next, that during a particular loading step, corresponding to{,} say, the $k$th plastic correction iteration, the local deformed configuration is described by the metric tensor $\hat{\textbf{C}}_k$ and the  measure of plastic strain is $\textbf{m}_k$. Suppose also that the reduced metric tensor 
 $\hat {\textbf{C}}^e_k=\textbf{m}_k^{\top}\hat {\textbf{C}}_k\textbf{m}_k$, 
which is the image of $\hat{\textbf{C}}_k$ in the FED, is in compliance with the `elastic stability' condition \eqref{eq37c3}. Finally, assume that the associated displacement field does not ensure elastic  equilibrium and an additional `elastic' equilibration (elastic correction) step is required.  
 
 This step produces a  new displacement field solving the elastic problem at fixed $\textbf{m}_k${,} which also generates a new local value of the metric tensor $\hat{\textbf{C}}_{k+1}$. Using the known value of $\textbf{m}_k$ we can compute the new  value of the reduced metric tensor $\hat{\textbf{C}}^e_{k+1}$ and check   whether the `yield condition' is respected,  i.e. that  the metric tensor $\hat{\textbf{C}}^e_{k+1}$ is indeed  inside the FED  limits given  by condition \eqref{eq37c3}. If this condition is violated, the local deformed configuration described by the metric tensor $\hat{\textbf{C}}^e_k$  lies outside the FED and one needs to activate a plastic correction by updating the plasticity measure from $\textbf{m}_{k}$ to  $\textbf{m}_{k+1}$. The choice of the  slip-plane to  be activated and  of the direction of the slip to be taken {depends} on the answer to the question which part of the boundary of the FED  has been crossed. This choice ultimately controls the new value of {the plastic} strain measure $ \textbf{m}_{k+1}$. 

The  above abstract  discussion  can be re-stated in the form of  an iterative algorithm. An elementary step of this algorithm can be written in the form:  
 \begin{equation}
 \label{eq371}
 \textbf{m}_{k} \to \textbf{m}_{k+1}=\textbf{m}_{k}{\textbf{m}}^r_{k}, 
 \end{equation}
where 

\begin{equation}
\begin{split}
 {\textbf{m}}^r_{k}=\begin{cases}\begin{bmatrix}
1 & \mp 1\\
0 & 1 
\end{bmatrix}, \ \pm 2\hat{C}^{e}_{k, 12}>\hat{C}^{e}_{k, 11} \ \text{and} \ \hat{C}^{e}_{k,22}>\hat {C}^{e}_{k,11} \\ \\
\begin{bmatrix}
1 & 0\\
\mp 1 & 1 
\end{bmatrix}, \  \pm 2\hat{C}^{e}_{k,12}>\hat{C}^{e,}_{k,22} \ \text{and} \ \hat{C}^{e}_{k, 11}>\hat {C}^{e}_{k, 22}.\end{cases}
\end{split}
\label{eq37}
\end{equation}
As an illustration of these general relations we now consider  an example. Suppose that the update is performed using an elementary lattice-invariant shear of the form
\begin{equation}
{\textbf{m}}^r_k=\begin{bmatrix}
1 & -1\\
0 & 1 
\end{bmatrix}  
\label{eq38a}
\end{equation}
which corresponds to a horizontal {(clockwise)} shear. 
Then $\hat{\textbf{C}}^{e }_{k+1,11}=\hat{\textbf{C}}^{e }_{k,11}, \
2\hat{\textbf{C}}^{e }_{k+1,12}=-2\hat{\textbf{C}}^{e }_{k,11}+
2\hat{\textbf{C}}^{e }_{k,12}$ and the new reduced metric tensor is  characterized by the following relations
\begin{equation}
\begin{split}
2\hat{\textbf{C}}^{e}_{k,12}>\hat{\textbf{C}}^{e}_{k,11}\Rightarrow 2\hat{\textbf{C}}^{e}_{k+1,12}>-\hat{\textbf{C}}^{e}_{k,11}  = -\hat{\textbf{C}}^{e}_{k+1,11}\ \Rightarrow 2\hat{\textbf{C}}^{e}_{k+1,12}>-\hat{\textbf{C}}^{e}_{k+1,11} \\
2\hat{\textbf{C}}^{e}_{k,12}<3\hat{\textbf{C}}^{e}_{k,11}  \Rightarrow 2\hat{\textbf{C}}^{e}_{k+1,12}<\hat{\textbf{C}}^{e}_{k,11} = \hat{\textbf{C}}^{e}_{k+1,11} \ \Rightarrow 2\hat{\textbf{C}}^{e}_{k+1,12}<\hat{\textbf{C}}^{e}_{k+1,11}
\end{split}
\label{eq38b}
\end{equation} 
In other words, if before the plastic correction we had
$\hat{\textbf{C}}^{e}_{k,11}<\hat{\textbf{C}}^{e}_{k,22},\hat{\textbf{C}}^{e}_{k,11}<2\hat{\textbf{C}}^{e}_{k,12}<3\hat{\textbf{C}}^{e}_{k,11}$,
and the conditions \eqref{eq37c3}  were not satisfied, then after the plastic correction we have  
$|2\hat{\textbf{C}}^{e }_{k+1,12}|<\hat{\textbf{C}}^{e}_{k+1,11}$,
which  implies that the `yield condition' is now satisfied. This indicates  that the quantized plastic  correction was performed adequately and that the discrete (quasi-)automaton has reached a stable fixed point.  It is now straightforward to provide  similar illustrations    when `plastic correction' implies  {the} crossing  of the  other three smooth sub-boundaries of the FED (and {involves the} three other elementary matrices $\textbf{m}^r_k$ in (\ref{eq37})).

The quasi-automaton  epitomized in (\ref{eq371}, \ref{eq37}) can be viewed as the  analogue of the conventional incremental `flow rule' postulated phenomenologically  in classical CP.  Two observations are in order. Note first that (\ref{eq371}, \ref{eq37}) is not a real automaton formulated in terms of the `condensed' variables only because the corresponding thresholds are not explicitly expressible in terms of the components of the integer-valued matrices $ {\textbf{m}}$. {Such} incomplete separability of {the} elastic and plastic problem{s} is {characteristic also for} the continuum CP approach. Note next that in our particular  setting the elementary `plastic corrections' performed by the quasi-automaton involve two types of binary choices: of the slip system and   of the slip direction.  Both are governed exclusively by the structure of the energy landscape. Instead,  in continuum CP such choices are not always unique (equivalent slip systems) and have to be made based on {\emph{ad hoc}} phenomenological constitutive assumptions {\cite{ Forest1998,Busso2005, Hartley2020, Dequiedt2023}}.

\paragraph{Configurational energy landscape} As we have already explained, 
in the `purely elastic' version of the MTM  the discrete $\text{SL}(2,\mathbb{Z})$ symmetry generates a natural tessellation of the configurational space of metric tensors into periodicity domains. In view of such (infinite)  periodicity, the generic metric $\textbf{C}$ with $\text{det}(\textbf{C})=1$ can be mapped onto the FED and the corresponding elastic response can always {be} described in terms of the reduced deformation measure $ \textbf{C}^{e}=\textbf{m}^{\top} \textbf{C}\textbf{m}$. To construct  the implied  (infinitely) periodic configurational energy density  landscape in the space of normalized metric tensors  $$\hat {\textbf{C}}=J^{-1}\textbf{C},$$  where   $J=\sqrt{\text{det}(\textbf{C})}$ describes the volumetric deformation, one needs to know the function $\psi$ only inside {the} FED. Then it can be simply extended beyond this FED {based on the} periodicity.

With this idea in {mind}, we have chosen, {for our numerical illustrations, to model the} elastic response inside the FED by  the  simplest expression for the function $\psi(\textbf{C}^e)$ {respecting} the point group symmetry of the square lattice {and having several desirable properties as detailed below}:
  \begin{equation}
 \begin{split}
 \psi(\textbf{C}^e)=  \frac{1}{2}\frac{K_{11}-K_{12}}{2}\left(\frac{C^e_{11}-C^e_{22}}{2}\right)^2+\frac{1}{2}K_{44}(C^e_{12})^2+\frac{1}{2}\frac{K_{11}+K_{12}}{2}(J-1)^2. 
\end{split}
\label{eq21}
\end{equation}
{With} $J$ fixed, the elastic response is physically linear. {The} parameters $K_{11},K_{12},K_{44}$ represent the standard cubic linear elastic moduli.
 In applications, one also encounters {combinations} of the same linear elastic constants{, such as} 
\begin{equation} \label{C1}
\kappa = (K_{11}+K_{12})/2>0,\xi = (K_{11}-K_{12})/2>0 ,  \eta = K_{44}>0.
\end{equation}
 {The} dimensionless ratio 
 \begin{equation} \label{C2}
\mathcal{A}= \eta/\xi >0
\end{equation}
 is known in linear elasticity of cubic crystals as the Zener parameter \cite{Walpole,Zener}. While the simple formula \eqref{eq21} for the energy inside the FED was chosen to boost the numerical efficiency of the resulting algorithm, other expressions, having similar general properties, could be used as well, see  \cite{Roberta, BaggioST2023b} where {at least three other possible choices are discussed}.

\begin{figure}[htpb!]
\centering
\centerline{
\includegraphics[scale=0.55,trim={0cm 0cm 0cm 0cm},clip]{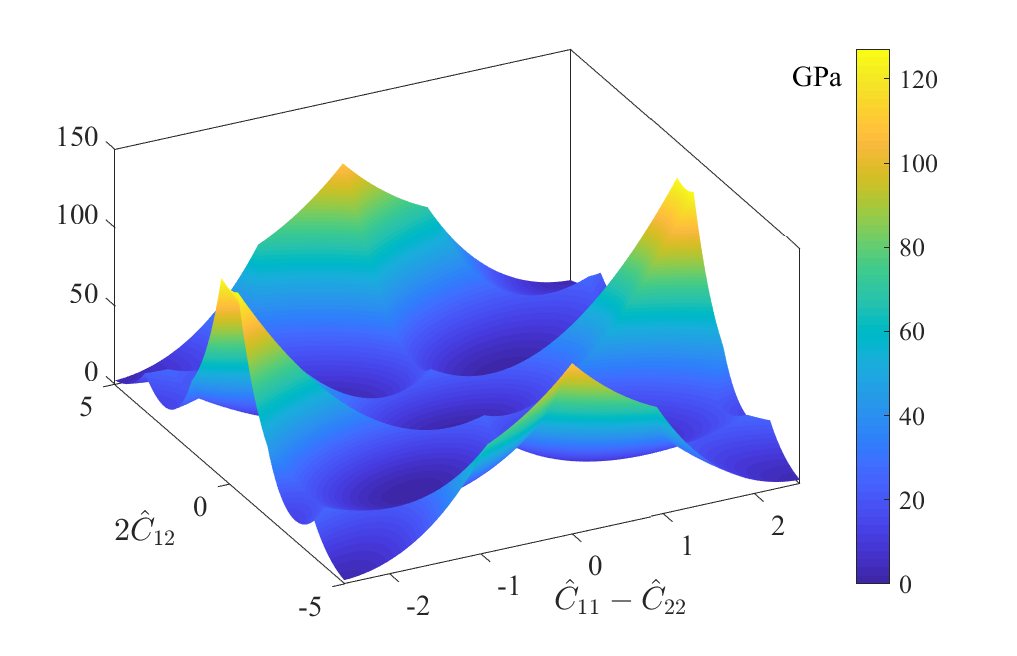}}
\caption{Elastic energy density $\eqref{eq21}$ for {$J=1$} extended periodically beyond the FED 
with the parameter choice  
$\xi=159.37\text{GPa}$ and $\eta=160.72\text{GPa}$, suitable for the description of ambient Tungsten \cite{Tungsten}.}
\label{FigEnergyMap1}
\end{figure}

In Fig. \ref{FigEnergyMap1} we illustrate the   periodic continuation  of the  energy density  \eqref{eq21} in the space of normalized metric tensors  $\hat {\textbf{C}}$, obtained  through piece-wise smooth   periodic  extension  from  inside the FED to the entire configurational space. The material described by the energy  density \eqref{eq21} is elastically  `well behaved'  locally{,} as the associated equilibrium equations can be shown to satisfy the conditions of strong ellipticity inside the FED. Globally,  the ensuing   energy density is, of course, no{t} rank{-}one convex{,} which is the reason we use it below only in a regularized setting. 

Note that to emphasize the `purely elastic' character of the resulting MTM  and to ensure that the  global energy is single{-}valued, the individual `parabolic'  branches of the energy density, shown in  Fig. \ref{FigEnergyMap1}, were truncated outside each replica of the FED. However, as we have explained in the discussion of the toy model, the new MTM framework, focusing on the evolution of the plastic strain measure $\textbf{m}$, suggests  that {the} individual `parabolas'  should not be truncated {outside} the limits of the corresponding FEDs.  

The problem is that in {the} nonlinear setting the reliance on such {a} multivalued energy density {function} may lead to artefacts. For instance, as we show below, additional constrains should be imposed on the coefficients in \eqref{eq21} to ensure that an elementary plastic correction in the resulting mutivalued energy setting is always dissipative.

\paragraph{Dissipativity} Suppose that at a certain iteration involving plastic correction, the values of $\textbf{m}$ and $\hat{\textbf{C}}$ are such that $2\hat{C}^{e}_{12}>\hat{C}^{e}_{11}, \hat{C}^{e}_{11} <\hat{C}^{e}_{22}$ and therefore the conditions  in \eqref{eq37c3} are violated; here $\hat{\textbf{C}}^{e}=\textbf{m}^{\top}\hat{\textbf{C}}\textbf{m}$ is  the reduced (projected onto the FED) metric in a given iteration of the plastic correction. We suppose further that for small enough loading increments the metric $\hat{\textbf{C}}^e$ violates the yield conditions only slightly.

A subsequent plastic correction produces the metric $\hat{\textbf{C}}^{e'}=\textbf{m}^{r \top}\hat{\textbf{C}}^{e}\textbf{m}^r$ with  $\textbf{m}^r$   given in \eqref{eq38a} and therefore   
$\hat{C}_{11}^{e'}=\hat{C}_{11}^{e},2\hat{C}_{12}^{e'}=-2\hat{C}_{11}^{e}+2\hat{C}_{12}^{e}, \ \hat{C}_{22}^{e'}=\hat{C}_{22}^{e}+\hat{C}_{11}^{e}-2\hat{C}_{12}^{e}$.
We can now  introduce new variables $(x,y)$ and $(x',y')$ such that 
$0<y'=2\hat{C}_{12}^{e'}=2\hat{C}_{12}^{'}-2\hat{C}_{11}^{e}<y= 2\hat{C}_{12}^{e}$ and $  0>x'= \hat{C}_{11}^{e'}-\hat{C}_{22}^{e'}=2\hat{C}_{12}^{e}-\hat{C}_{22}^{e}>x=\hat{C}_{11}^{e}-\hat{C}_{22}^{e}.$
Therefore, the plastic correction  transforms the point $(x,y)$ into the point $(x',y')$ such that $x'^2<x^2,y'^2<y^2$. This plastic correction is dissipative if 
$\psi(x'^2,y'^2)<\psi(x^2,y^2)$
for all appropriate $x,x',y,y'$. Assuming that the energy density is smooth inside the FED, we can then write the conditions 
$   \partial\psi(x^2,y^2)/\partial x^2 =\mathcal{C}_1(x^2,y^2)>0,   \partial\psi(x^2,y^2)/\partial y^2=\mathcal{C}_2(x^2,y^2)>0$,
 as sufficient for plastic dissipativity. In addition to the case   $x<0,y>0$, considered above, we could also similarly consider the cases $x<0,y<0$, $x>0,y<0$ and $x>0,y>0$, and for all of them  the obtained sufficient  conditions   guarantee   dissipativity.  For instance, the dissipativity  is ensured if the energy density satisfies the conditions   $\mathcal{C}_1(x^2,y^2)=\text{const} >0,\mathcal{C}_2(x^2,y^2)=\text{const}>0$. In the case of the energy density \eqref{eq21} these constraints reduce to inequalities to be satisfied by the elastic moduli, namely, $\xi=8\mathcal{C}_1>0,\eta=8\mathcal{C}_2>0$ and $\mathcal{A}=\mathcal{C}_1/\mathcal{C}_2>0$. In our numerical experiments  we used the elastic parameters $K_{11}=523.27\text{GPa},K_{12}=204.53\text{GPa}$,  $K_{44}=160.72\text{GPa}$ corrersponding to Tungsten at ambient conditions \cite{Tungsten}, and  evidently compatible with the  plastic dissipativaty constraints.


\paragraph{Numerical algorithm} In  the absence of either PDE-based or ODE-based formulations of the MTM, the discrete computational scheme, {in} itself, should be considered as a crucial element of the proposed mesoscopic approach to crystal plasticity. 

The first step in designing such a scheme  is to incorporate the  globally periodic multivalued energy density into a spatially discretized  framework. The discretization, representing physical regularization of the model,  is obtained by introducing finite elements (FE) whose (coarse-grained, mesoscopic) mechanical  response is assumed to be affine. 

In our specific setting the FEM discretization is achieved by introducing a square grid of computational nodes, aligned with the presumed material crystallographic symmetry axes of the 2D crystal. Each square of the grid is then divided by the principal diagonal into two right isosceles triangular elements. 

Note that the ensuing  DE formulation contains an internal length scale $h_0$, the element spacing, which is viewed as a physical parameter. The physical fields are then interpolated inside the elements using the most basic piecewise linear, tent basis functions.

The \emph{elastic}  problem then reduces to finding the discrete displacement field $\textbf{u}(\textbf{x})=\textbf{y}(\textbf{x})-\textbf{x}$ minimizing the total elastic energy of the system under the  boundary conditions $ \textbf{u}|_{\textbf{x}=\textbf{x}_b} =
 \textbf{u}_b  (\alpha)$, where $\alpha$ is the loading parameter. 
 The interpolated displacement field can be written in the form $\textbf{u}(\textbf{x} )=\underset{i}\Sigma{\phi_i(\textbf{x})\textbf{U}_{i}}$, where $\textbf{x}$ are reference coordinates, $\phi_i(\textbf{x})=\int{\mathbb{D}_{ij}d\textbf{x}_j}$ are the piecewise-linear, compact-support shape functions and $\mathbb{D}_{ij}$ are matrices defined by 
$$\begin{pmatrix} F_{11} \\ F_{12} \\ F_{21} \\ F_{22} \end{pmatrix} =\begin{pmatrix} 1 \\ 0 \\ 0 \\1 \end{pmatrix}+\mathbb{D}\begin{pmatrix} U^{(1)}_{1} \\ U^{(1)}_{2} \\ U^{(2)}_{1} \\ U^{(2)}_{2} \\ U^{(3)}_{1} \\ U^{(3)}_{2} \end{pmatrix}, \ \ {\mathbb{D}}=h_0^{-1}\begin{pmatrix} -1 & 0 &1 &0 &0 &0 \\ 0 &0 &1 &0 &-1 &0 \\ 0 &-1 &0 &1 &0 &0 \\0 &0 &0 &1 &0 &-1 \end{pmatrix}.$$  
Denoting the nodal displacements (column) vector by $\bar{\textbf{U}}=[\textbf{U}_{1}^{\top},\textbf{U}_{2}^{\top},...]^{\top}$, we can write the
total elastic free energy  as a sum of the energies of the elements: 
$
\bar\Psi (\bar{\textbf{U}}) =\sum\limits_{i,j}{ V_i\psi\left(\mathbb{D}_{ij}\textbf{U}_{j}, \textbf{m}_i\right)}
$, where $V_i=h_0^2/2$  and $\textbf{U}_{i}$ are nodal row vectors. In what follows  the regularizing (cut-off) parameter $h_0$  is assumed to be fixed  so that the question of mesh refinement is not posed;  the optimal choice for the  parameter {$h_0$} will be discussed elsewhere.
 In all our  numerical experiments we assumed that a linear displacement field compatible with \eqref{eq21fh} and parametrized by $\alpha$ is prescribed on the boundary of the domain (hard device). The boundary conditions are then $
 \textbf{u}_b =
  [\textbf {F} (\alpha) - \textbf {I}]\textbf{x}_b $.
  
In the description above we implicitly assumed that the original problem could be  split into \emph{elastic} and \emph{plastic} problems.
As we have seen the  \emph{elastic problem} pre-supposes that the  set of integer-valued plastic strain variables $\textbf{m}_i$ is known. Then, indeed the solution of the \emph{elastic problem}  {consists in finding} at each value of the loading parameter  the energy-minimizing discrete set of displacement variables $\textbf{U}_{i}$, for $i=1,...,n_xn_y$, representing the deformed locations of the nodes of the FE mesh. However, the set of plastic variables $\textbf{m}_i$ should also emerge as an outcome of energy minimization  
modulo the condition that the entries in $\textbf{m}_i$ always remain integer-valued with det$(\textbf{m})=1$, and that the plastic update only occurs if yield conditions applied to $\textbf{U}_{i},\textbf{m}_i$ are violated. The challenge is to design an algorithm where  the corresponding  \emph{plastic problem} is maximally separated from the elastic one.


In our algorithm the parallel energy-minimization problems for $\textbf{U}_{i}$ and $\textbf{m}_i$ are solved at each increment of the loading in a sequence of two-step operations. 

At the first, \emph{predictor} stage, 
the {response} is assumed to be {elastic and use is made of} the tangent moduli calculated from the solution for the previous loading increment. {T}he plastic strain $\textbf{m}_i$ at this step is kept equal to the converged value from the previous increment of the loading. 

At the second, \emph{corrector} stage we check in which elements the yield conditions were violated, perform plastic reduction there, and then solve the elastic problem again for the nodes associated with those elements  in a fixed-boundary setting. 
Such refinement of the solution is performed iteratively, and at every iteration a new improved update for the plastic variables $\textbf{m}_i$ is obtained, along with an improved update for the `elastic variables' $\textbf{U}_i$ (during each of such incremental refinements of the solution, plastic {reduction} is performed {when needed}, until the `yield condition'  \eqref{eq37c3} is satisfied everywhere). 

More precisely, if in a given element the value of $\textbf{C}_e$, calculated based on  the current value of $\textbf{m}$, {falls} outside the FED, the $\textbf{m}$ matrix is updated. 
Then, {(first-order)} elastic energy minimization is performed by updating the global nodal displacements vector. Since the minimization is of {the} first order, a correction to the displacement field is chosen {to be} proportional to the vector of residual forces on the nodes (force mismatches at the nodes between adjacent elements). The corresponding {pre-factor} (step-size) is found by sub-iterations to {ensure} the maximum energy decrease. 

One noteworthy feature of the proposed approach is that the energy minimization procedure is implemented in a single loop over all the unstable elements and therefore there is no ambiguity regarding the order of the updates. Another one is that the coupling between the elements is only activated during the elastic predictor step -- we discuss this in \ref{AppendixA}, where {the} numerical implementation of the  algorithm, involving a hybrid Gauss-Newton--Cauchy energy{-}minimization scheme, is presented in full detail. 
 
Here we only mention that the additional numerical acceleration vis-a-vis the `purely elastic' approach, achieved in the proposed algorithm, is {due} to the use {of an} elastic energy density functional that is `well-behaved' inside the FED. This allows the minimization algorithm to converge with only a small number of iterations{,} as the elastic problem is effectively solved `on the fly'. Moreover, the piece-wise smooth structure of the configurational energy landscape allows one to {avoid navigation through a} complex configuration of saddle points.  Instead, we use an efficient  nonsmooth discrete mapping algorithm. 

Further acceleration {is achieved by the use of a} Cauchy update {during plastic corrections (which would otherwise take a significant portion of the computational effort in each increment)} instead of the LBFGS update, as in the purely elastic approach. This allows one to recalculate at each {iteration} only those components of the energy gradient and only those contributions to the total energy that are associated with the elements that have just {undergone} plastic correction. In view of the achieved {nearly} complete separation between the continuous elastic and discrete plastic problems, we {regard} the proposed algorithm as a (quasi-)automaton. We discuss the   computational efficiency of the {developed} numerical method  in more quantitative terms in \ref{AppendixB}.

 \paragraph{Quenched disorder}  {In view of the massive degeneracy} associated with the possibility of collective crossing of the `ridges' of the energy landscape (the boundaries of the FED)  in pristine   crystals, some additional regularization is called-for.  The problem  is resolved in the proposed algorithm through the  introduction of small quenched disorder. 
 
 To minimize the impact of such disorder, we randomized only the displacements on the boundary using the decomposition  $\Delta\bar{\textbf{U}}_b=\Delta\bar{\textbf{U}}_b^{(o)}+\Delta\bar{\textbf{U}}_b^{(d)}$, where the first term is the deterministic contribution 
\begin{equation}
\begin{split}
\Delta\bar{\textbf{U}}_b^{(o)}=\Delta {\textbf{F}}\textbf{x}_b=\Delta\alpha\begin{bmatrix} 0 &1 \\ 0 & 0\end{bmatrix}\textbf{x}_b,
\end{split}
\label{eqappld2}
\end{equation}
while the second term, described by the relations
 \begin{equation}
^{(i)} \Delta\bar{\textbf{U}}^{(d)}_{b_{\text{bot/top}}}\cdot \hat{\textbf{e}}_1=0, \  ^{(i)}\Delta\bar{\textbf{U}}^{(d)}_{b_{\text{left/right}}}\cdot \hat{\textbf{e}}_2=0
, \ ^{(i)}\Delta\bar{\textbf{U}}^{(d)}_{b_{\text{left/right}}}\cdot \hat{\textbf{e}}_1=0,  ^{(i)}\Delta\bar{\textbf{U}}^{(d)}_{b_{\text{bot/top}}}\cdot \hat{\textbf{e}}_2=\delta_d^{(i)},
\label{eqRandomParam}
\end{equation}
depends on the random parameters $\delta _d^{(i)}$.  The latter were sampled for each node $i$ independently, using a Gaussian distribution with a zero mean and standard deviation $\mathfrak{s}$ which  was chosen sufficiently small ($\mathfrak{s}/h_0=10^{-9}$)  to exclude effects {that} could be interpreted as physically-significant {\cite{Umut2011,Peng}}.



To make our assumptions in \eqref{eqRandomParam} more transparent, we assume that a square crystal is deformed with four rigid plates (lines) of equal {dimensions} moving as a  parallelogram (the bottom side being fixed).  
{To} mimick surface roughness along the sides of the sample, we assumed that the disorder affects only {the} components of the displacement field $^{(i)}\Delta\bar{\textbf{U}}^{(d)}_{b}$ oriented along the normal to the boundaries. Also, for simplicity, the implied quenched disorder \eqref{eqRandomParam} was introduced only along the top and bottom (horizontal) boundaries (for it to be normal both to the surface and to the displacements). Since the corresponding two (top and bottom) loading  plates were assumed to have `bumps' on their surface, inducing compression, at the nodes at which the generated random displacements turned out to be tensile, the random components of  $^{(i)}\Delta\bar{\textbf{U}}^{(d)}_{b}$ were replaced by zeros. 

Another factor which we attempted to take into account is the macroscopic nature of the quenched disorder.  
This consideration requires the spatial resolution of the disorder to remain below the mesoscopic features of the model characterized by the parameter $h_0$. In this respect, imposing random fluctuations on all boundary nodes would be in contradiction with the  idea that the disorder is macroscopic. Therefore, the random corrections to the imposed boundary displacement field  were neglected on some boundary nodes (by keeping only those values that were larger than the standard deviation). 

\paragraph{{Link to classical CP}} 
To interpret the results of our meso-scopic numerical experiments in macroscopic terms we need a quantitative basis for comparison of {the} MTM with CP. The first question is whether in the re-formulated MTM the increment of plastic strain, serving as the analogue of the plastic velocity gradient $$\textbf{L}_p=\dot{\textbf{F}}_p{\textbf{F}}_p^{-1},$$ can be additively decomposed into rank-one components. In other words, the question is whether the proposed kinematic description is compatible with the CP ansatz 

\begin{equation} \label{22}
{\textbf{L}}_p=\sum\limits_{j=1}^{N_s}{\dot\gamma_j}\textbf{s}^{(j)}\left[\textbf{n}^{(j)}\right]^{\top}, 
\end{equation} 
where the unit vectors $\textbf{s}_j$ and $\textbf{n}_j$ represent the tangent and normal vectors to crystallograpically-specific slip planes. Observe first that in the re-formulated MTM the analogue of the plastic velocity gradient is the quantized increment of plastic strain. Indeed, if we write $$\textbf{L}_p^{(k)} \Delta{t}= -\left(\textbf{m}^{(k)}\right)^{-1}\Delta{\textbf{m}}^{(k)},$$ 
where $$\Delta{\textbf{m}}^{(k)}={\textbf{m}}^{(k+1)}-{\textbf{m}}^{(k)},$$ then using   (\ref{eq37}) we obtain    $\textbf{L}_p^{(k)} \Delta{t}= -{\textbf{m}}_r^{(k)}+\textbf{I}$. 
We can then express   $\Delta{\textbf{m}}_r^{(k)}\triangleq{\textbf{m}}_r^{(k)}-\textbf{I}$ as a sum of two rank-one tensors: 
\begin{equation}
\begin{split}
\Delta{{\textbf{m}}_r^{(k)}}=-\textbf{L}_p^{(k)} \Delta{t}=-\Delta\gamma_{12}^{(k)}\begin{bmatrix} 0 &1 \\0 & 0 \end{bmatrix}-\Delta\gamma_{21}^{(k)}\begin{bmatrix} 0 &0 \\1 & 0 \end{bmatrix}.
\end{split}
\label{eq26b2}
\end{equation}
The next step is to combine  the individual iterations constituting a plastic correction.  Under the assumption that it is comprised of  $k_*>1$ fast time steps 
we can write $$ \Delta{{\textbf{m}}_r^{(k_*)}}=\prod_{k=1}^{k*}{\textbf{m}}_r^{(k)}-\textbf{I}.$$ Therefore in slow time $$\ \textbf{L}_p=\lim_{\Delta t \to 0} \left[-\frac{1}{\Delta{t}}\left[\prod_{k=1}^{k*}{\textbf{m}}_r^{(k)}-\textbf{I}\right]\right],$$ where  $\Delta t=\sum_{k=0}^{k*}{\Delta t_k}$. 
Therefore, in view of  \eqref{eq26b2}  we can write 
\begin{equation}
\label{eqLp}
{\textbf{L}}_p= {\dot\gamma_1}\begin{pmatrix} 1 \\ 0\end{pmatrix} \begin{pmatrix} 0 & 1\end{pmatrix}+ {\dot\gamma_2}\begin{pmatrix} 0 \\ 1\end{pmatrix} \begin{pmatrix} 1 & 0\end{pmatrix},
\end{equation}
where the pre-factors $\dot\gamma_j$ characterize the rates of the corresponding plastic slips. The ansatz \eqref{eqLp} is in agreement with the classical CP representation of plastic slip rate \eqref{22} given that in 2D square lattices  $s^{(j)}_i=\delta_{ji}$ and $n^{(j)}_i=1-\delta_{ji}$. 
%

\paragraph{{Coarse-grained response}} 

Assume again that our computational domain, representing a  sample,  is a  macrosopic square containing   $N$ mesoscopic square finite elements.  This macroscopic square  is  deformed in simple shear  to form   a parallelogram  and we can   interpret   the resulting total energy per volume of the  sample as the coarse-grained   elastic energy density associated with such  shear deformation.

Since energy is material additive (and the elements are identical in Lagrangian volume) we can obtain the coarse-grained energy density simply by averaging the energy densities of the $N$ finite elements: \begin{equation}
 \langle\psi\rangle=\frac{1}{N}\underset{i}{\sum}{\psi_i}.
 \label{eqCGae1}
\end{equation}
Assuming elastic response  and using definitions \eqref{eq21}, \eqref{C1} and \eqref{C2}, we can express the measure of the macrosopic geometric simple shear through the coarse-grained stored energy denisty \eqref{eqCGae1} as follows:  
\begin{equation}
\begin{split}
\alpha_e = \sqrt{4\sqrt{\frac{1}{4}\mathcal{A}^2+\frac{1}{2\xi}\langle\psi\rangle}-2\mathcal{A}}
\end{split}
\label{eqCGae}
\end{equation}
Note that expression \eqref{eqCGae} can be used to represent the elastic component of the macroscopic strain even outside the purely elastic regime. 
Next, applying the multiplicative elastoplastic decomposition to the entire computational domain, and substituting the assumption of simple horizontal shear, one obtains the macroscopic plasticity measure as
\begin{equation}
\begin{split}
 \langle- m_{12}\rangle =
  \alpha - \alpha_e 
\end{split}
\label{eqCGap}
\end{equation}
where $\alpha$ is the macroscopic applied shear strain measure.

We can also express the relevant component of the macroscopic shear stress as the derivative of the macroscopic energy with respect to the macroscopic strain
\begin{equation}
\begin{split}
\langle P_{12} \rangle =\left.\frac{\partial\langle\psi\rangle}{\partial\alpha}\right|_{\alpha_e}.
\end{split}
\label{eqCGP12psie}
\end{equation}
We observe  that this derivative should  be calculated along an elastic branch, which means as the limit of a finite difference, with no plastic events in between. 

Using such macroscopic measures we will obtain from our mesoscopic model the effective macrosocpic constitutive relations, which can already allow comparison with experimental results.

\section{Case study}

We now discuss the outcomes of our computational experiments{,} which were performed under several realizations of quenched disorder. {Some of the} results are presented {after} averaging over the disorder.

An initially-square domain was divided into $N=n \times n $ square cells with the linear side scale $h_0=1\mu m$. 
The spatial resolution was set by the system size $n=100$.  
 The affine deformation \eqref{eq21fh}
was applied incrementally to the nodes on the boundary of the computational domain (the four edges of the square computational  box).
The loading was  updated with $\alpha_{l+1}=\alpha_l+\Delta\alpha, $ starting from $
\alpha_0=0 $ (unloaded sample) up to $\alpha=1$ (developed  plasticity), with the {value} of the increment  $\Delta{\alpha}= 2\times10^{-4}$ operative after the yield at $\alpha=0.5$. {Such a} value of the increment was found {to be} sufficient{ly small} to both fully resolve intermittent plastic avalanches and provide the convergence of the numerical algorithm.

 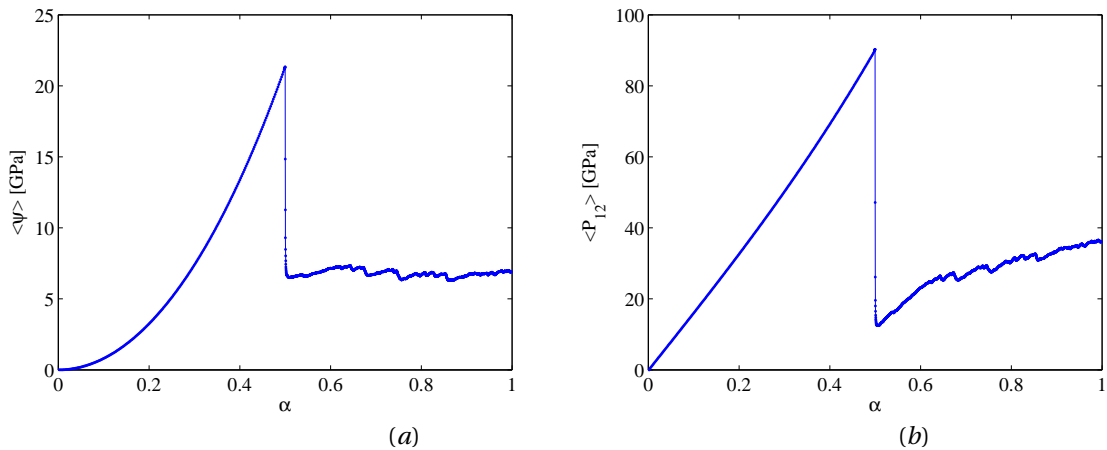
\begin{figure}[h!]
\begin{center}
\includegraphics[scale=0.52,trim={0cm 0cm 0cm 0cm},clip]{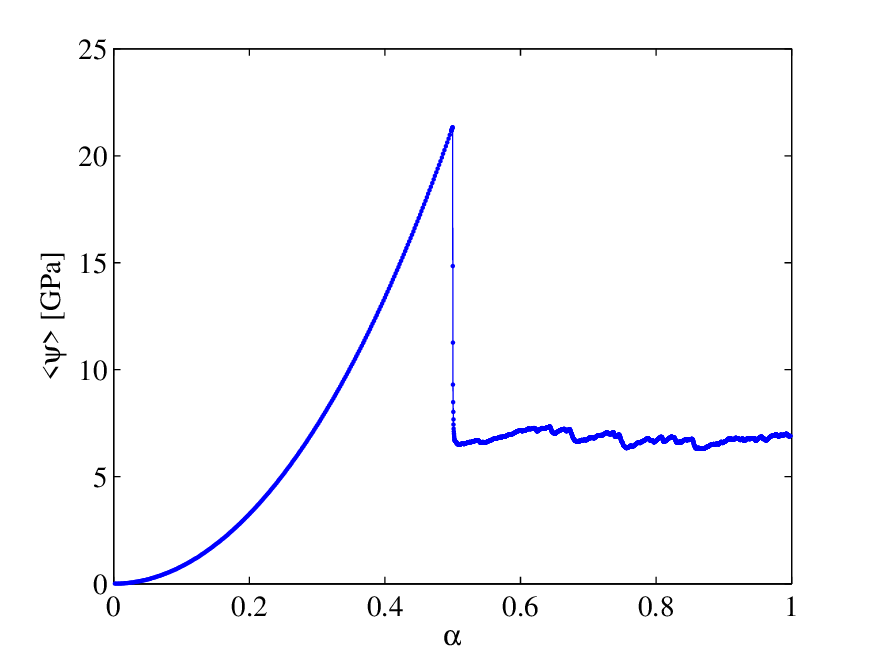}
\includegraphics[scale=0.52,trim={0cm 0cm 0cm 0cm},clip]{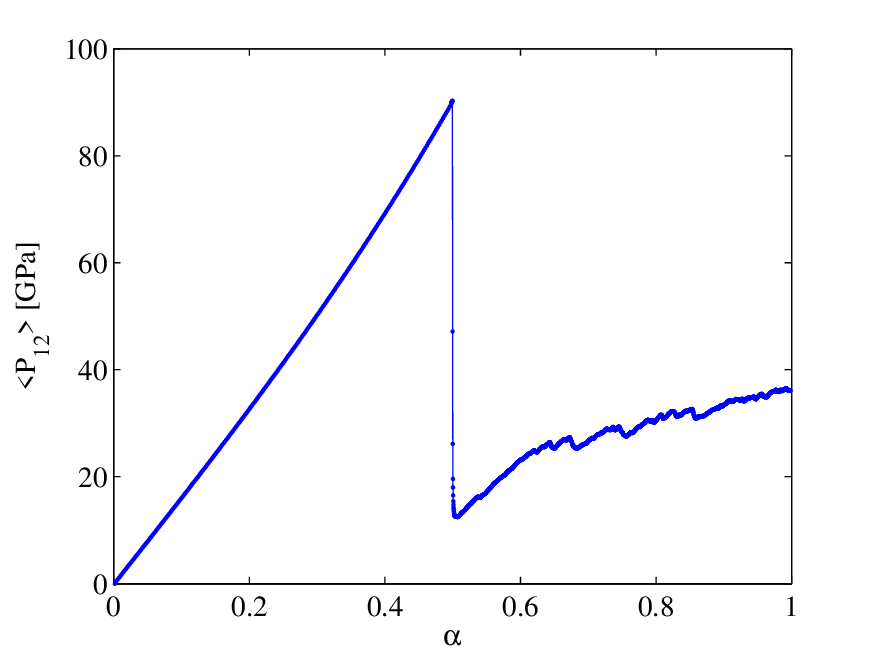}
   \begin{picture}(0,0)
  \put(-290,-10){$(a)$ }
  \put(-99,-10){$(b)$ }
  \end{picture}
\end{center}
{\caption{\label{FigProfs} Macroscapic constitutive response:  (a) coarse-grained free-energy density versus applied macroscopic strain (scaling with the loading parameter $\alpha$),
(b) the coarse-grained shear stress versus $\alpha$.
}}

\end{figure}


\paragraph{Macroscopic constitutive response} In Fig. \ref{FigProfs}(a) we show the strain dependence of the  coarse-grained energy density  $\langle\psi\rangle$. The loading parameter $\alpha$, introduced in  \eqref{eq21fh} and \eqref{eqappld2}, is changing quasistatically and at each value of this parameter a locally stable equilibrium state is reached  as a result of  energy minimization with respect to elastic and plastic degrees of freedom. 
For the same loading protocol we show in Fig. \ref{FigProfs}(b) the concurrent strain-induced evolution of the coarse-grained {1st} Piola-Kirchhoff stress tensor represented by the component $\langle P_{12}\rangle$, which is conjugate to the varying component of the applied distortion \eqref{eqappld2}. 
 
According to Fig. \ref{FigProfs}(a,b), in the interval $0<\alpha<0.5$ the response is elastic and the global deformation is (almost) affine, with all the elements following essentially  identical trajectories in the configurational space of metric tensors (while remaining inside the FED). During this deformation stage, the fluctuations are effectively absent and the classical continuum elasticity theory is fully applicable. Despite the overall nonlinearity of the model, this purely elastic response along the simple shear deformation path is almost linear. The quenched disorder is apparently too small to cause any tangible plastic deformation.  Such a stage  is realistic only for specially grown pristine crystals and its main role here is to \emph{prepare} a more generic state of the crystal with a meaningful  configuration of crystal defects.

The role of such preparation is played by a system-size dislocations avalanche converting a singular (cold) configuration into a generic (hot) one. Such an avalanche takes place at $\alpha=0.5$ when the configurational points collectively reach the boundary of the FED  and the crystal undergoes  massive instability. The system is frustrated due to the  necessity to perform plastic correction simultaneously in almost all elements, and in our algorithm the associated massive correction is performed  in a single loop over all the elements.

The system-size avalanche at $\alpha=0.5$ proceeds (in fast time) through the nucleation of a large number of dislocations, which partially annihilate and partially self-organize inside the computational domain,  eventually  forming a complex post-avalanche spatial distribution of mutually locked dislocations. During this process the stress drops and a large amount of energy is  dissipated. The dissipated {energy} is not accounted for directly, as we neglect the possibility of its transport by waves and keep the processes of thermalization and heat conduction unresolved.


As the loading process proceeds  beyond $\alpha=0.5$,  we observe in Fig. \ref{FigProfs}(a,b) the emergence of pronounced fluctuations. Detailed analysis shows that intermittent  dislocation avalanches intermix with extended periods of purely elastic behavior. Despite the overall macroscopic hardening, the elastic energy remains practically constant, as almost all the work of the loading device, accumulated during the elastic stages, is effectively dissipated during intermittent avalanches, see Fig. \ref{FigProfs}(a). The avalanches take the form {of} broadly distributed plastic events. The fact that after each of these events the energy is practically recovered during quiescent periods is in agreement with the general observation that the stored energy of cold work is extremely small. 

\begin{figure}[htpb!]
\begin{center}
\includegraphics[scale=0.5 ,trim={0cm 0cm 0cm 0cm},clip]{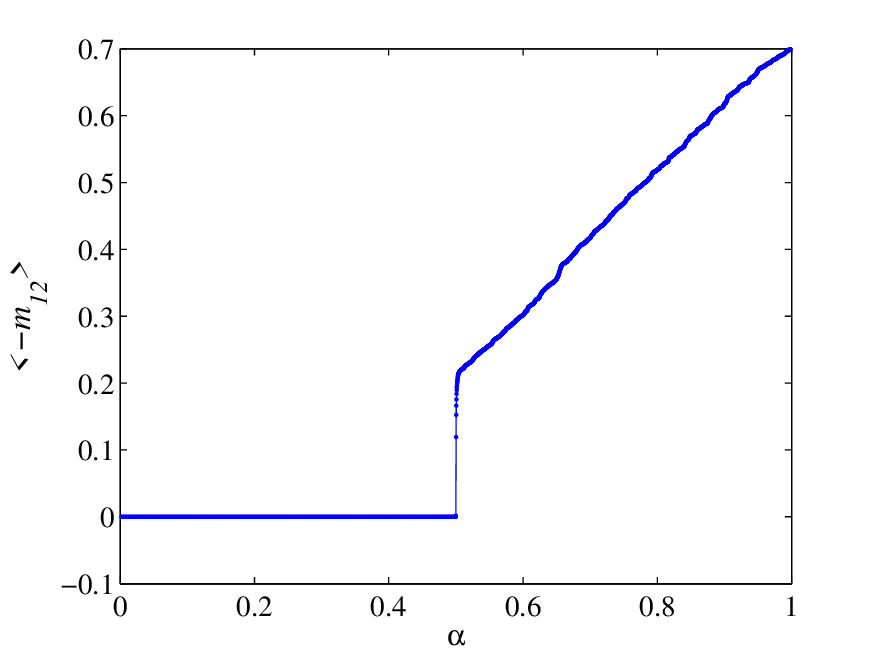}
  \begin{picture}(0,0)
  \end{picture}
\end{center}
\caption{
  Coarse-grained measure of plasticity ($\textbf{m}$) versus the loading parameter $\alpha$ }
\label{FigProfs1}
\end{figure}

As our re-formulation of the MTM gives us direct access to plastic strain, it is reasonable to raise the question of its dependence on the loading. A natural coarse-grained measure of plasticity attuned to our particular loading protocol is again  $\langle-m_{12}\rangle$. As one can learn from Fig. \ref{FigProfs1}, showing the dependence of this plastic strain measure on the loading parameter $\alpha$, no plastic deformation takes place till the system reaches the boundary of the FED. After the system-size avalanche at $\alpha=0.5$,   plastic strain  grows almost linearly with $\alpha$. This power law correlation is another indication that the steady-hardening regime is (almost) scale free. The dip  experienced by the function $\langle-m_{12}\rangle(\alpha)$ around the system-size avalanche at $\alpha=0.5$  is a signature of a major restructuring in the highly ordered pristine crystal, giving rise to a network of grain boundaries and  preparing the crystal in this way for the subsequent `mature' scale-free plastic flow.

\begin{figure}[htpb!]
{
\begin{center}
\includegraphics[scale=0.54 ,trim={0cm 0cm 0cm 0cm},clip]{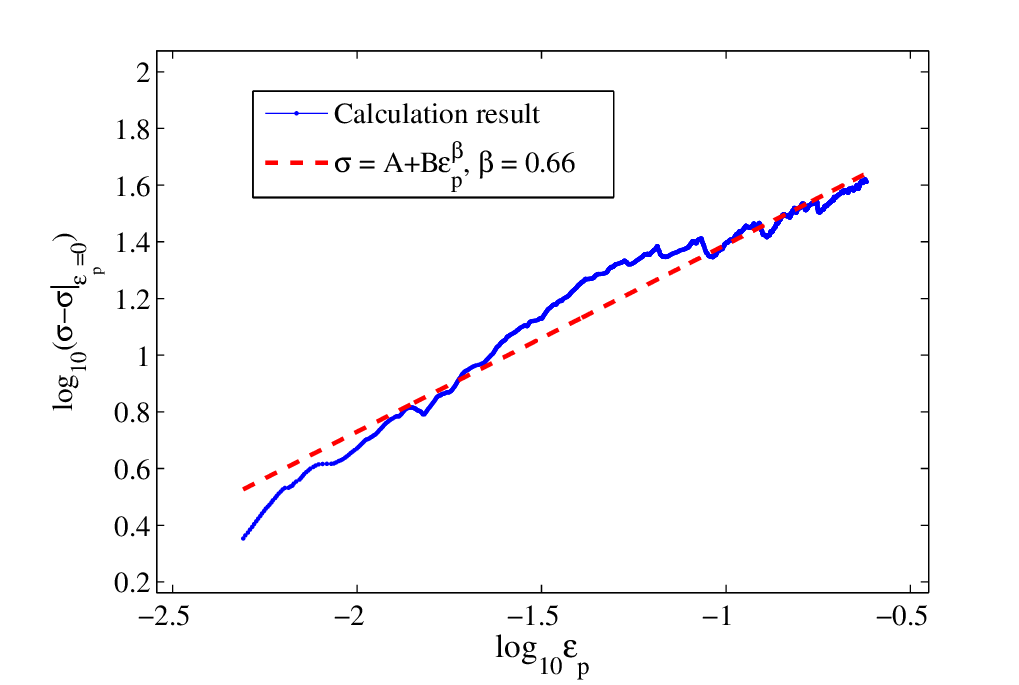}  
\end{center}
}
\caption{\label{FigPowerLawmZAlog} Effective (von-Mises) stress vs. effective plastic strain (blue solid line with a  dot marker) with a power-law fit for the trend (red dash), on a log-log scale -- showing almost two decades of power-law behavior, the optimal fit producing the exponent $\beta=0.66$}
\end{figure}

To characterize the after-yield ductile hardening regime more precisely, we show in Fig. \ref{FigPowerLawmZAlog} the relation between the averaged effective stress $ \sigma \approx \sqrt{3}\langle P_{12}\rangle$ and the averaged effective small plastic strain $\epsilon_p = \frac{1}{2}\langle- m_{12}\rangle$. This relation is presented in log-log coordinates to show that it is in agreement with the well known phenomenological Johnson-Cook correlation, describing the dependence of the effective (von Mises) stress on plastic strain, 
$
\sigma =A+B\varepsilon_p^\beta,
$ which is often used to rationalize the results of physical experiments on polycrystalline-metal plasticity \cite{JC}. As we show below, the system-size avalanche practically turns the initial pristine single-crystal into an effective polycrystal, with no characteristic grain size, and the presence of such a  power-law type correlation provides an evidence for the developing scale-free structure.

Interestingly, as our numerical experiments are targeting Tungsten-like 2D crystals, the obtained hardening exponent $\beta \approx  0.66$ (see Fig. \ref{FigPowerLawmZAlog}) can be favorably compared to the exponent $\beta=0.63$ obtained experimentally for polycrystalline Tungsten \cite{Goh}.

We should note that the reason for our choosing of specifically Tungsten parameters is related to the fact that crystal Tungsten is one of the most elastically-isotropic elements, and therefore one can expect a 2D description be more representative of 3D behavior than for other elements, especially when comparison to 3D experiments is concerned. For less isotropic materials it is possible that complex 3D modes become the energetically favorable ones and they would be those that would underlie the mechanism responsible for observed response.

For small quenched disorder, usually, after the first (big) yield the emerging coarse-grained stress is relatively small. Only after some additional (plastic) deformation the stress rises to the level normally observed in experiments (for zero measured macroscopic plastic strain). Therefore, a correct coarse-grained plastic strain would need to be measured relative to a value for which the microstructure is already randomized enough to represent real metals, corresponding to some given value of stress. We therefore define the plastic strain measure to be compared to the experimental one as plastic strain starting from the point where it starts to grow approximately linearly with the applied global strain. For the performed calculations on the 2D crystal with Tungsten elastic constants, we get $\epsilon_p^*\approx 0.12$. Consequently, we set $\varepsilon_p=\epsilon_p-\epsilon_p^*$.

We terminated the  loading protocol at $\alpha=1.0$ under the assumption that the plastic flow, observed in the range $ 0.5<\alpha<1.0$,  is steady, the system having reached a non-equilibrium steady state (NESS). This, however, may be a transient regime and new system-size avalanches may be expected as the loading extends  beyond $\alpha=1.0$. Such avalanches may be associated with the formation of shear bands, a major restructuring of  dislocation cell structures and even the formation of cracks. These and other intriguing aspects of `mature' plastic flows are left for future studies.

\paragraph{Temporal  intermittency} As we have seen, in the regime of steady state plastic flow the system exhibits temporal fluctuations. We now discuss the statistical distribution of the corresponding plastic avalanches.  

\begin{figure}[h!]
\begin{center}
{
\includegraphics[scale=0.54 ,trim={0cm 0cm 0cm 0cm},clip]{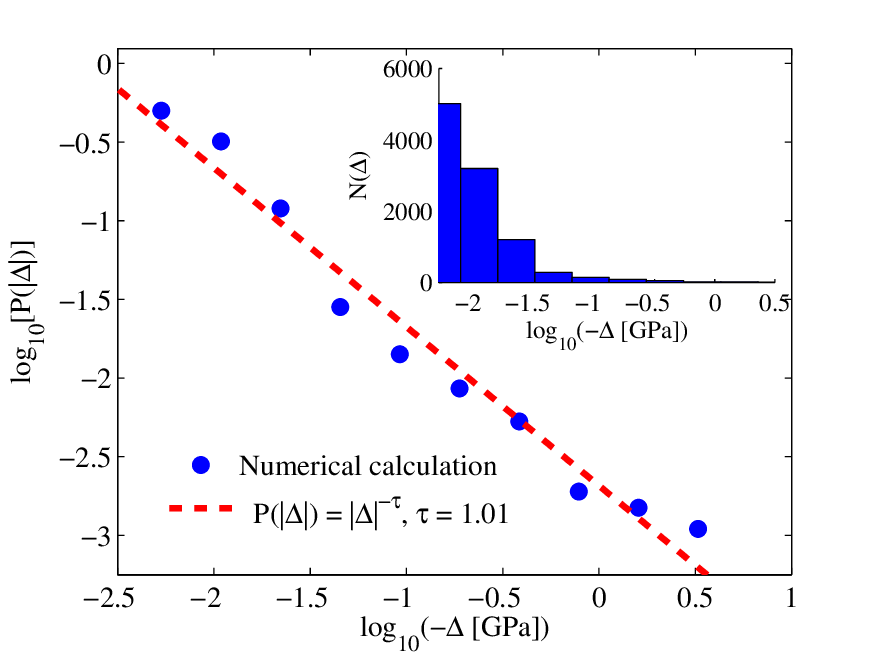}
}
\end{center}
\caption{Probability for energy-density drop $\Delta$ in a certain range of energy-density-drop magnitudes versus the middles of those ranges, on a log-log scale, with a linear-fit (red dashed line).  The total number of avalanches in the power-law tail is 10000 (obtained form 20 simulations), equal to the number of computational nodes, 10 boxes are used in the histogram, with a minimum of 10 avalanches in each box (the obtained exponent is locally insensitive to a change in the binning)}
\label{FigPowerLaw}
\end{figure}

The nearly 1000 energy drops shown in Fig. \ref{FigProfs} come in a broad range  of magnitudes. In Fig. \ref{FigPowerLaw} we quantify these fluctuations by showing (using a log-log scale) the probability distribution of $\Delta \langle\psi\rangle$ to which we refer simply as $\Delta $ (the inset shows the actual  histogram), based on a total of around 10000 relatively large avalanches (from the high-energy tail of the distribution), obtained from 20 different runs, each with a different sample of quenched boundary disorder (with the same statistical parameters). The obtained distribution   spans  more than 2.5 decades; small and large scale cut-offs are not shown.  A power law  fit $P(\Delta) \sim \Delta^{-\tau}$ with exponent  $\tau=1.01$ produces a good  approximation. It is again suggestive of the successful self-organization of the system towards scale free NESS reminiscent of developed turbulence  (see \cite{BaggioST2023b} for additional arguments supporting this analogy). Note that the value of the power law exponent  $\tau=1$ is a signature of the archetypically ``wild'' plasticity in the sense of \cite{Weiss2015}.

It is interesting that almost the same value of the power-law exponent ($\tau \approx 1$) was obtained in discrete dislocation dynamics (DDD) studies, which exhibited steady plastic flow in numerical experiments conducted on disorder-free 2D samples containing a fixed number of preexisting dislocations \cite{Alava,Alava2}. The exponent $\tau\approx  1$ was also recorded in numerical experiments involving pristine crystals conducted using the scalar version of the MTM \cite{Peng}. In the literature on crystal plasticity the exponent $\tau=1$ is usually linked  with either dislocation jamming or self-induced glassiness \cite{Ovaska,Lehtinen,XZhang,Ruscher}. The reason is that this value of the exponent has previously emerged in a fully analytical mean-field theory of spin glasses, where it was associated with marginal stability \cite{Pazmandi,Doussal,Franz}. Based on this analogy, we can argue that for low quenched disorder our system self-generates, through the first system-size avalanche, a `mature' self-induced disorder, which appears to be bringing the system from a stable (elastic) to a marginally stable
 (glassy) state. The latter is known to emerge as a result of hierarchical (ultrametric) self-organization in the phase space  
\cite{Berthier2019,Muller2015}. Using the same argument one can rationalize the fact that the exponent $\tau\approx 1$ has also been recorded in numerical studies of quasi-elastic flow regimes in structural glasses \cite{Shang,Tyukodi,Ferrero}.  
More generally, the exponent $\tau=1$ can be viewed as an indicator of the fact that in the course of self-organization the system has posited itself at the threshold of stability. Consequently, this value of the exponent can be seen as a marker of robust criticality, observable over a range of parameter values. Such criticality is  different from the more conventional tuned criticality, occuring, for instance, when the system exhibits an isolated critical point in the parameter space \cite{Christensen2005,Sornette2006}.

\begin{figure}[h!]
\begin{center}
{
\includegraphics[scale=0.28 ,trim={0cm 2.2cm 0cm 1cm},clip]{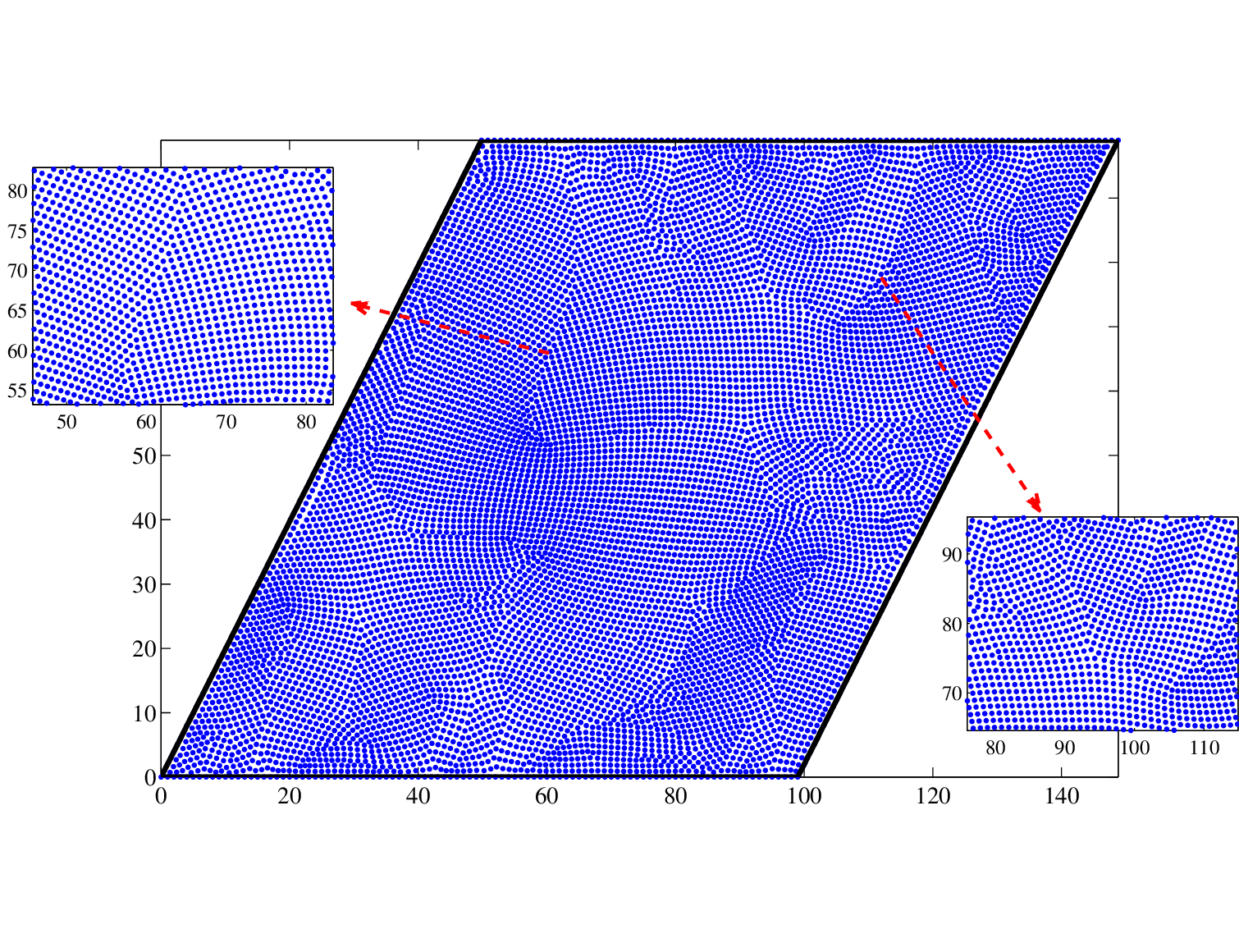}
\includegraphics[scale=0.28 ,trim={0cm 0.7cm 0cm 0cm},clip]{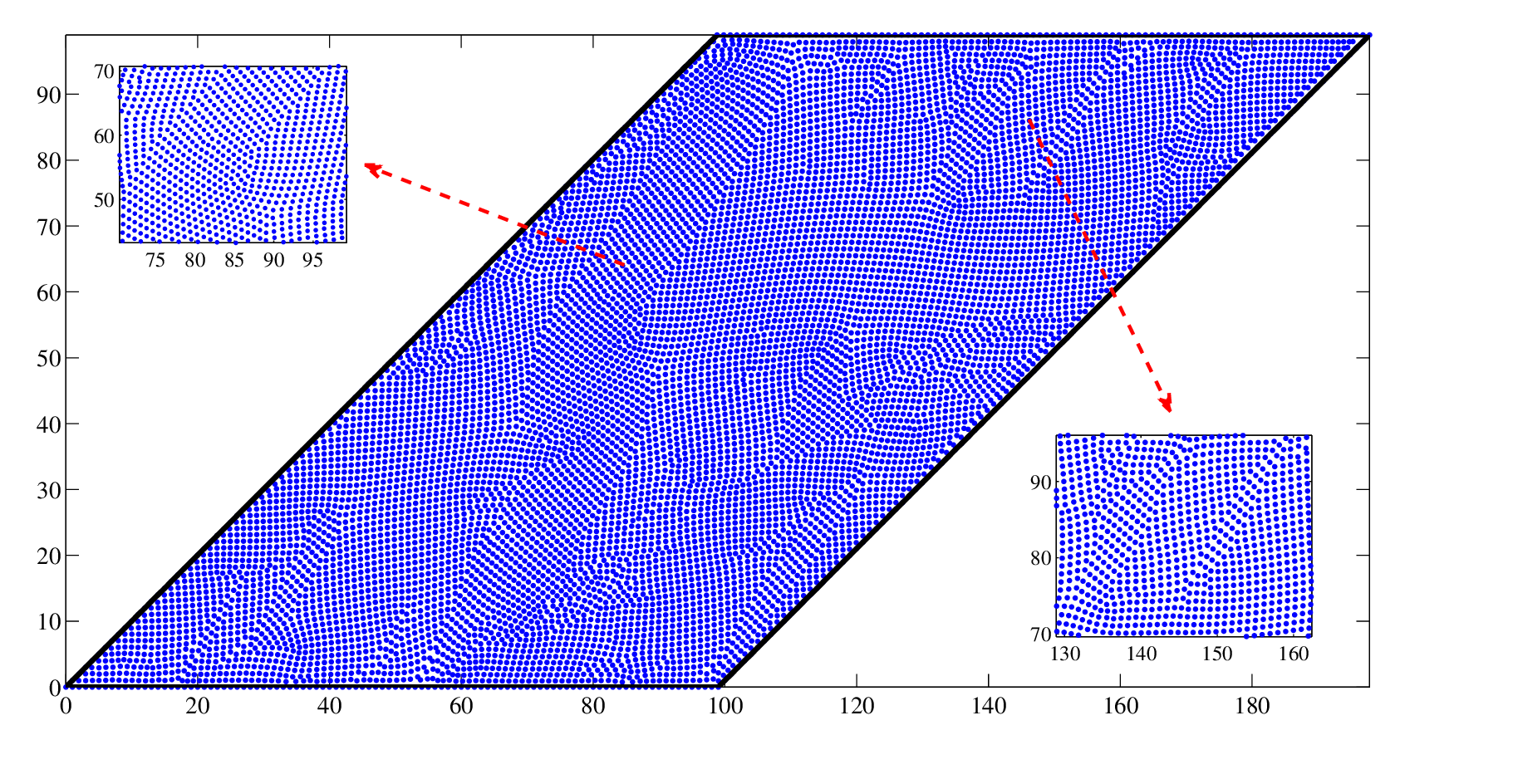}
}
 \begin{picture}(0,0)
\put(-137,-5){${(b)}$ }
\put(-362,-5){${(a)}$ }
  \end{picture}
\end{center}
\caption{\label{FigUs} (a) The positions of the nodes of the computational Lagrangian grid right after the first system-size avalanche showing the emergence of a multigrain, polycrystal-type pattern; (b) the configuration of the nodes at the final state of deformation, showing the emergence of shear bands.
}
\end{figure}
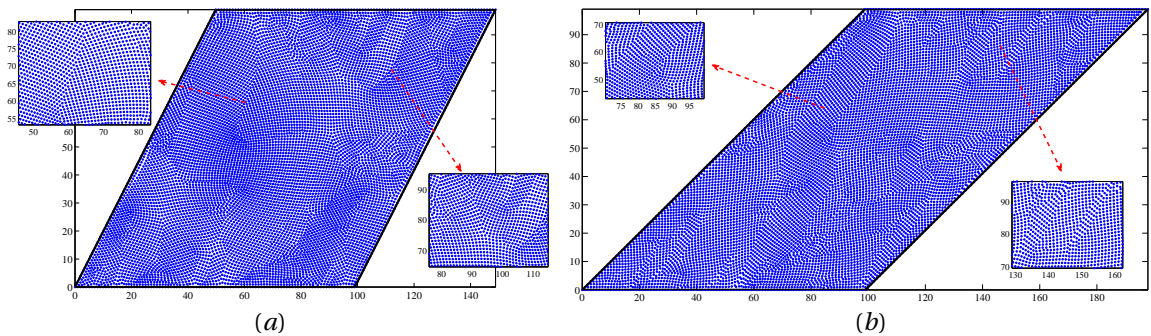

\paragraph{Spatial correlations}  Our numerical experiments also allow one to visualize the emerging dislocation configurations and trace their development along the various stages of the overall deformation. In  Fig. \ref{FigUs} (a) we show the configuration of the computational nodes right after the occurrence of the first system-size avalanche taking place at $\alpha=0.5$. 

On can see again that as a result of the symmetry-breaking system-size plastic event, the affine state is destroyed through massive collective dislocations nucleation. Numerous successive slip events, developing in the fast time during {this} avalanche, transform a pristine crystal into a mixture of apparently randomly oriented grains. Each grain represents an almost unloaded (but rotated) version of the original square lattice and different grains are separated by dislocations-rich grain boundaries which carry almost all the elastic energy \cite{BaggioST2023b}. 

As an  example, consider the two large grains in the center of Fig. \ref{FigUs} (a), which are separated by an almost straight high-angle grain boundary {(}its crystallographic nature will be explored in detail in a separate paper). More complex patterns develop near the  edges of the square domain, which serve as sources of  strain concentration and contribute to heterogeneous nucleation of dislocations \cite{BaggioST2023a}. The typical deformed configuration in the regime of developed plastic yield, $0.5<\alpha<1.0$, is shown in Figure \ref{FigUs}(b), which corresponds to $\alpha=1$. It suggests that the post system-size avalanche grain structure evolves through  smaller avalanches, increasing the complexity but not producing systematic configuration refinement. Some of the grains develop into elongated bands inside which the lattice retains its nearly-square structure. The bands are separated from the rest of the lattice by dislocation-rich zones whose inner structure is finely tuned to the misorientation of the neighboring lattice patches. Overall, the size distribution of grain sizes remains broad and is characterized by long power-law tails. The apparently scale-free plastic flow maintains its statistical signature despite the presence of stress-hardening and recurrent local restructuring. 

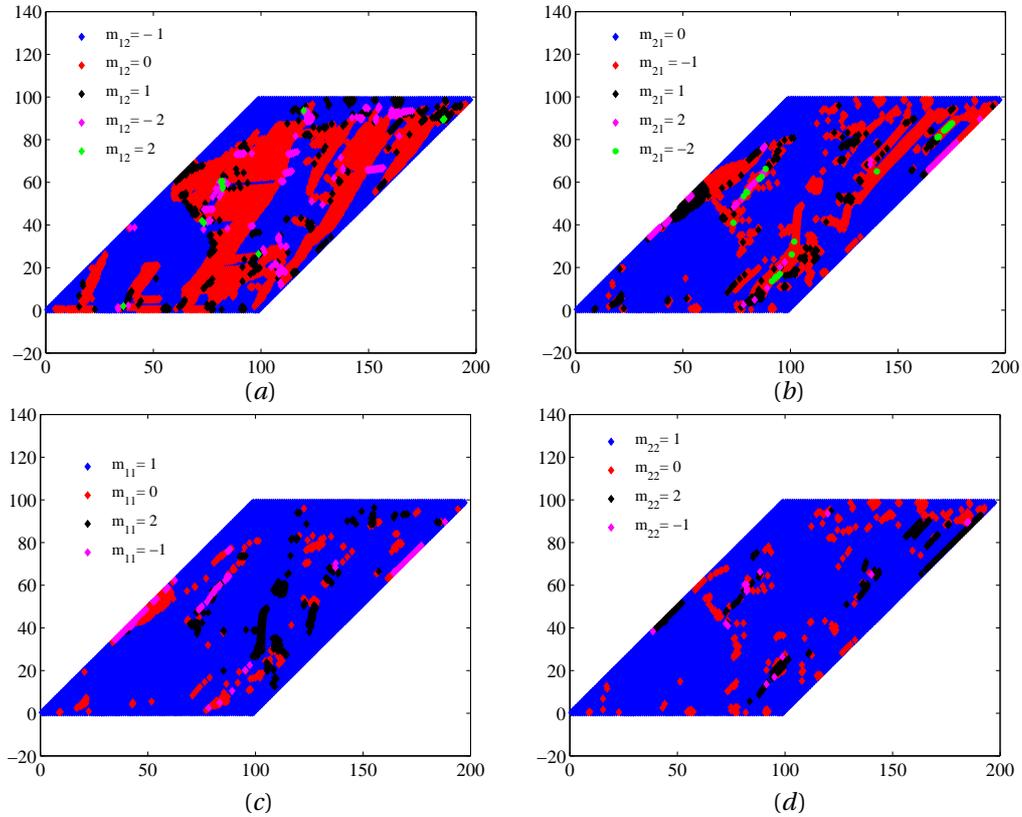
\begin{figure}[h!]
\begin{center}
{
\includegraphics[scale=0.5,trim={1cm 0.5cm 0cm 0cm},clip]{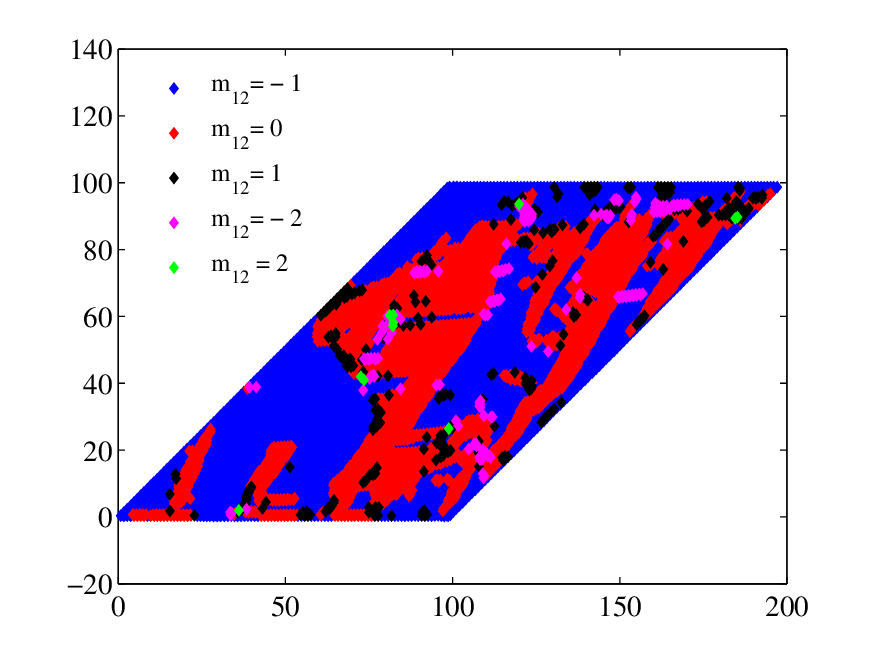}
\includegraphics[scale=0.5,trim={1cm 0.5cm 0cm 0cm},clip]{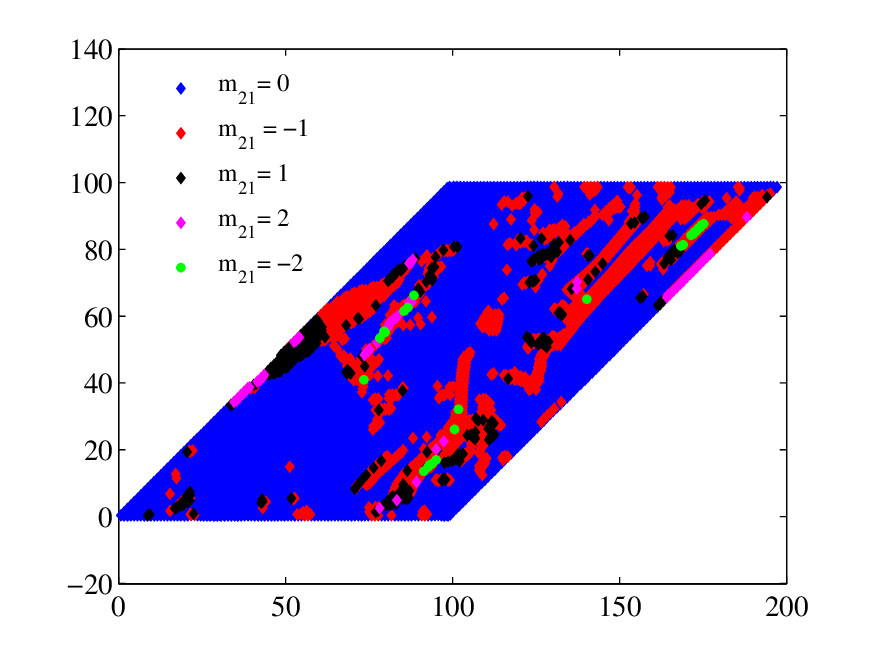}
\includegraphics[scale=0.5,trim={1cm 0.5cm 0cm 0cm},clip]{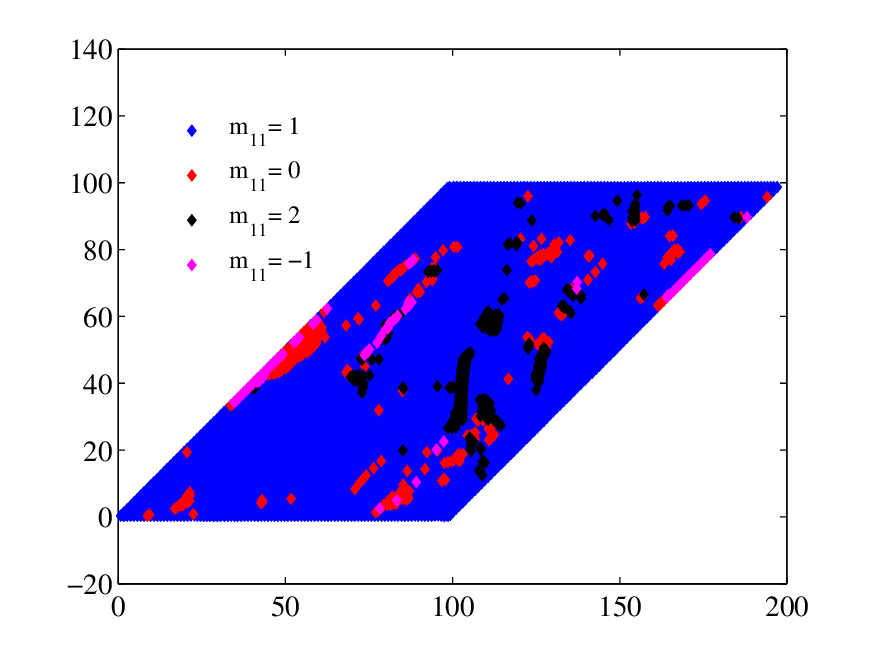}
\includegraphics[scale=0.5,trim={1cm 0.5cm 0cm 0cm},clip]{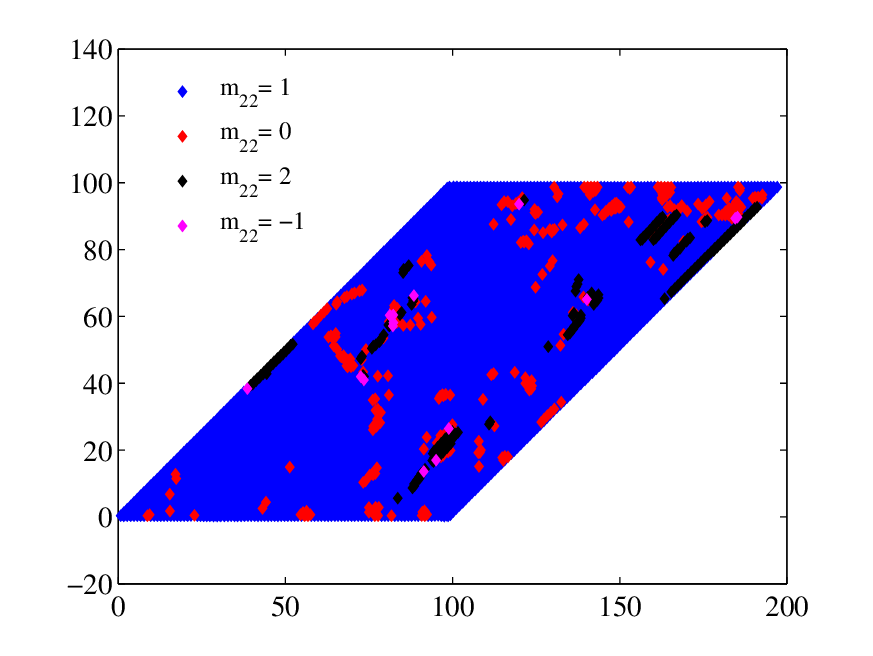}
}
 \begin{picture}(0,0)
 \put(-308,145){$(a)$ }
   \put(-110,145){$(b)$ }
  \put(-308,-10){$(c)$ }
   \put(-110,-10){$(d)$ }
  \end{picture}
\end{center}
\caption{\label{Fig2slips1E} Maps of the four different components of the plastic strain  measure \textbf{m}, namely, 
(a)  $m_{12}$, (b) $m_{21}$, (c) $m_{11}$, and (d) $m_{22}$. }
\end{figure}

Additional insight can be gained from  the maps showing the distribution of the plastic strain measure   $\textbf{m} (\textbf{x})$   at the final  state where $\alpha=1$, see Fig.  \ref{Fig2slips1E}. We observe that the dominant plasticity mode is the  horizontal simple shear aligned with the loading. However, such single-slip-system type plasticity affects  only about   three quarters of the elements, see Fig.  \ref{Fig2slips1E}(a).  In view of  self induced disorder, amplified by the system size avalanche, the second main slip system is activated as well, see Fig.  \ref{Fig2slips1E}(b). Non-simple shear modes are activated too, but only within very few elements, see Fig.  \ref{Fig2slips1E}(c,d).

A meaningful measure of the observed spatial complexity emerges as we look at the occurrence of various types of plastic-slip events (increments of plastic deformation), focusing on the distribution of individual elements that have experienced {a} particular number of slips.



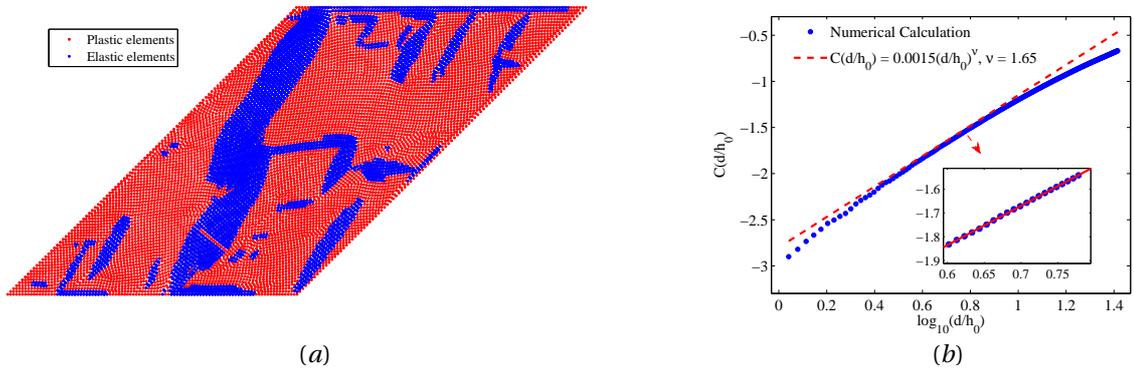
\begin{figure}[h!]
\begin{center}
{
\includegraphics[scale=0.36 ,trim={3cm 3.5cm 2cm 0cm},clip]{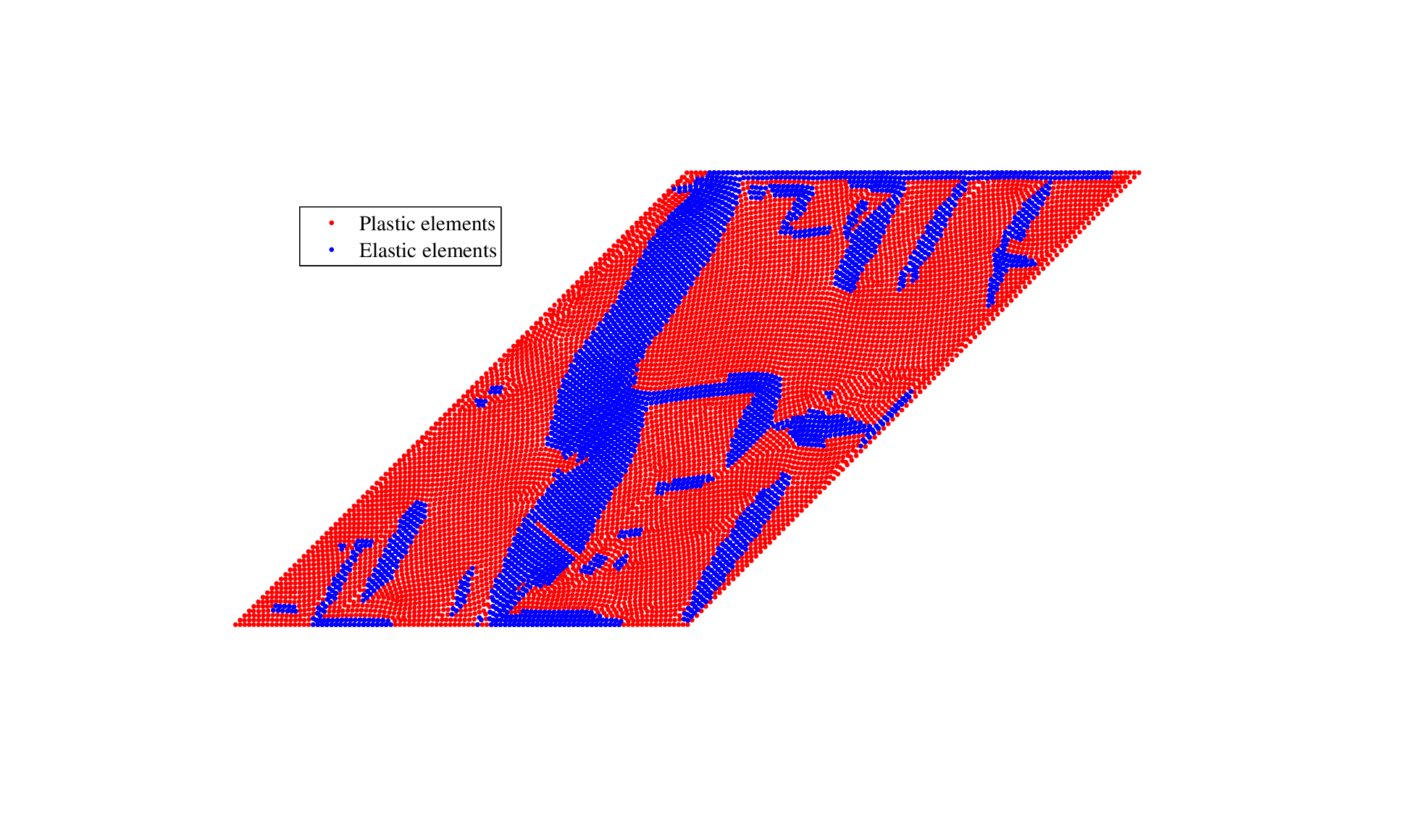}
\includegraphics[scale=0.412,trim={0cm 0cm 1cm 0cm},clip]{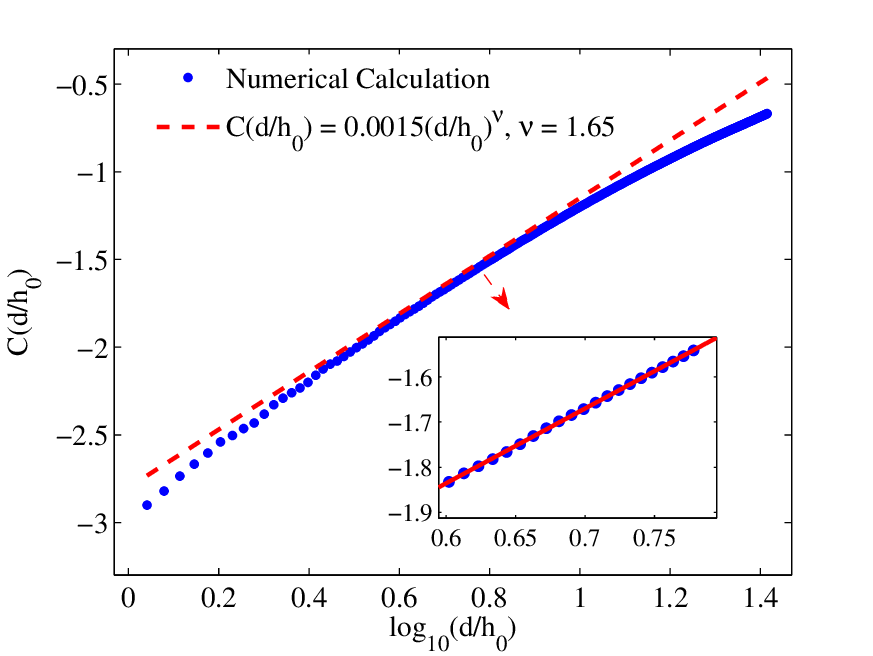}
}
 \begin{picture}(0,0)
  \put(-320,-10){$(a)$ }
   \put(-83,-10){$(b)$ }
  \end{picture}
\end{center}
\caption{\label{Fig2slips1} 
(a)  The centers of the elastic elements that had experienced no plastic slip during the deformation (blue dots) and the centers of elastic elements that have experienced at least one plastic slip (red dots) in the final state; (b) a log-log plot showing the  correlation function (blue dots) along with a linear fit (red dash) revealing the value of the fractal dimension of the distribution of zero-slip elements, $\nu=1.65$.}
\end{figure}
 
Our numerical experiments show that by the final stage of deformation (100\% geometric shear strain),  about 4500 out of almost 20000 elements have experienced no plastic slip at all. This means that plasticity have spread over most the sample and most of the elements have experienced at least one slip event. The analysis also shows that at the end ($\alpha=1$) only a small number of elements have experienced two or more consecutive plastic slip events.

These observations suggest that a relevant measure of the emerging spatial complexity (at this stage of the deformation) can be obtained if we focus on elements that have experienced no slip. Those are  illustrated in Fig. \ref{Fig2slips1}(a), where we show the spatial distribution of those 4500 elements{,} which
  exhibits a considerable degree of clusterization. The latter can be quantified if we evaluate the corresponding spatial correlation:     
 \begin{equation}
 C(\zeta) \triangleq \frac{1}{N_0(N_0-1)} \sum\limits_{j=1}^{N_0}\sum\limits_{k\neq j}^{N_0} \mathcal{H}\left(\zeta-|\textbf{x}_j-\textbf{x}_k|\right), 
 \label{eqCI1}
 \end{equation}
where $\mathcal{H}(\cdot)$ is the Heaviside step function, $\textbf{x}_i$ are the coordinates of the  zero-slip elements and  $N_{0}$ is the total number of elements in the computational domain, which experienced no plastic slip. A log-log plot of the function in \eqref{eqCI1}, revealing the emerging correlations, is presented in Fig. \ref{Fig2slips1}(b).  The correlation dimension is computed from the slope inside the range where the correlation distance $d$ is small enough for the  chosen measure to be meaningful, but large enough with respect to the discretization resolution (between 3 to 6 finite-element side lengths). The computed value of the fractal dimension,  $\nu=1.65$, lies  very close to the range [1.66,1.82] observed in the scalar model  \cite{Umut2011}. It also falls inside the range [1.62,1.68] obtained in DDD analysis of FCC crystals \cite{Zaiser1998}.  The obtained evidence for scale-free spatial fluctuations  is, again, indicative of complex hierarchical self-organization behavior, at least in the intermediate range of scales, bounded on one side by the size of the system and on the other side by the cut-off scale.  

\begin{figure}[h!]
\begin{center}
\includegraphics[scale=0.33]{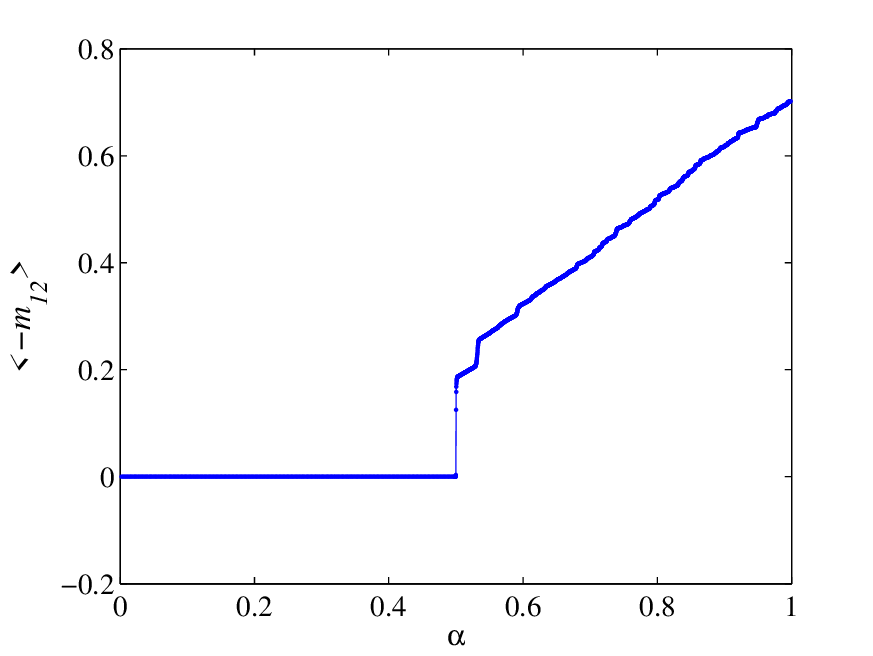}
\includegraphics[scale=0.315]{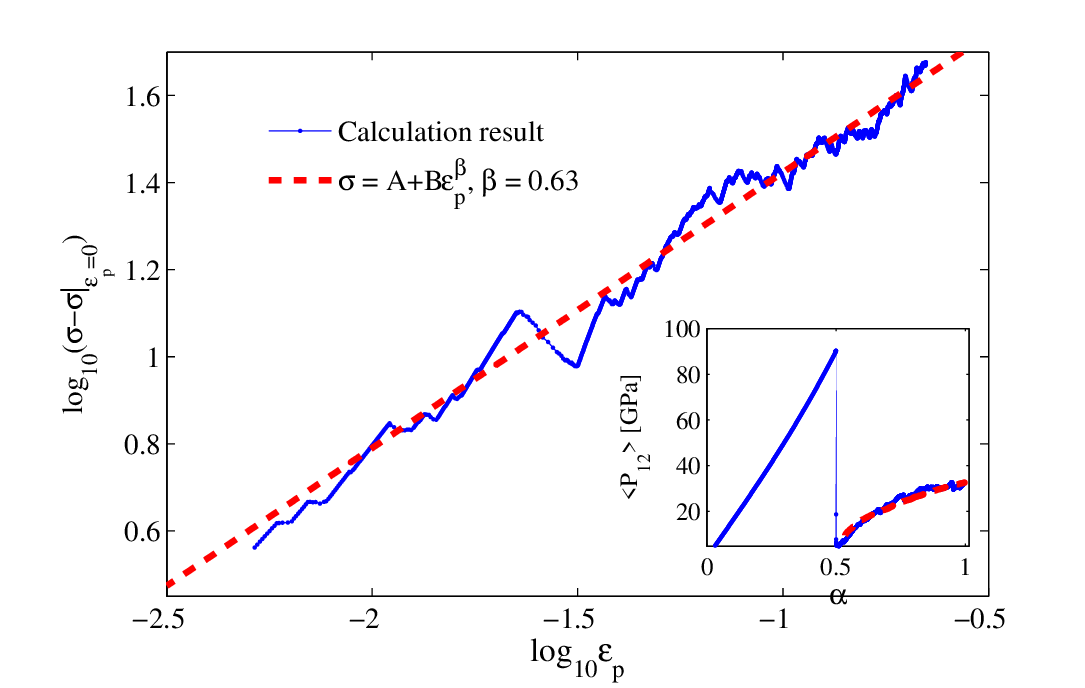}
\includegraphics[scale=0.328]{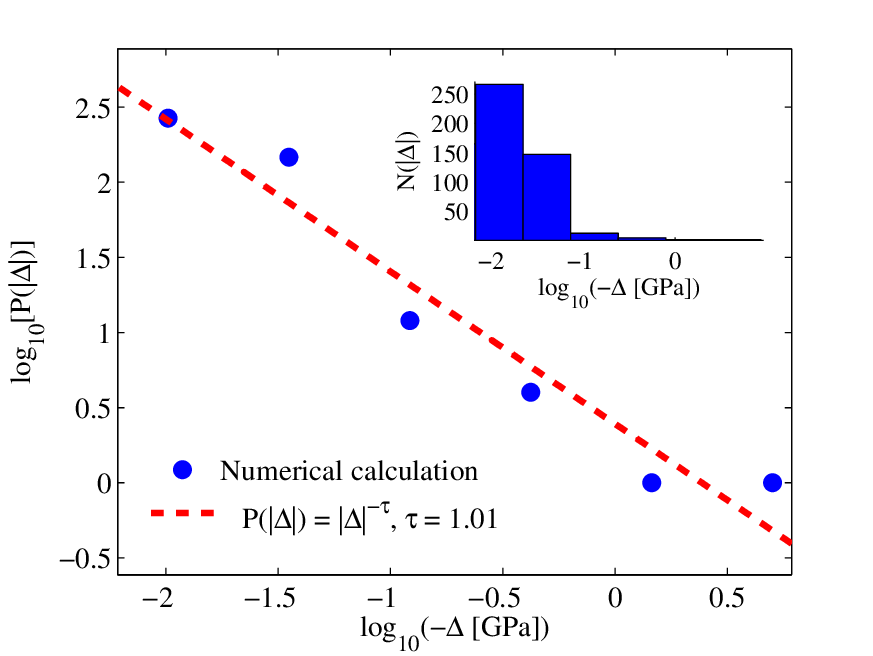}
   \begin{picture}(0,0)
  \put(-381,-10){$(a)$}
   \put(-224,-10){$(b)$}
  \put(-75,-10){$(c)$ }
  \end{picture}
\end{center}
{\caption{\label{FigProfs502} Selected computational results for a smaller system, with $N=50\times50$ nodes: (a) coarse-grained plasticity measure shear component versus applied macroscopic strain $\alpha$,
(b) macroscopic effective stress--small plastic strain relation on a log-log scale, showing almost two decades of power-law behavior with an exponent coinciding with the experimentally measured one. The inset shows the macroscopic stress--strain curve with the power-law approximation for the hardening part after the system size avalanche; (c) Probability distribution of energy-density drops $\Delta$ on a log-log scale with a linear fit (red dashed line), showing the exponent of 1.01, exactly like in the $N=100\times100$ case.
}}
\end{figure}
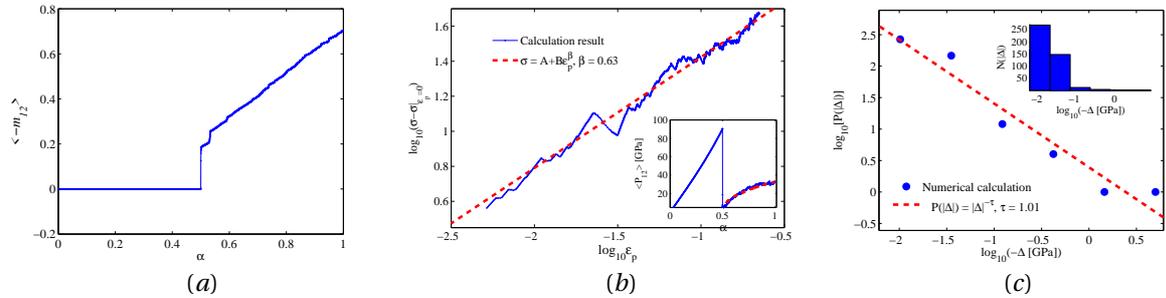

\paragraph{Different system size}  To show that the main statistical characteristics of the plastic flow, identified in this study,  are not affected by the system size, we present below some results of numerical experiments conducted for a smaller system, with $50\times50$ nodes, but with the same value of the internal length scale, $h_0$, see  Fig. \ref{FigProfs502}. We observe  that while the statistical characteristics are expectedly less pronounced, the coarse-grained quantities and the fluctuation-statistics exponents computed for this smaller system agree fairly well with the ones obtained as shown above for the larger system with the $100\times100$ nodes. Indeed, in both cases the pre and post yield coarse-grained energy is around 20-25GPa and 5-10GPa, respectively, and the maximum post yield coarse-grained stress is around 30-40GPa. Similarly, in both cases the initial post-yield and the final coarse-grained geometric plastic shear strain is around 0.2 and 0.7, respectively. The plastic-strain hardening exponent for both system sizes is in the range 0.63--0.66 and the avalanche sizes statistics is characterized by exactly the same tail exponent of 1.01.  Similarly, the correlation dimension of the elastic zones pattern is size-independent, remaining in both cases at the value 1.65. It is indeed rather remarkable that all the aforementioned information can be obtained from a comparatively small system, with only $50\times50$ nodes, using an algorithm that runs for only half an hour on a standard $i$7 desktop computer. 

Needless to say that if instead of decreasing the system size we had increased it, the statistical quality of the results would have improved considerably, albeit at the price of considerable increase of the computational cost.




\section{Conclusions}
\label{sec6}

The paper offers  a  fundamental  re-formulation of the  MTM (mesoscopic tensorial model),    a novel mesoscopic computational approach  to  crystal plasticity. The unique innovating nature  of  the MTM  is in operating with the macro-scale notions of stress and strain  while carrying some essential elements of the micro-scale physics of crystal lattices.  
 
The MTM represents a crystal as an interacting collection of homogeneously deforming meso-scopic elastic elements whose nonlinear   response is governed by a globally periodic elastic potential. The  periodicity is designed   to  represent   the full tensorial symmetry of Bravais lattices. On one side, the MTM  can be viewed as a discretized continuum theory of  the Landau type, with a tensorial strain as the order parameter and with  lattice-invariant shears representing the  minima of the equivalent energy wells. On the other side, it is expected to mimic the  friction-type dissipation, which in conventional continuum theories of plasticity is  associated  phenomenolgically with active  plastic mechanisms. Instead of being postulated, the implied dissipative structure   emerges  in the MTM naturally as an outcome of  a homogenized description of a succession of instabilities experienced by an  overdamped system on  a rugged energy landscape. 

Similarly to the Ginzburg-Landau theory, the MTM carries a regularizing parameter, which is, however, not associated with a gradient term in the energy, but rather arrives as a length scale representing the finite size of the mesoscopic elastic elements. This length scale  describes  the physical cut-off distance below which deformation is considered homogeneous (affine).  It is viewed as a physically informed internal parameter and  should not be confused with the mesh-size scale characterizing  conventional discretization approximations.

An important aspect of the MTM is the precise account of geometric nonlinearity. The  advantage of incorporating finite-strain kinematics into crystal  plasticity theory is the possibility    to adequately distinguish between different crystal-invariant shears while accounting  faithfully  for finite rotations. It is rather remarkable  that those traditional aspects of nonlinear elasticity  ended up being so crucial for  plasticity theory.  

A highly promising feature of the MTM as a candidate to emerge as  a fundamental and essentially  irreducible mesoscopic  description of crystal plasticity, is that it is  practically free of phenomenological assumptions. Even though it relies  only on the specifics of the interatomic potential, which is in principle sufficient for reconstructing the energy landscape, its global periodicity and the corresponding organization of the energy wells in the configurational space of metric tensors is universal, as it only depends on the symmetry of the crystal lattice. The precise nature of these equivalent wells and the value of the cut-off length scale appear to be irrelevant for the statistical structure of the resulting plastic flows. In this respect the MTM can be interpreted as the analogue of the Navier Stokes equations for crystal plasticity, carrying in the simplest form the essential  nonlinearity, which turns out to be purely geometric.

The principal conceptual advance of the present paper in the context of the MTM framework  is the \emph{explicit} introduction of plastic strain. In its original formulation, the MTM is effectively a discretized nonlinear elasticity theory where plastic strain appears only \emph{implicitly}. In the proposed re-formulation of the theory, plastic strain enters the model directly, in the form of a locally-defined integer-valued uni-modular matrix. A particular group-structure of the set of such matrices provides a geometrically-precise and crystalographically-specific description of lattice-invariant shears.  

 
The main goal  of the proposed re-formulation is  to build  a  conceptual bridge between the MTM and the continuum  CP theory. In the latter, the explicit distinction between elastic and plastic contributions to deformation allows one to effectively separate the conservative elastic prediction problem from the dissipative plastic correction problem. Moreover, in the CP, the conventional assumption  that the elasticity is linear  allows one to solve the elastic  problem  `on the fly', while  delegating all the complexity  to   a separate  nonlinear `condensed' problem describing  the evolution of the plastic strain. With the present paper, the same possibility  becomes available  inside the   MTM framework. Furthermore,   given the quantized  nature of  plastic deformation in the reformulated MTM, one can, in principle, relegate the evolution of plastic strain to a discrete automaton by minimizing out the elastic fields.  In such an automaton, the driven overdamped   dynamics in a rugged energy landscape would be effectively represented  by a succession of discontinuous advances of quantized plastic strain. As we show, however, the complete separation of the elastic problem in the full geometrically nonlinear setting is problematic and, therefore, in the general case one can expect to formulate the reduced plastic problem at best as a quasi-automaton. 


  
To reveal the inner mechanism of the re-formulated MTM, we opened the paper with a presentation of the  fully analytical study of a zero-dimensional prototypical `purely elastic' model of the depinning type. The mathematical transparency of such a model reveals why the evolution of the strain field in the MTM under quasi-static loading necessarily leads to a succession of quasi-elastic continuous stages interrupted by instantaneous quasi-plastic avalanches. It also explains how in the continuum limit, by accumulating infinitely many infinitely small avalanches of this type, one can obtain the characteristic  rate-independent dissipation postulated in CP phenomenologically. We show that the analysis of this toy model naturally leads to {the} idea of separation between the elastic and plastic components of the deformation. Even more importantly, it provides the intuitive rationale for the ultimate quantization of plastic strain. Needless to say that, in view of its oversimplified nature, the zero-dimensional model remain{s}, at most, suggestive. In particular, it is unable to predict either the spatial or the temporal complexity of the observed plastic avalanches. It suggests then that  a fully  geometrically-faithful analogue should be developed and studied to capture the critical nature of plastic flows in crystals. The goal of the present paper was to show that the proposed re-formulation of the MTM is perfectly capable  to serve this purpose. 


The idea of solving, inside the MTM framework, {the} elastic and plastic problems separately was implemented in the new numerical algorithm detailed in the paper. An important technical challenge was to speedup the original, `purely elastic', numerical algorithm, currently used in various applications of the MTM. From the computational perspective, the acceleration was achieved as a result of the transition from a soft-spin to a hard-spin formulation, which became possible when the description of the purely elastic part of the model was  maximally simplified. 

The main outcome of the proposed re-formulation of the MTM was  an effective enslavement of continuous elastic deformation to the discrete evolution of the matrix-valued spin-field. While the proposed algorithm can indeed be viewed as producing a sequence of load-driven updates of such quantized plastic strain, as we have already mentioned, the geometric nonlinearity of the model implies that each of the updates necessarily contains an embedded implicit nonlinear energy-minimization step. In the adopted setting this step is practically straightforward due to the close to quadratic variation of the chosen elastic energy functional inside its periodicity domain. 

Nevertheless, the solution still cannot be expressed in terms of an explicit `elastic propagator' and therefore, as we have already explained,  one can qualify the resulting discrete algorithm  only as a   quasi-automaton. In fact, we made it almost an automaton, since, with the chosen elastic energy function,   a single Gauss-Newton minimization step (before plastic correction) is  almost always sufficient for the convergence of the elastic problem. After a plastic correction a single (quickest-descent) Cauchy step is employed for the same purpose, which provides significant speedup, as a Cauchy step allows performing only local updates of the displacement field in each quasi-automaton iteration.  

An important  part of the paper is dedicated to the detailed analysis of the   observations   obtained in a set of numerical experiments.  In these experiments, in which we studied the mechanical response of pure 2D crystals with square symmetry, subjected to simple shear{,} the main target was the emerging complexity resulting from the collective evolution of a large number of dislocations. The obtained results suggest that the proposed  version of the MTM is capable of adequately reproducing the statistical structure of both temporal and spatial plastic fluctuations observed in  several other independent numerical studies. This  task is notoriously challenging due to the complex interplay in such a co-evolution of  short and long range interactions among individual dislocations. The latter  is what ultimately  lies behind self-organization towards hierarchical structures and  the associated intermittent dynamics.

Finally, we should mention that the use of the accelerated numerical approach was crucial for acquiring large-enough statistics to have convincing evidence that plastic fluctuations are indeed scale-free in a sufficient range of scales. The achieved computational speedup now paves the way towards using the MTM for the modeling of the distinctions between plastic fluctuations in 3D crystals with different crystallographic symmetries (e.g. HCP, FCC and BCC) and obtaining in this way the quantitative predictions that can lend direct comparison with physical experiments.

\begin{section}{Acknowledgements}

The authors are grateful to S. Conti and G. Zanzotto for helpful discussions at the initial stage of the project. Constructive practical suggestions by U. Salman  given during the course of the project are deeply appreciated. We also thank  S. Patinet for  a careful reading of the manuscript. The authors appreciate the Agence National de la Recherche (ANR), grant No. 17-CE08-0047, for the financial support.
\end{section}

\appendix

\renewcommand\thefigure{A.\arabic{figure}}
\setcounter{figure}{0}

\section{Numerical algorithm}
\label{AppendixA}

As described in the main text, at each increment of the loading the energy-minimization problem for {$\textbf{U}_{j}$ at the nodes} and  $\textbf{m}_i$ {in the elements} is solved in two steps. 

At the first step, 
the elastic response is assumed to be linear, described by {the} tangent moduli calculated based on the solution at the previous loading increment; the set of plastic strains $\textbf{m}_i$ at this step is kept equal to the stabilized value from the previous loading increment. The implied first-order Taylor-series expansion for the gradient of the energy in nodal displacements can be presented  in the form: 
$$
\left. d\bar\Psi/d\bar{\textbf{U}}\right|_{\bar{\textbf{U}}^*}=\left. d\bar\Psi/ d\bar{\textbf{U}}\right|_{\bar{\textbf{U}}_l}+ \mathring{\textbf{H}}^{(l)} (\bar{\textbf{U}}^*- \bar{\textbf{U}}_l),
$$
where $l$ denotes the previous loading increment for which a stable solution is known 
and 
 $$\mathring{\textbf{H}}^{(l)}=\left.\frac{\partial^2\bar\Psi}{\partial\bar{\textbf{U}}\partial\bar{\textbf{U}}^{\top}}\right|_{\bar{\textbf{U}}^{\circ}_l}.$$ The ensuing quadratic energy minimization problem reduces to the solution of a standard linear system of equations 
\begin{equation}
\begin{split}
\mathring{\textbf{H}}^{(l)}(\Delta\bar{\textbf{U}}^*-\Delta\bar{\textbf{U}}_b)=\textbf{b}_l, \ \textbf{b}_l\triangleq -\textbf{g}_l-\mathring{\textbf{H}}^{(l)}\Delta\bar{\textbf{U}}_b, \ \Delta\bar{\textbf{U}}_*\triangleq \bar{\textbf{U}}^*- \bar{\textbf{U}}_l,
\textbf{g}_l\triangleq \left. d\bar\Psi/ d\bar{\textbf{U}}\right|_{\bar{\textbf{U}}_l}.
\end{split}
\label{eqInc2}
\end{equation}
The crucial issue here is the location of the point at which the Hessian is calculated.

%
Recall that for a function $\bar\Psi(\bar{\textbf{U}})$ whose Hessian matrix is semi-positive-definite (SPD) everywhere, the solution of the linear system \eqref{eqInc2} always minimizes the energy. However, an objective energy density functional in principle cannot be convex everywhere and thus the Hessian of the elastic energy would in general have {some} negative eigenvalues.  In such {cases}, Newton's method with its first-order Taylor approximation would not necessarily bring the state closer to a solution of the equilibrium equations.  
  
To remedy the problem we use a version of the Gauss-Newton {approach} \cite{GaussNewton}, which in general implies omitting second derivatives of a quadratic form of a set of nonconvex functions of one variable. In our case this means that instead of the local Hessian we would use a the Hessian calculated at the bottom of the closest energy well, where such second derivatives vanish automatically. 

To be more specific, since under this choice one has $J=1$ and $C^e_{11}-C^e_{22}=C^e_{12}=0$, the non-SPD {contributions to the Hessian} are effectively eliminated. {Moreover}, since the Hessian is symmetric, the linear equations representing mechanical equilibrium can be solved by using the conjugate gradient (CG) method \cite{ConjGrad}, with its optimal convergence characteristics.  

Note, however, that in view of  the objectivity{-}induced degeneracy of the bottom of the elastic energy well, one must specify how exactly the approximate  Hessian should be calculated. Recall first that elastic distortion can be written in the form $\textbf{F}_e=\textbf{F}\textbf{m}$. Using  polar decomposition we can then write  $\textbf{F}\textbf{m}=\textbf{R}_e\textbf{U}_e=\textbf{R}_e \textbf{C}_e^{1/2} $. 
 The implied elastic rotation is thus  $\textbf{R}_e=\textbf{F}\textbf{m}\textbf{C}_e^{-1/2}$.
In our {numerical} algorithm we choose the Hessian to be calculated at the location ({on} the orbit representing the bottom of the energy well) corresponding to the current value of the plasticity measure  $\textbf{m}$ and to the current value of the elastic rotation $\textbf{R}_e$. In other words, 
 the state at which the Gauss-Newton update is calculated, is chosen to be $$\mathring{\textbf{F}}
 =\textbf{R}_e\textbf{F}_p$$ which gives for the Hessian:
\begin{equation}
\mathring{\textbf{H}}=\textbf{H}|_{\textbf{F}=\mathring{\textbf{F}}}, \ \mathring{\textbf{F}}= \textbf{F}\textbf{m} \textbf{C}_e^{-1/2}\textbf{m}^{-1}.  
\label{eqHessGN}
\end{equation}
The corresponding value of $\bar{\textbf{U}}^{\circ}_l$ is {then} defined through $$\mathring{\textbf{F}}_{(i)}=\underset{j}{\sum}{\mathbb{D}_{ij}\bar{\textbf{U}}^{\circ,(j)}_l},$$ where $i$ denotes an element number and $j$ denotes a node number. 
Similarly, the contributions of the nodes to the Hessian matrix  is given by $$\mathring{\textbf{H}}_{(j,j')}^{(l)}=\underset{i,i'}{\sum}{V_i\mathbb{D}^{\top}_{ij}{\mathring{\textbf{Q}}}^{(i,i')}_l}\mathbb{D}_{i'j'}.$$ 
While now  equations (\ref{eqInc2}) are well-posed at the internal nodes, at the boundary nodes the left-hand side may still vanish.
This is corrected  by using the augmented entries: 
\begin{equation}
\begin{split}
\bar{\textbf{b}}_l\triangleq \begin{cases} \textbf{b}_l, \textbf{x} \in \Omega \\ 0, \textbf{x} \in \partial \Omega \end{cases}, \ \bar{{H}}^{(l)}_{ij}\triangleq \begin{cases} \mathring{H}^{(l)}_{ij}, \textbf{x} \in \Omega \\ 0, \textbf{x} \in \partial \Omega, \ i \neq j \\ 1, \ \textbf{x}\in \partial{\Omega}, \ i=j \end{cases}
\end{split}
\label{eqInc3}
\end{equation}
where $\Omega$ and $\partial\Omega$ are the sets of internal and boundary nodes, respectively. Then the  elastic predictor can be written in the form
$$
\bar{\textbf{U}}^*=\bar{\textbf{U}}_l+\Delta\bar{\textbf{U}}_b+\bar{\textbf{H}}_{l}^{-1}\bar{\textbf{b}}_l
,$$  where the entries $\Delta\bar{\textbf{U}}_b$ are presumed to be parametrized  through $ \alpha$.

{At the second step of the algorithm, plastic correction} to the predicted displacement-field-increment is sought in a fixed-boundary setting. The predictor step should generally be large enough to drive some elements out of the elastic {domain}. It is then that the energy is first minimized further by the operation of plastic reduction, during which $\textbf{m}$ is updated for elements not complying with \eqref{eq37c3}.
After this initial plastic reduction, the elastic problem is solved again for the nodes associated with the elements having just undergone active plastic reduction. The solution of this elastic problem consists in updating the displacement increment for the aforementioned nodes using a Cauchy maximum-descent step. Then, again, plastic reduction is performed for (some of) the elements associated with the nodes the displacement increment of which was just updated.
After the yield conditions \eqref{eq37c3} have been satisfied for all the elements, final equilibrization is attained by a Gauss-Newton predictor step from the subsequent loading increment, which provides convergence for small-enough loading increments. The efficiency of the Gauss-Newton--Cauchy solver owes in part to the fact the energy-density functional as chosen for the crystal is close enough to a quadratic one, and any distinctly nonquadratic features (local nonconvexity) are circumvented by the calculation of the Hessian at the bottoms of the energy wells.

To give more details, in the chosen piecewise smooth setting, during the plastic correction steps, the elements have to travel through `ridges' of the energy landscape, occasionally reaching  states where one cannot rely on tangent moduli. In such cases our numerical code activates a first-order Cauchy optimization algorithm \cite{Cauchy}. The corresponding gradient of the energy is pre-multiplied by a (positive) scalar (step-size), which is chosen to ensure the largest possible energy decrease. 
 The corresponding algorithm can be written in the following form:  
\begin{equation}
\begin{split}
\bar{\textbf{U}}^{(k=1)}_{l+1}=\bar{\textbf{U}}^* , \ \bar{\textbf{U}}^{(k+1)}_{l+1}= \ \begin{cases} \bar{\textbf{U}}^{(k)}_{l+1}+ \textbf{d}^{(k)}_{l+1}, \textbf{x} \in \Omega \\ \bar{\textbf{U}}^{(k)}_{l+1}, \textbf{x} \in \partial \Omega \end{cases} , \ \textbf{d}^{(k)}_{l+1}=-\Delta t^{(k)}_{l+1}{\textbf{g}}^{(k)}_{l+1}.
\end{split}
\label{eqInc5}
\end{equation}
It should be noted that a positive step-size producing a decrease in the energy always exists. Indeed, for a small-enough positive value of the step-size, one can always use a second-order Taylor-series expansion to obtain
\begin{equation}
\begin{split}
\hat\Psi^{(k+1)}_{l+1}\to \hat\Psi^{(k)}_{l+1}+\left[{\textbf{g}}^{(k)}_{l+1}\right]^{\top} \left[\hat{\textbf{U}}^{(k+1)}_{l+1}-\hat{\textbf{U}}^{(k)}_{l+1}\right]+\frac{1}{2} \left[\hat{\textbf{U}}^{(k+1)}_{l+1}-\hat{\textbf{U}}^{(k)}_{l+1}\right]^{\top}\bar{\textbf{H}}^{(k)}_{l+1}\left[\hat{\textbf{U}}^{(k+1)}_{l+1}-\hat{\textbf{U}}^{(k)}_{l+1}\right]=\\=\hat\Psi^{(k)}_{l+1}-\Delta t^{(k)}_{l+1}\left|{\textbf{g}}^{(k)}_{l+1}\right|^2+\frac{1}{2}\left[\Delta t^{(k)}_{l+1}\right]^2\left[{\textbf{g}}^{(k)}_{l+1}\right]^{\top}\bar{\textbf{H}}^{(k)}_{l+1}{\textbf{g}}^{(k)}_{l+1}<\hat\Psi^{(k)}_{l+1} \\ \forall \ \  0<\Delta t^{(k)}_{l+1}\ll \Delta t_{cr}^{(k,l)}=\frac{\left|{\textbf{g}}^{(k)}_{l+1}\right|^2}{\frac{1}{2}\left|\left[{\textbf{g}}^{(k)}_{l+1}\right]^{\top}\bar{\textbf{H}}^{(k)}_{l+1}{\textbf{g}}^{(k)}_{l+1}\right|}
\end{split}
\label{eqInc6}
\end{equation}
In order to determine  the right  step-size to employ in Eq. (\ref{eqInc5}), an iterative process has to be used, with sub-iterations $q$. During those sub-iterations, the gradient $\textbf{g}^{(k)}_{l+1}$ is kept fixed, and only the step size decreases, starting from a specified initial value. During the sub-iterations, the total elastic free energy $^{(q)}\hat\Psi^{(k+1)}_{l+1}$ is calculated until it becomes smaller than $\hat\Psi^{(k)}_{l+1}$. As in the Armijo-Goldstein algorithm, an exponentially decreasing step-size is chosen, following the formula:
\begin{equation}
\begin{split}
^{(q)}\Delta t^{(k)}_{l+1}=\frac{\Delta t_{cr}^{(k',l')}}{2^q}, \ \ q = \underset{q'}{\text{argmin}}^{(q')}\hat\Psi^{(k+1)}_{l+1}, \ k'\in[1,k], \ l'\in[1,l]
\end{split}
\label{eqInc7}
\end{equation}
where the choice of $k',l'$ can generally be optimized for, with fewer sub-iterations for $k'=k,l'=l$ and fewer matrix multiplications for $k'=l'=1$ (which was used in the case study in this work).

\begin{figure*}[htbp]
\centering
\centerline{
\includegraphics[scale=0.9,trim={0cm 15cm 0cm 2cm},clip]{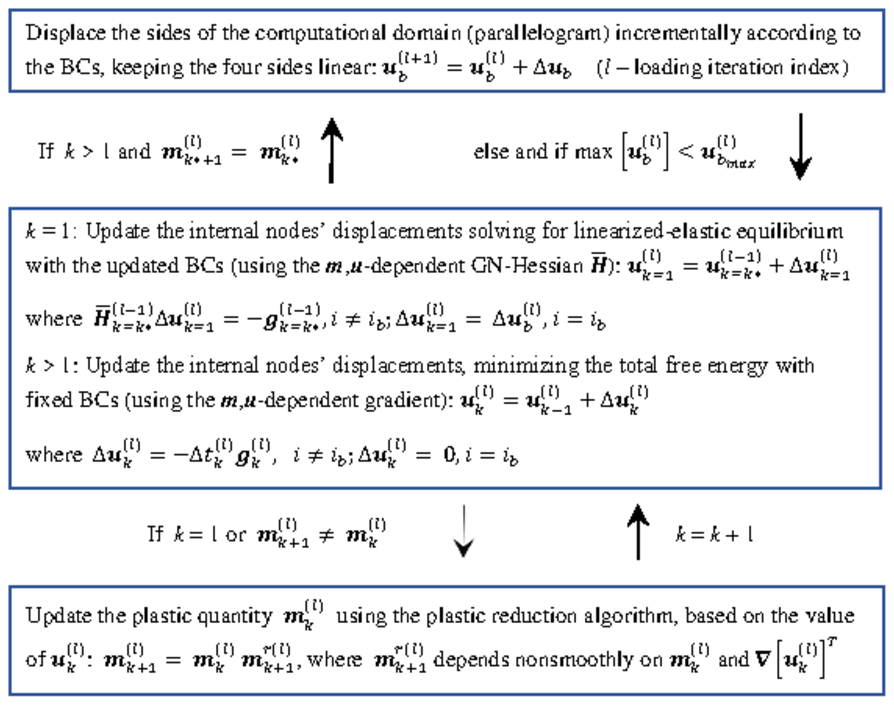}}
\caption{A flow chart of the solution algorithm}
\label{FigFlow}
\end{figure*}

Our Fig. \ref{FigFlow} presents the detailed flow chart of the resulting solution algorithm (save for the details of the step-size determining procedure).

To summarize, in the proposed  numerical approach for the corrector stage we first decrease the energy while  tolerating possible exit from the `elastic' domain for some elements. Only after the energy is minimized  elastically and the maximal energy-decreasing step-size is found, the plastic reduction is performed, further reducing  the energy. Then, another Cauchy step is attempted and if it does not drive any element out of the FED, the iterative procedure is considered  terminated. The  consecutive loading increment is then applied.

The described basic first-order optimization scheme can be considered efficiently accelerated in the following sense. The evaluation of the gradient vector is computer-time-consuming, as it requires a loop over all the elements, which adds linear complexity to the algorithm. A significant boost in computational efficiency (run-time) was  obtained by avoiding performing a loop on all the elements. Thus, we are  taking advantage of the fact that  the  plastic-corrector Cauchy-step can be activated only at the nodes  shared by elements in which $\textbf{m}$ has changed in the previous iteration. Similar observations concern the computation of the total elastic energy
which has to be recomputed for every sub-iteration  for the determination of the step size.
In the proposed accelerated approach the update for the energy in each iteration was computed by adding   the increment in a loop  only over the elements that have just undergone a plastic slip. 



For the convenience of the readers we finish this section by giving the explicit expressions for the four components of the Piola-Kirchhoff stress tensor (calculated at a given ${\textbf{F}}$), which were used to compute the residual nodal forces, and further the components of the gradient of the total energy (Cauchy directions,  ${\textbf{g}}_{(j)}^{(l)}=\underset{i}{\sum}{V_i\mathbb{D}^{\top}_{ij}{{\textbf{P}}}^{(i)}_l}$):  
 
\begin{equation}
\begin{split}
P_{11}=\kappa F_{22}(F_{11}F_{22}-F_{12}F_{21}-1)+ \\+\frac{\xi }{4}\left[(m_{11}^2-m_{12}^2)(F_{11}^2+F_{21}^2)+\right. 2(m_{11}m_{21}-m_{12}m_{22})(F_{11}F_{12}+F_{21}F_{22})+\\ \left.+(m_{21}^2-m_{22}^2)(F_{12}^2+F_{22}^2)\right]\left[2F_{11}(m_{11}^2-m_{12}^2) \right.  \left.+2F_{12}(m_{11}m_{21}-m_{12}m_{22})\right]+\\+\frac{\eta }{4}\left [2m_{11}m_{12}(F_{11}^2+F_{21}^2)+\right. 2(m_{12}m_{21}+m_{11}m_{22})(F_{11}F_{12}+F_{21}F_{22})+\\ \left.+2m_{21}m_{22}(F_{12}^2+F_{22}^2)\right]\left[4F_{11m_{11}m_{12}}+\right.  \left. 2F_{12}(m_{12}m_{21}+m_{11}m_{22})\right]
\end{split}
\label{eq21mFd1}
\end{equation}
\begin{equation}
\begin{split}
P_{12}=-\kappa F_{21}(F_{11}F_{22}-F_{12}F_{21}-1)+ \\+\frac{\xi }{4}\left[(m_{11}^2-m_{12}^2)(F_{11}^2+F_{21}^2)\right.  +2(m_{11}m_{21}-m_{12}m_{22})(F_{11}F_{12}+F_{21}F_{22})+\\ \left.+(m_{21}^2-m_{22}^2)(F_{12}^2+F_{22}^2)\right]\left[2F_{12}(m_{21}^2-m_{22}^2) \right.  \left.+2F_{11}(m_{11}m_{21}-m_{12}m_{22})\right]+\\+\frac{\eta }{4}\left [2m_{11}m_{12}(F_{11}^2+F_{21}^2)+\right.2(m_{12}m_{21}+m_{11}m_{22})(F_{11}F_{12}+F_{21}F_{22})+\\ \left.+2m_{21}m_{22}(F_{12}^2+F_{22}^2)\right]\left[4F_{12}m_{21}m_{22}+\right.  \left. 2F_{11}(m_{12}m_{21}+m_{11}m_{22})\right]
\end{split}
\label{eq21mFd2}
\end{equation}

\begin{equation}
\begin{split}
P_{21}=-\kappa F_{12}(F_{11}F_{22}-F_{12}F_{21}-1)+ \\+\frac{\xi }{4}\left[(m_{11}^2-m_{12}^2)(F_{11}^2+F_{21}^2)+\right. 2(m_{11}m_{21}-m_{12}m_{22})(F_{11}F_{12}+F_{21}F_{22})+\\ \left.+(m_{21}^2-m_{22}^2)(F_{12}^2+F_{22}^2)\right]\left[2F_{21}(m_{11}^2-m_{12}^2)\right.  \left.+2F_{22}(m_{11}m_{21}-m_{12}m_{22})\right]+\\+\frac{\eta }{4}\left [2m_{11}m_{12}(F_{11}^2+F_{21}^2)+\right.2(m_{12}m_{21}+m_{11}m_{22})(F_{11}F_{12}+F_{21}F_{22})+\\ \left.+2m_{21}m_{22}(F_{12}^2+F_{22}^2)\right]\left[4F_{21}m_{11}m_{12}\right. \left.+2F_{22}(m_{12}m_{21}+m_{11}m_{22})\right]
\end{split}
\label{eq21mFd3}
\end{equation}

\begin{equation}
\begin{split}
P_{22}=\kappa F_{11}(F_{11}F_{22}-F_{12}F_{21}-1)+ \\+\frac{\xi }{4}\left[(m_{11}^2-m_{12}^2)(F_{11}^2+F_{21}^2)+\right. 2(m_{11}m_{21}-m_{12}m_{22})(F_{11}F_{12}+F_{21}F_{22})+\\ \left.+(m_{21}^2-m_{22}^2)(F_{12}^2+F_{22}^2)\right]\left[2F_{22}(m_{21}^2-m_{22}^2)\right. \left.+2F_{21}(m_{11}m_{21}-m_{12}m_{22})\right]+\\+\frac{\eta }{4}\left [2m_{11}m_{12}(F_{11}^2+F_{21}^2)+\right. 2(m_{12}m_{21}+m_{11}m_{22})(F_{11}F_{12}+F_{21}F_{22})+\\ \left.+2m_{21}m_{22}(F_{12}^2+F_{22}^2)\right]\left[4F_{22}m_{21}m_{22}\right.\left.+2F_{21}(m_{12}m_{21}+m_{11}m_{22})\right]
\end{split}
\label{eq21mFd4}
\end{equation}

\bigskip

We also present below explicit formulas for the ten second derivatives required for the calculation of the  Gauss-Newton Hessian matrix evaluated at  $\mathring{\textbf{F}}$:  

\begin{equation}
\begin{split}
\mathring{Q}_{11}=\kappa \mathring{F}_{22}^2+\xi \left[\mathring{F}_{11}(m_{11}^2-m_{12}^2) \right. \left.+\mathring{F}_{12}(m_{11}m_{21}-m_{12}m_{22})\right]^2
+\\+\frac{\xi }{2}(m_{11}^2-m_{12}^2)\left[(m_{11}^2-m_{12}^2)(\mathring{F}_{11}^2+\mathring{F}_{21}^2)+\right. 2(m_{11}m_{21}-m_{12}m_{22})(\mathring{F}_{11}\mathring{F}_{12}+\mathring{F}_{21}\mathring{F}_{22})+\\ \left.+(m_{21}^2-m_{22}^2)(\mathring{F}_{12}^2+\mathring{F}_{22}^2)\right]+2\eta m_{11}m_{12}\left [m_{11}m_{12}(\mathring{F}_{11}^2+\mathring{F}_{21}^2)+\right. (m_{12}m_{21}+m_{11}m_{22})(\mathring{F}_{11}\mathring{F}_{12}+\mathring{F}_{21}\mathring{F}_{22})+\\ \left.+m_{21}m_{22}(\mathring{F}_{12}^2+\mathring{F}_{22}^2)\right]
+\eta \left[2\mathring{F}_{11}m_{11}m_{12}\right.  \left. +\mathring{F}_{12}(m_{12}m_{21}+m_{11}m_{22})\right]^2
\end{split}
\label{eq21mFd21111}
\end{equation}

\begin{equation}
\begin{split}
\mathring{Q}_{12}=-\kappa \mathring{F}_{21}\mathring{F}_{22}+  \\ +\xi \left[\mathring{F}_{11}(m_{11}m_{21}-m_{12}m_{22})\right.  \left.+\mathring{F}_{12}(m_{21}^2-m_{22}^2)\right] \left[\mathring{F}_{11}(m_{11}^2-m_{12}^2) \right.  \left.+\mathring{F}_{12}(m_{11}m_{21}-m_{12}m_{22})\right]\\
+\frac{\xi }{2}\left[(m_{11}^2-m_{12}^2)(\mathring{F}_{11}^2+\mathring{F}_{21}^2)+\right. 2(m_{11}m_{21}-m_{12}m_{22})(\mathring{F}_{11}\mathring{F}_{12}+\mathring{F}_{21}\mathring{F}_{22})+\\+ \left.(m_{21}^2-m_{22}^2)(\mathring{F}_{12}^2+\mathring{F}_{22}^2)\right](m_{11}m_{21}-m_{12}m_{22})+
\\
+\eta\left [\mathring{F}_{11}(m_{12}m_{21}+m_{11}m_{22})+\right. \left. 2\mathring{F}_{12}m_{21}m_{22}\right]\left[2\mathring{F}_{11}m_{11}m_{12}+\right. \left. \mathring{F}_{12}(m_{12}m_{21}+m_{11}m_{22})\right]+
\\
+\eta\left [m_{11}m_{12}(\mathring{F}_{11}^2+\mathring{F}_{21}^2)+\right. (m_{12}m_{21}+m_{11}m_{22})(\mathring{F}_{11}\mathring{F}_{12}+\mathring{F}_{21}\mathring{F}_{22})+\\ \left. +m_{21}m_{22}(\mathring{F}_{12}^2+\mathring{F}_{22}^2)\right](m_{12}m_{21}+m_{11}m_{22})
\end{split}
\label{eq21mFd21112}
\end{equation}

\begin{equation}
\begin{split}
\mathring{Q}_{13}=-\kappa \mathring{F}_{12}\mathring{F}_{22}+ \xi\left[\mathring{F}_{21}(m_{11}^2-m_{12}^2)\right. \\ \left.+\mathring{F}_{22}(m_{11}m_{21}-m_{12}m_{22})\right]\left[\mathring{F}_{11}(m_{11}^2-m_{12}^2) \right. \left.+\mathring{F}_{12}(m_{11}m_{21}-m_{12}m_{22})\right]+\\ +\eta \left [2\mathring{F}_{21}m_{11}m_{12}\right. \left.+\mathring{F}_{22}(m_{12}m_{21}+m_{11}m_{22})\right]\left[2\mathring{F}_{11}m_{11}m_{12}+\right.  \left. \mathring{F}_{12}(m_{12}m_{21}+m_{11}m_{22})\right]
\end{split}
\label{eq21mFd21121}
\end{equation}

\begin{equation}
\begin{split}
\mathring{Q}_{14}=\kappa (2 \mathring{F}_{11}\mathring{F}_{22} -\mathring{F}_{12}\mathring{F}_{21}-1)\\ +\xi\left[\mathring{F}_{21}(m_{11}m_{21}-m_{12}m_{22})+\mathring{F}_{22}(m_{21}^2-m_{22}^2)\right]\left[\mathring{F}_{11}(m_{11}^2-m_{12}^2)+ \mathring{F}_{12}(m_{11}m_{21}-m_{12}m_{22})\right]\\+\eta \left [\mathring{F}_{21}(m_{12}m_{21}+m_{11}m_{22})+2\mathring{F}_{22}m_{21}m_{22}\right] \left[2\mathring{F}_{11}m_{11}m_{12}+\mathring{F}_{12}(m_{12}m_{21}+m_{11}m_{22})\right]
\end{split}
\label{eq21mFd21122}
\end{equation}

\begin{equation}
\begin{split}
\mathring{Q}_{22}=\kappa \mathring{F}_{21}^2+ \xi \left[\mathring{F}_{12}(m_{21}^2-m_{22}^2) \right. \left.+\mathring{F}_{11}(m_{11}m_{21}-m_{12}m_{22})\right]^2+
 \\+\frac{\xi }{2}(m_{21}^2-m_{22}^2)\left[(m_{11}^2-m_{12}^2)(\mathring{F}_{11}^2+\mathring{F}_{21}^2)+\right. \\+2(m_{11}m_{21}-m_{12}m_{22})(\mathring{F}_{11}\mathring{F}_{12}+\mathring{F}_{21}\mathring{F}_{22}) \left.+(m_{21}^2-m_{22}^2)(\mathring{F}_{12}^2+\mathring{F}_{22}^2)\right]+
\\+\eta\left[2\mathring{F}_{12}m_{21}m_{22} +\mathring{F}_{11}(m_{12}m_{21}+m_{11}m_{22})\right]^2
+2\eta m_{21}m_{22}\left [m_{11}m_{12}(\mathring{F}_{11}^2+\mathring{F}_{21}^2)+\right.\\+(m_{12}m_{21}+m_{11}m_{22})(\mathring{F}_{11}\mathring{F}_{12}+\mathring{F}_{21}\mathring{F}_{22})+ \left. m_{21}m_{22}(\mathring{F}_{12}^2+\mathring{F}_{22}^2)\right]
\end{split}
\label{eq21mFd21212}
\end{equation}

\begin{equation}
\begin{split}
\mathring{Q}_{23}=-\kappa (\mathring{F}_{11}\mathring{F}_{22}-2\mathring{F}_{12}\mathring{F}_{21}-1)+ \\ +\xi \left[\mathring{F}_{21}(m_{11}^2-m_{12}^2)+\mathring{F}_{22}(m_{11}m_{21}-m_{12}m_{22})\right]  \left[\mathring{F}_{12}(m_{21}^2-m_{22}^2)+ \mathring{F}_{11}(m_{11}m_{21}-m_{12}m_{22})\right] \\+\eta\left [2\mathring{F}_{21}m_{11}m_{12}+\mathring{F}_{22}(m_{12}m_{21}+m_{11}m_{22})\right]  \left[2\mathring{F}_{12}m_{21}m_{22}+\mathring{F}_{11}(m_{12}m_{21}+m_{11}m_{22})\right]
\end{split}
\label{eq21mFd21221}
\end{equation}

\begin{equation}
\begin{split}
\mathring{Q}_{24}=-\kappa \mathring{F}_{11}\mathring{F}_{21}+ \\+\xi\left[\mathring{F}_{21}(m_{11}m_{21}-m_{12}m_{22})+\mathring{F}_{22}(m_{21}^2-m_{22}^2)\right] \left[\mathring{F}_{12}(m_{21}^2-m_{22}^2)+ \mathring{F}_{11}(m_{11}m_{21}-m_{12}m_{22})\right] \\+\eta\left [\mathring{F}_{21}(m_{12}m_{21}+m_{11}m_{22})+2\mathring{F}_{22}m_{21}m_{22}\right]  \left[2\mathring{F}_{12}m_{21}m_{22}+\mathring{F}_{11}(m_{12}m_{21}+m_{11}m_{22})\right]
\end{split}
\label{eq21mFd21222}
\end{equation}

\begin{equation}
\begin{split}
\mathring{Q}_{33}=\kappa \mathring{F}_{12}^2+\xi\left[\mathring{F}_{21}(m_{11}^2-m_{12}^2)\right. \left.+\mathring{F}_{22}(m_{11}m_{21}-m_{12}m_{22})\right]^2+
\\+\frac{\xi }{2}(m_{11}^2-m_{12}^2)\left[(m_{11}^2-m_{12}^2)(\mathring{F}_{11}^2+\mathring{F}_{21}^2)+\right. 2(m_{11}m_{21}-m_{12}m_{22})(\mathring{F}_{11}\mathring{F}_{12}+\mathring{F}_{21}\mathring{F}_{22})+\\ \left.+(m_{21}^2-m_{22}^2)(\mathring{F}_{12}^2+\mathring{F}_{22}^2)\right]+
\eta\left[2m_{11}m_{12}\mathring{F}_{21}\right. \left.+\mathring{F}_{22}(m_{12}m_{21}+m_{11}m_{22})\right]^2+
\\+2\eta m_{11}m_{12}\left [m_{11}m_{12}(\mathring{F}_{11}^2+\mathring{F}_{21}^2)+\right. (m_{12}m_{21}+m_{11}m_{22})(\mathring{F}_{11}\mathring{F}_{12}+\mathring{F}_{21}\mathring{F}_{22}) \left.+m_{21}m_{22}(\mathring{F}_{12}^2+\mathring{F}_{22}^2)\right]
\end{split}
\label{eq21mFd22121}
\end{equation}

\begin{equation}
\begin{split}
\mathring{Q}_{34}=-\kappa \mathring{F}_{11}\mathring{F}_{12}+ 
\\+\xi\left[\mathring{F}_{21}(m_{11}m_{21}-m_{12}m_{22})+\mathring{F}_{22}(m_{21}^2-m_{22}^2)\right] \left[\mathring{F}_{21}(m_{11}^2-m_{12}^2)+\mathring{F}_{22}(m_{11}m_{21}-m_{12}m_{22})\right]+
 \\+\frac{\xi }{2}(m_{11}m_{21}-m_{12}m_{22})\left[(m_{11}^2-m_{12}^2)(\mathring{F}_{11}^2+\mathring{F}_{21}^2)\right. +2(m_{11}m_{21}-m_{12}m_{22})(\mathring{F}_{11}\mathring{F}_{12}+\mathring{F}_{21}\mathring{F}_{22})+\\ \left.+(m_{21}^2-m_{22}^2)(\mathring{F}_{12}^2+\mathring{F}_{22}^2)\right]+
\\+\eta\left [\mathring{F}_{21}(m_{12}m_{21}+m_{11}m_{22})+2\mathring{F}_{22}m_{21}m_{22}\right] \left[2\mathring{F}_{21}m_{11}m_{12}+\mathring{F}_{22}(m_{12}m_{21}+m_{11}m_{22})\right]+
\\+\eta(m_{12}m_{21}+m_{11}m_{22})\left [m_{11}m_{12}(\mathring{F}_{11}^2+\mathring{F}_{21}^2)+\right. (m_{12}m_{21}+m_{11}m_{22})(\mathring{F}_{11}\mathring{F}_{12}+\mathring{F}_{21}\mathring{F}_{22})+\\ \left.+m_{21}m_{22}(\mathring{F}_{12}^2+\mathring{F}_{22}^2)\right]
\end{split}
\label{eq21mFd22122}
\end{equation}

\begin{equation}
\begin{split}
\mathring{Q}_{44}=\kappa \mathring{F}_{11}^2+\xi \left[\mathring{F}_{22}(m_{21}^2-m_{22}^2)\right.  \left.+\mathring{F}_{21}(m_{11}m_{21}-m_{12}m_{22})\right]^2+
\\+\frac{\xi }{2}(m_{21}^2-m_{22}^2)\left[(m_{11}^2-m_{12}^2)(\mathring{F}_{11}^2+\mathring{F}_{21}^2)+\right. 2(m_{11}m_{21}-m_{12}m_{22})(\mathring{F}_{11}\mathring{F}_{12}+\mathring{F}_{21}\mathring{F}_{22})+\\ \left.+(m_{21}^2-m_{22}^2)(\mathring{F}_{12}^2+\mathring{F}_{22}^2)\right]+
\eta\left[2\mathring{F}_{22}m_{21}m_{22}+\mathring{F}_{21}(m_{12}m_{21}+m_{11}m_{22})\right]^2+
\\+2\eta m_{21}m_{22}\left [m_{11}m_{12}(\mathring{F}_{11}^2+\mathring{F}_{21}^2)+\right. (m_{12}m_{21}+m_{11}m_{22})(\mathring{F}_{11}\mathring{F}_{12}+\mathring{F}_{21}\mathring{F}_{22}) \left.+m_{21}m_{22}(\mathring{F}_{12}^2+\mathring{F}_{22}^2)\right].
\end{split}
\label{eq21mFd22222}
\end{equation}

 \section{Algorithmic efficiency}
\label{AppendixB}
 
We begin with  the general  observation that, independently of the FE regularization, the main conceptual advantage of introducing the notion of plastic strain is the possibility to split a single incremental minimization problem of finding the deformation field $\textbf{y}(\textbf{x})$, in  a setting where 
  the associated globally periodic energy density \emph{is not rank-one convex}, into two problems: one, of finding an integer-valued field $\textbf{m}(\textbf{x})$  with det$(\textbf{m})=1$, and another one, of finding the continuous field $\textbf{y}(\textbf{x})$ whose metric tensor $\textbf{ C}^e (\textbf{x})$ (calculated based on $\textbf{m}(\textbf{x})$) is confined to the FED, where the energy \emph{is rank-one convex}.

While  the FE regularization  makes the problem well posed, the geometric nonlinearity, which cannot be neglected due to the ubiquitous presence of large rotations in elastoplastic flows \cite{BaggioST2023a, BaggioST2023b},  makes the numerical implementation of the whole algorithm nontrivial. More specifically, while one can schematically describe this algorithm  as a sequence of load-driven updates of the plastic variables $\textbf{m}$,   the apparent  discrete-automaton type  structure comes with a caveat that each update of $\textbf{m}$ contains an embedded elastic energy-minimization step.   Behind this incomplete separability of the elastic and plastic problems  is the unavoidable geometrical complexity of the tensorial elastic problem, whose  solution cannot be expressed in terms of an explicit `elastic propagator' as in the simpler scalar version of the MTM \cite{Umut2011,Umut2012,Zhangetal}. 

Still, since in the proposed setting the solution of the elastic problem  is manageable, due to  the close to quadratic behavior of the elastic energy inside the FED,  one can  favorably compare the algorithmic complexity of the proposed algorithm with that of the `purely elastic' algorithm used in \cite{Roberta,BaggioST2023a, SalmanBBZGT21,BaggioST2023b}. 

First, we note that in the `purely elastic' approach one has to rely on the computationally-expensive quasi-Newton solver, as the task is always to minimize a  highly non-convex function, which implies a large number of computational steps for each  increment of the loading parameter. To make an estimate of the computational cost associated with such an algorithm, we can assume that  the loading increment is $\Delta\alpha=10^{-6}$--$10^{-5}$ as in \cite{Roberta,BaggioST2023a, SalmanBBZGT21,BaggioST2023b}. Given that the adopted  quasi-Newton algorithm would have to deal  with  {t}ens of thousands of variables, no less than   $n_s=100$ solver steps would be usually needed \cite{NPJA20}. Since in each of these steps the gradient of the energy has to be calculated using a loop over all the elements, we obtain an estimate of around  $10^{7} aN $--$10^{8} aN $  operations, where $a$ is the number of basic operations needed to compute a single entry of the gradient and $N$ is the number of entries. The energy gradient has to be multiplied by the quasi-Newton matrix, which requires an update of the entire energy gradient with $pN$ operations where $p\sim n_s$. For $N=100\times 100=10^4$ this gives around $10^{13}a$--$10^{14}a$ operations along the   loading path  stretching up to 100\% geometric shear. 

Instead, using the proposed framework, we found empirically for the same type of problem, that the choice of the loading increment of $\Delta\alpha=2\times10^{-4}$ was sufficient. In each increment, at least for $N=100\times100$, the required total number of operations was around 1.5 times the number of operations in the corresponding linear elasticity problem (a single Gauss-Newton step for each increment) because the computational cost of plastic correction was approximately half of that of the elastic predictor. 
Note also that due to the chosen exponential decrease of the time-step with sub-iterations (say, when using line-search as in \cite{Armijo}), the additional computational time associated with such  sub-iterations is only logarithmic in the required number of sub-iterations, and  can be kept  bounded. Therefore, the overall gain in run-time due to the proposed accelerating measures is significant, and reaches  at least an order of magnitude, vis-a-vis the formal looping over all elements. 

Overall, we observed  (for the $N=100\times 100$ case) that the iterations associated with plastic corrections converged, on average, in about 30 function calls (automaton iterations plus step-size determining sub-iterations), with around 30 elements updated in each automaton step. This amounts to around $10^3 b$ operations for the plastic corrector stage, where $b \sim a $ is a constant characterizing the effort of  the computation of the Jacobian  and $a\approx 7$,  which translates to only about 20 microseconds on a simple 4-core \emph{i}7 desktop computer. An even more  detailed counting, which will be presented elsewhere,  suggests  that the proposed reformulation of the MTM framework can lead to an up to two orders of magnitude decrease in the computational time vis-a-vis  the more conventional, `purely elastic', MTM approach.



\end{document}